\documentclass[article]{aa}
\usepackage{graphicx}
\usepackage{mwe}
\usepackage{longtable}
\usepackage{siunitx}
\usepackage[table]{xcolor}
\usepackage{textcomp}
\usepackage{indentfirst} 

\usepackage{chemist}
\usepackage{multirow}
\usepackage{caption}
\usepackage[flushleft]{threeparttable}
\usepackage{subcaption}
\usepackage{float}
\usepackage{txfonts}
\titlerunning{}

\usepackage{natbib,twoopt}

\usepackage[colorlinks=true, pdfstartview=FitV, linkcolor=blue, citecolor=blue, urlcolor=blue, breaklinks=true]{hyperref} 
\bibpunct{(}{)}{;}{a}{}{,}             
\makeatletter
  \newcommandtwoopt{\citeads}[3][][]{\href{http://adsabs.harvard.edu/abs/#3}%
    {\def\hyper@linkstart##1##2{}%
     \let\hyper@linkend\@empty\citealp[#1][#2]{#3}}}
  \newcommandtwoopt{\citepads}[3][][]{\href{http://adsabs.harvard.edu/abs/#3}%
    {\def\hyper@linkstart##1##2{}%
     \let\hyper@linkend\@empty\citep[#1][#2]{#3}}}
  \newcommandtwoopt{\citetads}[3][][]{\href{http://adsabs.harvard.edu/abs/#3}%
    {\def\hyper@linkstart##1##2{}%
     \let\hyper@linkend\@empty\citet[#1][#2]{#3}}}
  \newcommandtwoopt{\citeyearads}[3][][]%
    {\href{http://adsabs.harvard.edu/abs/#3}
    {\def\hyper@linkstart##1##2{}%
     \let\hyper@linkend\@empty\citeyear[#1][#2]{#3}}}
\makeatother

\usepackage{color}

\usepackage{color}

\usepackage{cleveref}

\usepackage{chemist}
\usepackage{orcidlink}
\begin{document}
\title{Precise Fourier parameters of Cepheid Radial Velocity Curves\thanks{Full Tables \ref{Tab:data_ref}, \ref{Tab:data_funda_quality1}, \ref{Tab:data_funda_quality1a}, \ref{Tab:data_fourier_quality2}, \ref{Tab:data_fourier_quality1_overtone} and \ref{Tab:unknown} are only available at the CDS via anonymous ftp to cdsarc.u-strasbg.fr (130.79.128.5) or via
http://cdsarc.u-strasbg.fr/viz-bin/cat/J/A+A/}}

\titlerunning{Precise Fourier parameters of Cepheid Radial Velocity Curves}
\authorrunning{Hocd\'e et al. }
\author{V. Hocd\'e\inst{1}
\orcidlink{0000-0002-3643-0366}
\and P. Moskalik\inst{1}
\orcidlink{0000-0003-3142-0350}
\and N. A. Gorynya\inst{2,3}
\and R. Smolec\inst{1}
\orcidlink{0000-0001-7217-4884}
\and R. Singh Rathour\inst{1}
\orcidlink{0000-0002-7448-4285}
\and O. Ziółkowska\inst{1}
\orcidlink{0000-0002-0696-2839}
}

\institute{Nicolaus Copernicus Astronomical Centre, Polish Academy of Sciences, Bartycka 18, 00-716 Warszawa, Poland\\
email : \texttt{vhocde@camk.edu.pl}
\and Sternberg Astronomical Institute, Moscow State University, 119234, Moscow, Russia
\and Institute of Astronomy, Russian Academy of Sciences, 119017, Moscow, Russia
}

\date{Received ... ; accepted ...}
\abstract{Radial velocity (RV) curves of Classical Cepheids are indispensable in studying resonances between their pulsation modes. They allow precise determination of the resonant periods, which in turn helps to constrain fundamental parameters of these stars. The RV curves of Cepheids are also needed to identify their pulsation modes, which is essential for accurate calibration of the Cepheid period-luminosity (PL) relation. Finally, RV curves are necessary for the distance determination with the parallax-of-pulsation method.}
{The primary goal of this paper is to derive precise Fourier parameters of the radial velocity (RV) curves for fundamental and first-overtone Galactic Cepheids. The secondary objectives are then to analyze progression of the Fourier parameters up to the 7th harmonic as well as to propose identification of the pulsation modes of the stars.}{For each star, we carefully selected RV measurements available in the literature which yield the highest precision of Fourier parameters according to the procedure that follows. We performed a Fourier decomposition of the RV curves using unweighted least-square method. We corrected for zero-point differences between data sets. We subtracted RV modulations caused by binary motion and have removed other residual trends. Uncertainty of Fourier parameters are derived using the covariance matrix. Finally, we have displayed and analyzed qualitatively the progressions found for low- and for high-order Fourier parameters. We applied a standard identification of their pulsation mode based on their Fourier phase difference~$\phi_{21}$.} {We more than doubled the number of Cepheids with published Fourier parameters of their RV curves, determined together with their uncertainties. Our sample includes 178 fundamental-mode and 33 first overtone pulsators, as well as 7 additional Cepheids whose pulsation mode is uncertain or undetermined according to our criteria. For the fundamental-mode Cepheids, the precision of the obtained low-order Fourier phases is about 7 times better and of the low-order Fourier amplitude ratios about 25\% better as compared to precision achieved in previously published Fourier parameter surveys. The precision of the first-overtone Fourier parameters is improved by about 50\%. With highly accurate RV Fourier phases $\phi_{21}$, we are able to firmly identify V495 Cyg as a new first-overtone Cepheid. We confirm first-overtone nature of several other stars, tentatively identified as such with the shape of their lightcurves. We also show that $\alpha$ UMi should be definitely classified as a first-overtone pulsator. In 3 objects (VY Per, AQ Pup and QZ Nor) we found significant $\gamma$-velocity variations that we attribute to spectroscopic binarity after careful analysis RV measurements of these stars. Finally, the analysis of the fundamental mode Fourier parameters up to 7th order reveals very tight progression of Fourier phases for all pulsation periods. They are also tightly correlated with each other. For $P<10\,$day we find a very well defined upper limit for the Fourier amplitude $A_1$.}
{We have provided new precise Fourier parameters of Cepheid velocity curves, determined from RV measurements available in the literature together with unpublished data. The pulsation period coverage and the precision obtained, in particular for Fourier phase $\phi_{21}$, will be useful for studying the dynamics of Cepheid pulsations with the help of hydrodynamical models. Further radial velocity measurements from modern high-resolution spectroscopic instruments will be important to improve these results.}

\maketitle

\section{Introduction}\label{s_Introduction}
The shape of the radial velocity (RV) curve of Cepheids provides information on their stellar structure and dynamic. In the case of fundamental Cepheid light curves, \cite{Ludendorff1919,Hertzsprung1926} noticed that their shape varies systematically with the pulsation period. A secondary bump superimposed on the light curve appears first on its descending branch at a period of about $P\simeq 6\,$day. With increasing period, it moves toward maximum light, reaches the top at $P\simeq 10\,$day and then moves down the ascending branch at periods $P>10\,$day. This systematic behavior is called the \textit{Hertzsprung progression} or bump progression, and was later verified among Cepheids in the Magellanic Clouds and Andromeda Galaxy \citep{Parenago1936,Shapley1940,Payne-Gaposchkin1947}. Similar bump progression is observed for the RV curves \citep{Joy1937,LedouxWalraven1958,1983SimonTeays,kovacs90,Gorynya1998paper,Moskalik2000}.


The bump progression was first assumed to be caused by a pressure wave which propagates inward and reflects on the stellar core, then travels back to reach the surface and produces the observed bump \citep[also known as the echo model,][]{Whitney1956,christy68,Christy1975}.
Following hydrodynamical models of \cite{stobie69b}, \cite{SimonSchmidt1976} conjectured that the bump progression is due to the $P_2$/$P_0=0.5$ resonance between the fundamental and the second overtone pulsation modes characterized by a pulsation period of $P_0$ and $P_2$ respectively. Subsequently, the crucial role of the 2:1 resonance was confirmed with the extensive set of hydrodynamical models of \cite{Buchler1990,Moskalik1992}, and understood analytically with the amplitude equations formalism \citep{Buchler1986,kovacs1989}. Since the resonance is a dynamical phenomenon, RV curves are better suited for comparison with models than the light curves. Moreover, modelling light curves is subject to larger uncertainties because it is sensitive to the radiative transfer treatment in the outer layers of the model.
For example, RV curves of first-overtone Cepheids display a prominent resonance around 4.6$\,$day \citep{Kienzle1999}, in agreement with hydrodynamical models which predict a $P_4/P_1=0.5$ resonance between the fourth and the first overtone \citep{Kienzle1999,Feuchtinger2000}. In contrast, the light curves present a sharp feature at 3.2$\,$day which is not centered at the $P_4/P_1=0.5$ resonance \citep{Kienzle1999}. This demonstrates the importance of RV curves for studying dynamical effects.


\cite{SimonLee1981,1983SimonTeays} showed that the Fourier parameters can describe efficiently the shape of the light and RV curves. Therefore, they can be used to quantitatively compare the models and the observations. In particular, the Fourier phase $\phi_{21}$ of RV curve of the fundamental mode depends almost exclusively on the resonant period ratio $P_2$/$P_0$ and is numerically robust \citep{Buchler1990}. The same is true for first-overtone Cepheids, where the Fourier phase $\phi_{21}$ depends on $P_4$/$P_1$ \citep{Kienzle1999}. Thus, using observed RV curves it is possible to derive the period ratios for individual Cepheids, which in turn can constraint their internal structure as well as their Mass-Luminosity relation \citep{Moskalik1998}.

RV curves are also indispensable in the parallax-of-pulsation technique for individual distance determination of Cepheids \citep[also known as Baade-Wesselink (BW) method,][]{lindermann18,baade26,becker1940,wesselink46}. Although some authors emphasized the importance of homogeneous and consistent RV measurements for applying BW method 
\citep[see, for example][]{Anderson2018proceeding,borgniet2019}, in practice the community has to deal with multiple data set obtained from different instruments and cross-correlation measurement methods. Hence, a consistent method for fitting RV measurements of several data sets altogether would be valuable.

Therefore, obtaining precise Fourier parameters, and for a large sample of short and long-period Cepheids, is essential to reconstruct the Hertzsprung progression and explore it up to higher-order. Of paramount importance is the precision of observational $\phi_{21}$ from RV curves that not only impacts directly the error on the period ratio $P_2/P_1$ and $P_4/P_1$ \citep{Moskalik1998}, but is also a very useful tool to discriminate fundamental and first-overtone mode Cepheids, especially for pulsation period above 5$\,$days \citep{AntonelloPoretti1990}.

Owing to continuous effort of observers, the number of stars available to reconstruct the bump progression and the precision of Fourier parameters increased over time. Using a compilation of 9 different sources from the literature, \cite{1983SimonTeays} analyzed for the first time RV curves for 11 classical fundamental-mode and 1O-mode Cepheids. A compilation of Fourier parameter of 57 RV curves, with median uncertainties of $\sigma(\phi_{21})=0.18$, was later provided by \cite{kovacs90} who compiled RV data from 41 sources. \cite{Moskalik2000} also collected measurements for 131 Cepheids and reported a median uncertainty of $\sigma(\phi_{21})=0.08$. \cite{storm11a} provided Fourier parameters together with their uncertainty for 76 Galactic Cepheids using a compilation of 60 sources from the literature, providing an average uncertainty $\sigma(\phi_{21})=0.18$. In the case of first-overtone Cepheids, \cite{Kienzle1999} published Fourier parameters for 16 stars with a median uncertainty $\sigma(\phi_{21})=0.09$.

In recent years, many new high-quality RV observations have been collected \citep[see, e.g.][]{Anderson2016a,borgniet2019,Gallenne2019,Nardetto2023}. In this work, we take advantage of these recent measurements, in combination of an extensive dataset from the literature, to enlarge the number of Cepheids with known Fourier parameters and improve their precision. The paper is dedicated to the Galactic Classical (Population I) Cepheids, both pulsating either in the fundamental mode and in the first-overtone mode. We present our data set and our method for Fourier decomposition in Sect.~\ref{sect:fourier}. We present our results on precision of Fourier parameters in Sect.~\ref{sect:result_precision}, identification of first-overtone in Sect.~\ref{sect:1O} and binary Cepheids in Sect.~\ref{sect:binary_results}. We discuss the progression of the low- and high-order Fourier parameters with pulsation period as well as intrarelation of these parameters in Sect.~\ref{sect:discussion}. We conclude in Sect.~\ref{sect:conclusion}.


\section{Fourier decomposition of radial velocity curves}\label{sect:fourier}
\subsection{Data set of radial velocity measurements}\label{sect:data_set}
For each star we first collected the RV measurements available in the literature. All RV measurements used in this study are derived from high-resolution spectroscopy observations of metallic lines. Indeed, metallic lines represent a robust and long standing verified approach to obtain reliable RV measurement from the Cepheid photosphere. However, the use of \textit{Gaia} Radial Velocity Spectrometer (RVS) \citep{gaia2016,gaia2023} measurements based on Calcium infrared triplet lines \citep{Munari1999,Katz2023} is beyond the scope of this paper. Although \cite{Ripepi2023DR3} reported the reliability of the \textit{Gaia} RV measurements for Cepheids, dynamical effects observed in the Calcium triplet of several Galactic Cepheids  \citep{wallerstein2015,wallerstein2019,Hocde2020b} require more investigation to evaluate their impact on RV measurements compared to optical metallic spectral lines. Last, most of the RV measurements we collected were obtained from the cross-correlation technique \citep[optical and numerical,][]{Griffin1967,Baranne1979,Baranne1996} which is an important requirement for the application of the Baade-Wesselink method \citep{borgniet2019}.

Then, we included in our final sample only the Cepheids for which we derived the most precise Fourier parameters, as describes in the following sections.
This sample consists of 178 fundamental and 33 first overtone Cepheids and 7 Cepheids of uncertain or undetermined pulsation mode. The entire data set used in the analysis is composed of 82 different sources of radial velocity measurements which are listed in Table \ref{Tab:data_ref}.  For several Cepheids, the Fourier decomposition based on highest quality RV data has already been published by \cite{Kienzle1999}, \cite{Pont2001} and \cite{Baranowski2009}. For the sake of uniformity, we decided to reanalyze all these stars, supplementing the old data with newer measurements, if available (5 stars). The resulting Fourier parameters were found within the
errors to be the same as in the original papers. All these stars
are included in Table \ref{Tab:data_ref} and in following tables.
 In the next sections, we detail the steps of the Fourier decomposition, the different corrections applied and the quality classification of the fits.

\subsection{Fourier decomposition}
Precision on Fourier parameters depends primarily on the uncertainties of the RV observations, the number of data points, and the phase coverage. Collecting in the literature  datasets from various observations has the advantage of maximizing the number of data points and completing the phase coverage for many Cepheids. However, we have to deal with inhomogeneous uncertainty and datasets that can be far from each other in time. Indeed, there has been a significant improvement of the precision of RV measurements over time, from $\approx1\,$km/s in the 80's down to $\approx10\,$m/s for the most precise high-resolution spectrograph. Moreover, individual error determinations are rather uncertain before the 90's and are most often not provided but estimated from the dispersion of the measurements. On the other hand, definition of uncertainties varies widely in the literature and given formal error are often underestimated as compared to long-term uncertainties. Indeed, uncertainties of RV measurements depends on multiple factors including deviations in the adopted radial velocity of standard stars, misalignment of stars within the spectrograph slit, and potential discrepancies in the velocity scale between Cepheids and standard stars \citep{Barnes2005RV}. The precision is also strongly dependent on the choice of metallic lines, correlation template and the computation method used \citep{borgniet2019}. In order to avoid case-by-case assumption on the adopted uncertainties, we shall proceed with a unweighted least-square method which is also a standard approach in the field. We note that while heteroskedasticity of the dataset bias the \textit{precision} of the least-square estimator, the \textit{accuracy} of the fitted parameter is still preserved according to the Gauss-Markov theorem. Nevertheless, we show that the precision obtained are also robust in Sect.~\ref{sect:result_precision}, and, thus, that the assumption of homoskedasticity is also excellent.




We decomposed the RV curves into Fourier sine series by least-square fitting of the data (unweighted) with the following formula:
\begin{equation}\label{eq:fourier}
V_r(t) = A_0 + \sum_{k=1}^n A_k \mathrm{sin}[k \omega t +\phi_k]
\end{equation}
where $V_r(t)$ is the radial velocity measured at the time $t$, $A_k$ and $\phi_k$ are the amplitude and the phase of the $k$ Fourier component. Each harmonic is a multiple of the fundamental frequency $\omega=2\pi/P$. The pulsation period, $P$, is also optimized in the fitting process. We used the dimensionless Fourier parameters introduced by \cite{SimonLee1981}:
\begin{gather}
    R_{i1}=\frac{A_i}{A_1}\\
    \phi_{i1}=\phi_i - i \phi_1
\end{gather}
where $\phi_{i1}$ are then adjusted to lie between 0 and 2$\pi$. Choosing the appropriate order of the fit $n$ is important because a too low order may lead to higher uncertainties on the Fourier parameters, while the noise might be fitted if an order is too high. Many studies simply fix the order of the fit depending on the pulsation period of the star. We adopt a more adaptative approach where the order of the fit $n$ is iterated until $A_n/\sigma_{A_n}$>4. In some cases, we included one or two higher harmonics in order to stabilize the fit. This method is purely empirical and is based on practical results.

The standard deviation of the fit is given by
\begin{equation}\label{eq:sigma}
\sigma^2=\frac{1}{N-2n-2}\sum^N_{i=1}(V_r(t_i)-\hat{V}_i)^2
\end{equation}
where $N$ is the number of data points, $n$ the order of the fit, $V_r(t_i)$ and $\hat{V}_i$ are the radial velocity data and the Fourier model respectively. The derived standard deviation of the fit is then used to apply an iterative 3-$\sigma$ data clipping. In this step we remove data points deviating from the fit by more than 3$\sigma$. This is a standard outlier rejection technique, commonly used in the literature.

\subsection{Errors of Fourier parameters}
The standard deviation of the fit presented above is a good indicator for the goodness of fit although it does not reflect the uncertainty of the Fourier parameters. In fact, it can be misleading to interpret this quantity as an accuracy of Fourier parameters. For example, for low amplitude variations we may obtain a fit with very low dispersion, but uncertainties of the Fourier phases $\phi_{i1}$ might
 still be large. This may be so because harmonic components of the fit may be very low and difficult to measure (see e.g. Polaris). On the contrary, in case of high amplitude RV curves the dispersion of the fit may be higher, yet the precision of the Fourier parameters may be very good. In some cases, missing observational data in a specific phase range can lead to a small fit dispersion but rather uncertain Fourier parameters. Finally, we should note that precision of the Fourier parameters depends also on the number of measurements. Adding more and more datapoints of the same quality does not change the dispersion of the fit, but the errors of all Fourier parameters become progressively smaller.

The analytical formulae of the Fourier parameters obtained from the least-square method were derived by \cite{Petersen1986}.
This method was applied in comparable RV studies to derive Fourier parameter uncertainties \citep{kovacs90,Kienzle1999,Moskalik2000,Baranowski2009,storm11a} but also in the case of light-curves  \citep[see, e.g.][]{Deb2010}. In this approach, the errors of the Fourier amplitude ratios $\sigma(R_{k1})$ and phase differences $\sigma(\phi_{k1})$ at the order $k$ are obtained from propagation of the variance-covariance matrix of the fitted Fourier parameters \citep[see Eqs.~9 to 12 in][]{Petersen1986}.

In addition to this basic Fourier fitting method, we applied several corrections as described in the following.

\subsection{Zero-points correction}
As discussed in Sect.~\ref{sect:data_set}, we used data sets compiled from many different sources from the literature. Thus, zero-point offsets between data sets of the same star might be present, because of different instruments used, different choice of metallic lines and different methods of extracting RV from the spectra  \citep[see, e.g.,][]{Evans2015,borgniet2019,Nardetto2023}.
In order to correct for these effects, we treated zero-point differences between datasets as free parameters and determined them iteratively from the least-square fit. We note that zero-point corrections have been applied by other authors as well \citep[see, e.g.,][]{Herbig1952,EvansLloyd1982,KissVinko2000,Pont2001,anderson15a,breitfelder16,Torres2023}. Similar corrections have also been used when combining photometric data \citep[see, e.g.,][]{Coulson1985a,AntonelloPoretti1990}.  We note that applying zero-point correction is not suitable to detect time-variability of average velocities that would be caused by orbital motion. However, the \textit{systematic} detection of binary Cepheids is out of the scope this paper, and we only focus on removing orbital motion contribution as presented in the following section.

\subsection{Orbital motion correction}\label{sect:binary}
Since a significant fraction of Cepheids are binary or multiple systems, \citep{Evans2015,Szabados2003,Kervella2019}, some of the RV curves can be strongly affected by orbital motion. In this paper, we stress that we do not aim to resolve the Keplerian elements of the orbit, but simply to subtract the trend induced by the orbital motion. To this end, the orbital motion is modeled by an additional Fourier series, with the orbital period $P_\mathrm{orb}$. Thus, the fitting formula for a binary Cepheid consists of the two Fourier sums: one for pulsation variations (see Eq.\ref{eq:fourier}), and one additional for orbital motion. This procedure was used for example by \cite{kovacs90}. In this method, both pulsation and orbital motion are fitted simultaneously. As mentioned above, a Fourier sine series cannot determine the Keplerian elements of the orbit, but is naturally suitable for accurately determine the orbital period. Although this method is convenient for algorithmic purposes we note that we introduce additional degree of freedoms (DOFs) as compared to a Keplerian model. However, it must be kept in mind that the number of data points is significantly larger than the DOF for the Cepheids of our study, and, thus, this additional cost is largely mitigated. For example, in the case of SU Cyg we have 23 parameters, instead of 16 using Keplerian orbit, as compared to 331 datapoints, this yield an increase of the fit dispersion by only 0.8\%. For only three binaries this increase is slightly above 1\%, namely RX Cam (4.1\%), S Sge (1.6\%) and V350 Sgr (1.1\%). In exchange, we get a much simpler fitting formulae (linear in Fourier parameters). The orbital period, $P_\mathrm{orb}$, is either taken from the literature or determined from the fit. If the orbital period is consistent with the period given in the literature (which is usually the case), we choose the one with smaller formal error. We summarized the orbital period adopted in Table \ref{Tab:binary}. In Table \ref{Tab:binary_comp} we compare our new orbital periods with the periods previously publish in the literature.

\subsection{Trend correction}\label{sect:trend}
In some cases, after subtracting the pulsation variation from the RV curve, we observe a slow linear or parabolic trend in the residuals. Such a trend can be of instrumental origin or it could result from a slow orbital motion \citep[see, e.g.,][]{anderson15a}. In order to model both linear and parabolic trends in the data, we used a single Fourier component with a constant long period of $P_\mathrm{trend}=50000\,$day.  Alternatively, the trend can also be modeled by a polynomial function, but we chose the Fourier model for practical algorithmic purposes. Hence, Fourier sum for both pulsation variations and the trend are fitted simultaneously. For each star, we indicated if an additional trend is fitted (see Tables \ref{Tab:data_funda_quality1}, \ref{Tab:data_funda_quality1a}, \ref{Tab:data_fourier_quality2}, \ref{Tab:data_fourier_quality1_overtone} and \ref{Tab:unknown}).

\subsection{($O-C$) correction for long-period Cepheids}\label{sect:OC_correction}
The four Cepheids with the longest pulsation periods ($P>40\,$day), namely SV~Vul, GY~Sge, V1496~Aql and S~Vul, all display very fast, irregular $O-C$ variations, or equivalently, period variations \citep{Berdnikov2004,Berdnikov2007,Turner2009}. Phasing their radial velocity data with constant period will not yield satisfactory results. Therefore, we proceeded differently: we corrected for the $O-C$ variations by applying phase offsets (shifts), calculated for each season separately \citep[see, e.g.,][]{Berdnikov2004,anderson15a,kervella17}. A usual method is to use an $O-C$ diagram determined from {\it photometric} observations in order to calculate the required phase shift for each epoch \citep[see, e.g.,][]{kervella17}. Instead of this approach, we measured the phase shifts directly from the RV data for each individual dataset, similarly to \cite{anderson15a} in case of $\delta$ Cep. To this end, we first divided the data into seasons and chose the best sampled season as a reference. The pulsation period $P$ of the star was determined from this reference season. Then we applied individual phase shifts to all the remaining seasons. These phase shifts were determined for all datasets simultaneously by the least square method in which we minimized the dispersion of the overall Fourier fit. Our procedure of measuring the required $O-C$ shifts is in its idea equivalent to a Hertzsprung method \citep{Hertzsprung1919}. We note, that this approach is valid only if each seasonal chunk of data covers both the ascending branch and the descending branch of the RV curve. We applied this method only to the four long-period Cepheids mentioned above and to a single short period Cepheid V Cen, for which we had to combine data collected over many years (see individual fitting treatment in Sect.~\ref{sect:individual}).


\subsection{Quality classification of the fits}\label{sect:quality}
It is difficult to define a consistent criteria to quantify the quality of the fit for every star. Indeed, quality of the different fits depends on the number of measurements, coverage of the pulsation cycle, standard deviation of the fit and presence of instabilities.  Thus, we rather visually inspected each Fourier fit and we classified them into two main categories (see Table~\ref{Tab:quality}). Fourier parameters of highest quality radial velocity curves labeled "1"  are most accurate, as resulting from fits free of any significant instabilities and with an excellent phase coverage. By "instability" we mean a small artificial undulation of the RV curve, produced
by an ill-defined fit and having no physical meaning. In some cases the fits display slight instabilities, hence we assigned to them the quality flag "1a". Fourier fits classified as quality "2" have useful Fourier parameters 
but they can be affected by stronger instabilities, scatter or poorer
coverage of the pulsation cycle. For each star of quality "2" we provided
a flag which describes the type of deficiencies we identified (see Notes
to Tables \ref{Tab:data_fourier_quality2} and \ref{Tab:data_fourier_quality1_overtone}). We emphasize that this classification has to be considered only as an indication to guide the reader. 

\section{Results}\label{sect:result}
\subsection{New precise Fourier parameters}\label{sect:result_precision}
\begin{figure*} 
\begin{subfigure}{0.5\textwidth}
\includegraphics[width=\linewidth]{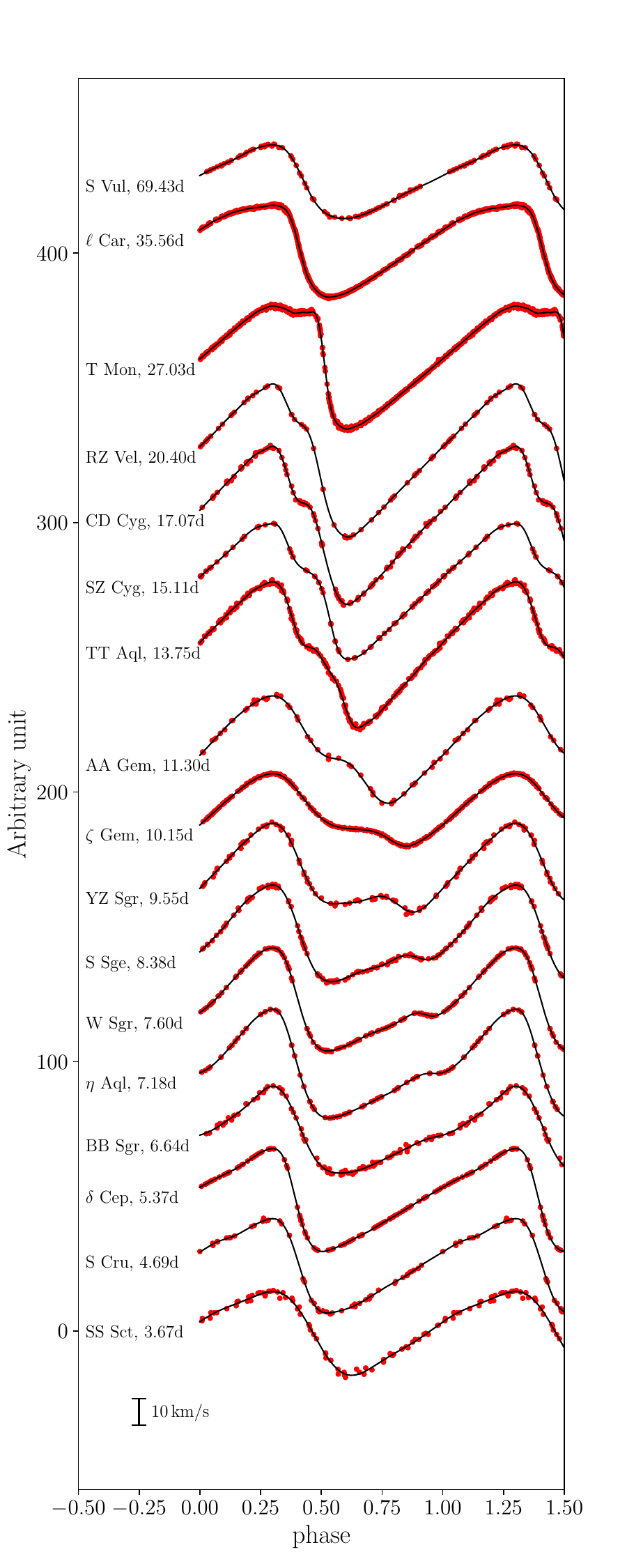}
\caption{} \label{fig:FU_progression}
\end{subfigure}\hspace*{\fill}
\begin{subfigure}{0.50\textwidth}
\includegraphics[width=\linewidth]{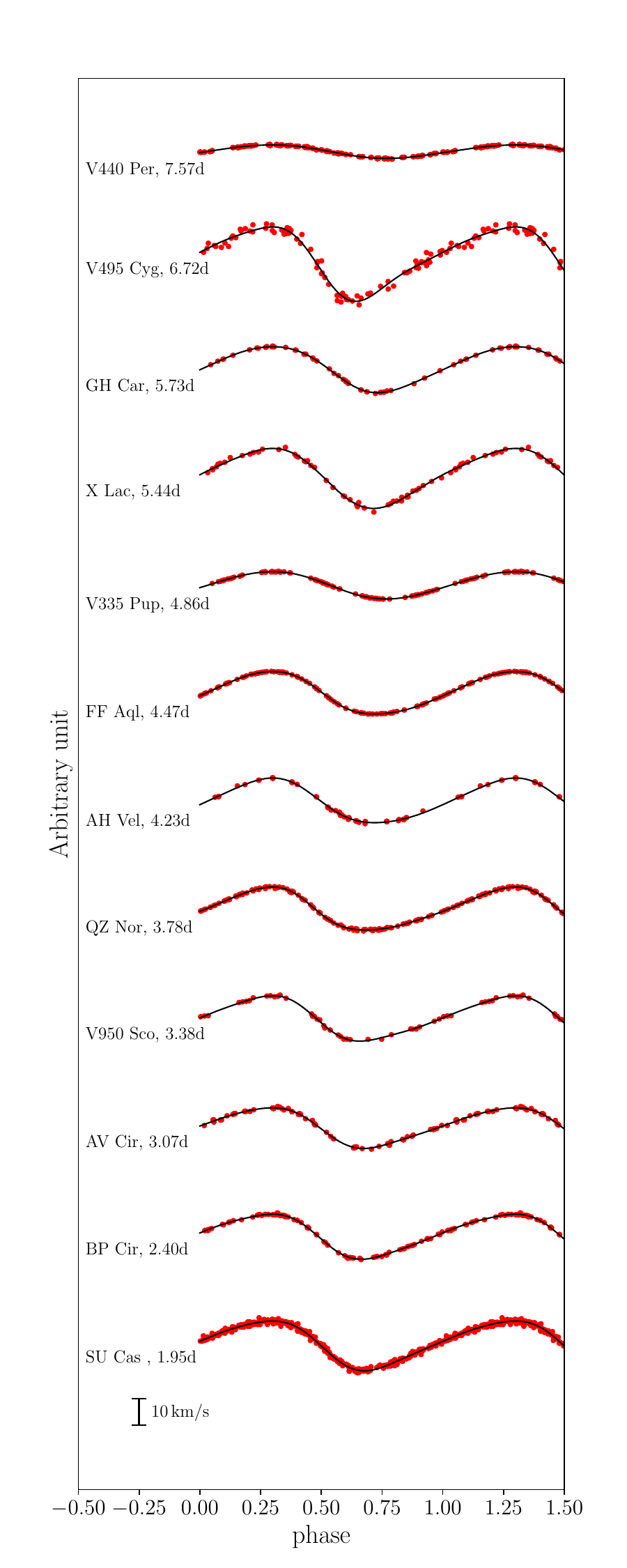}
\caption{} \label{fig:FO_progression}
\end{subfigure}

\caption{\small Radial velocity curves and Fourier fits for a set of fundamental-mode (a) and first-overtone Cepheids (b). The scale is indicated in the left bottom corner of the plots. \label{fig:Progression}}
\end{figure*}

\begin{figure}
\includegraphics[width={0.52\textwidth}]{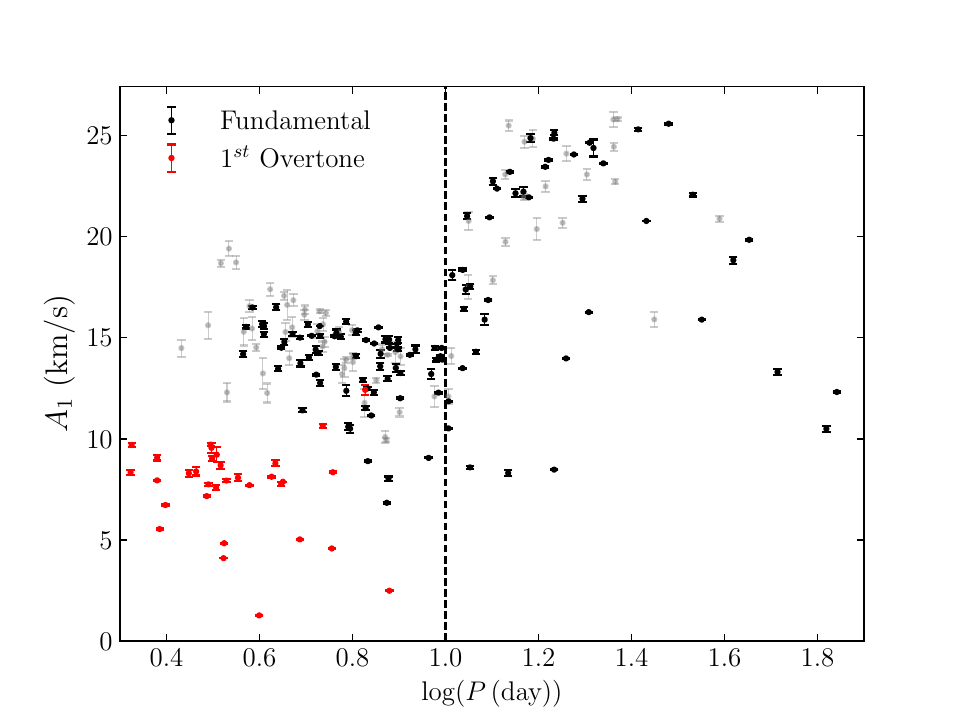}
\caption{\small First Fourier term, $A_1$, for fundamental-mode and first-overtone Cepheids. Fundamental-mode variables of Quality 1 and 1a are plotted with black points, while gray points are used for variables of Quality 2. A vertical dashed line indicates a period of 10 day and it is plotted to guide the eye in the resonance vicinity.\label{fig:A1}}
\end{figure}

\begin{figure*} 
\begin{subfigure}{0.5\textwidth}
\includegraphics[width=\linewidth]{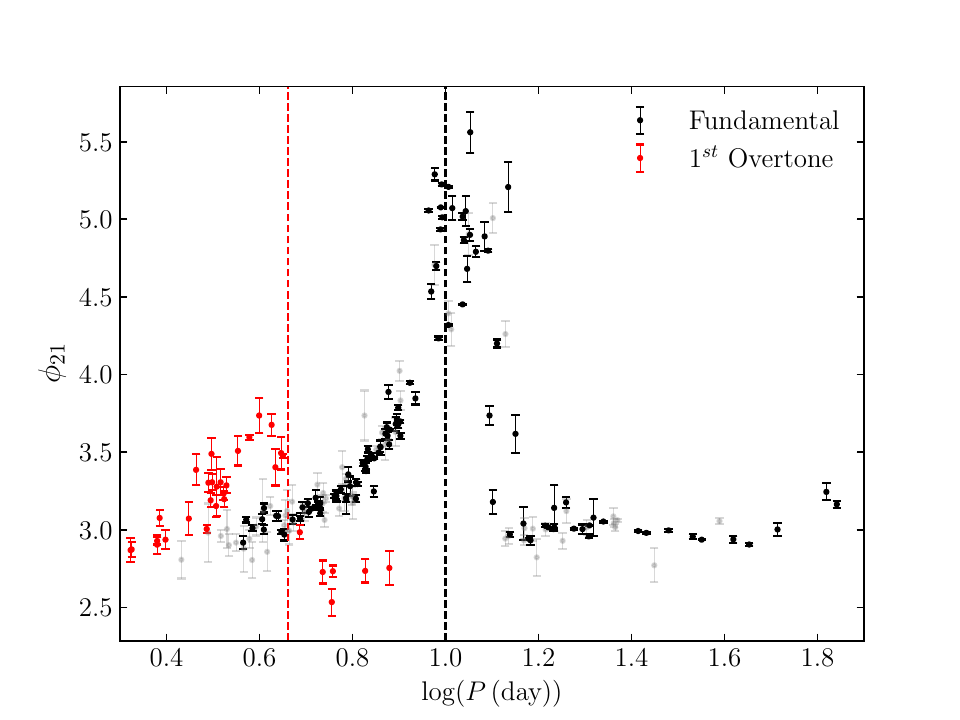}
\caption{} \label{fig:phi21}
\end{subfigure}\hspace*{\fill}
\begin{subfigure}{0.50\textwidth}
\includegraphics[width=\linewidth]{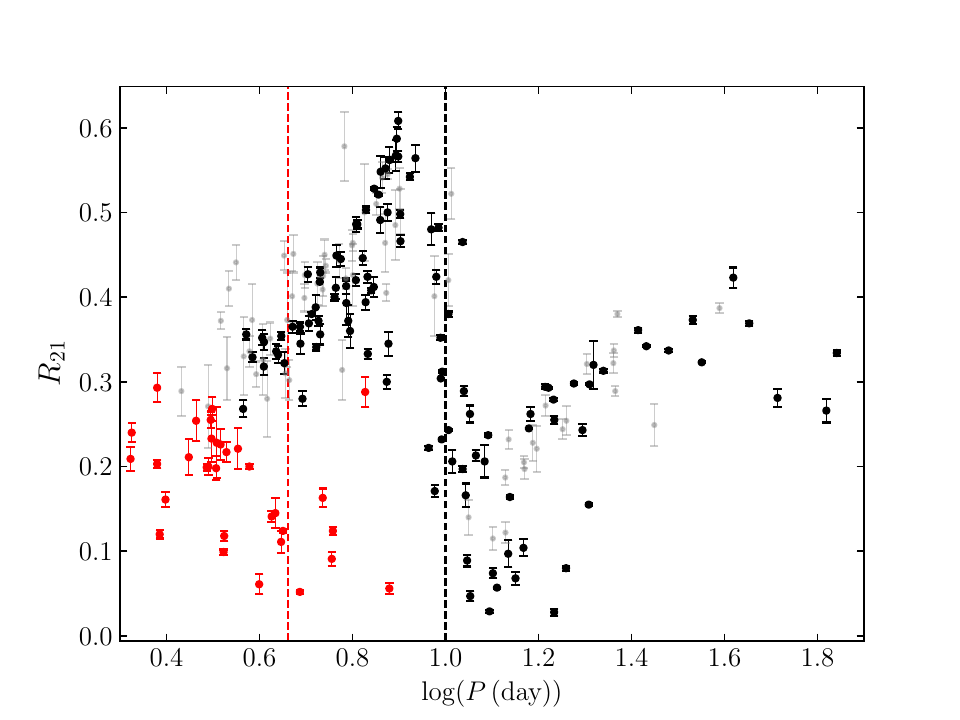}
\caption{} \label{fig:R21}
\end{subfigure}

\medskip
\begin{subfigure}{0.50\textwidth}
\includegraphics[width=\linewidth]{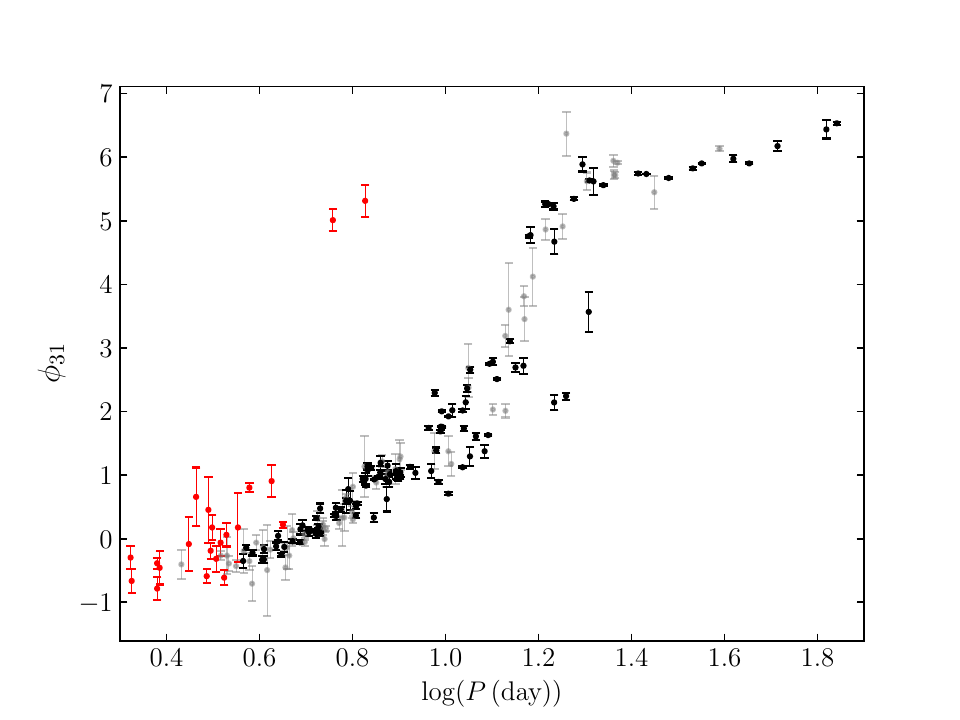}
\caption{} \label{fig:ephi31}
\end{subfigure}\hspace*{\fill}
\begin{subfigure}{0.50\textwidth}
\includegraphics[width=\linewidth]{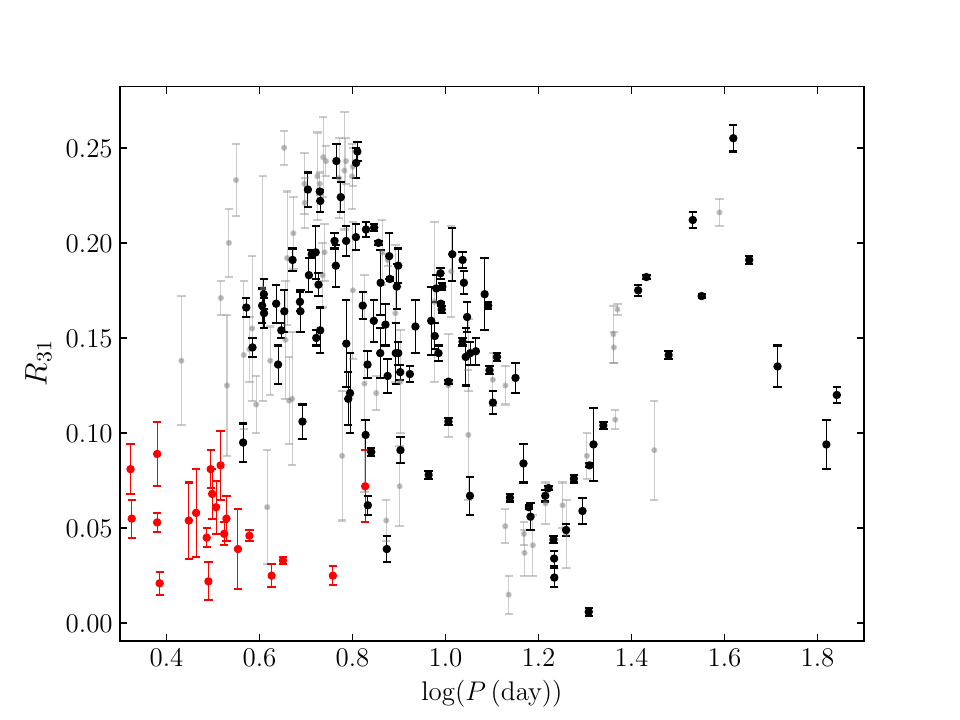}
\caption{} \label{fig:R31}
\end{subfigure}

\medskip
\begin{subfigure}{0.50\textwidth}
\includegraphics[width=\linewidth]{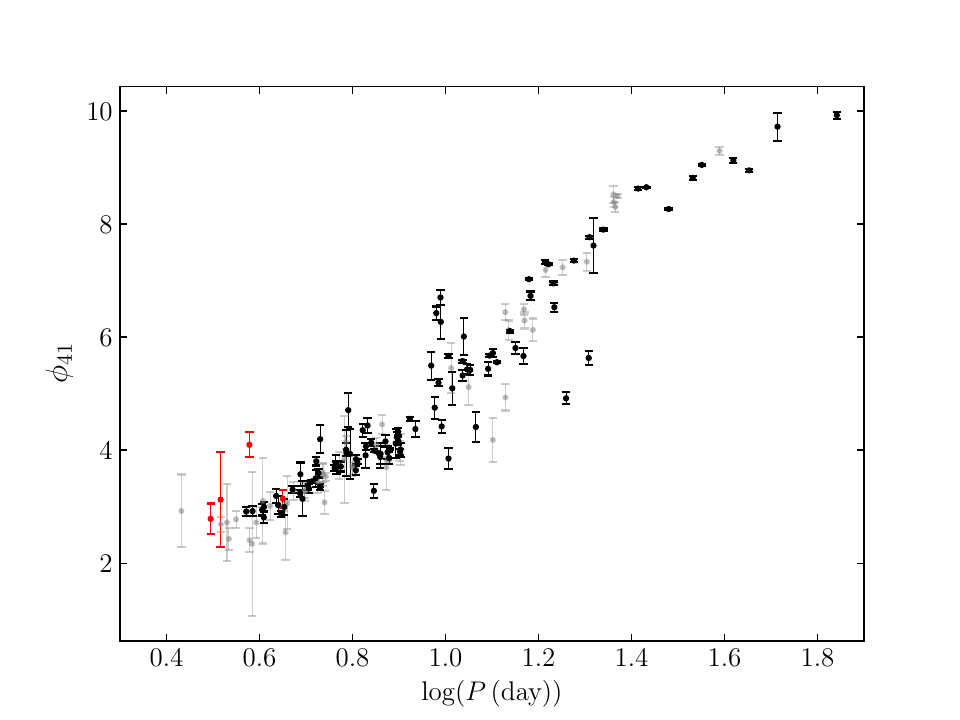}
\caption{} \label{fig:phi41}
\end{subfigure}\hspace*{\fill}
\begin{subfigure}{0.50\textwidth}
\includegraphics[width=\linewidth]{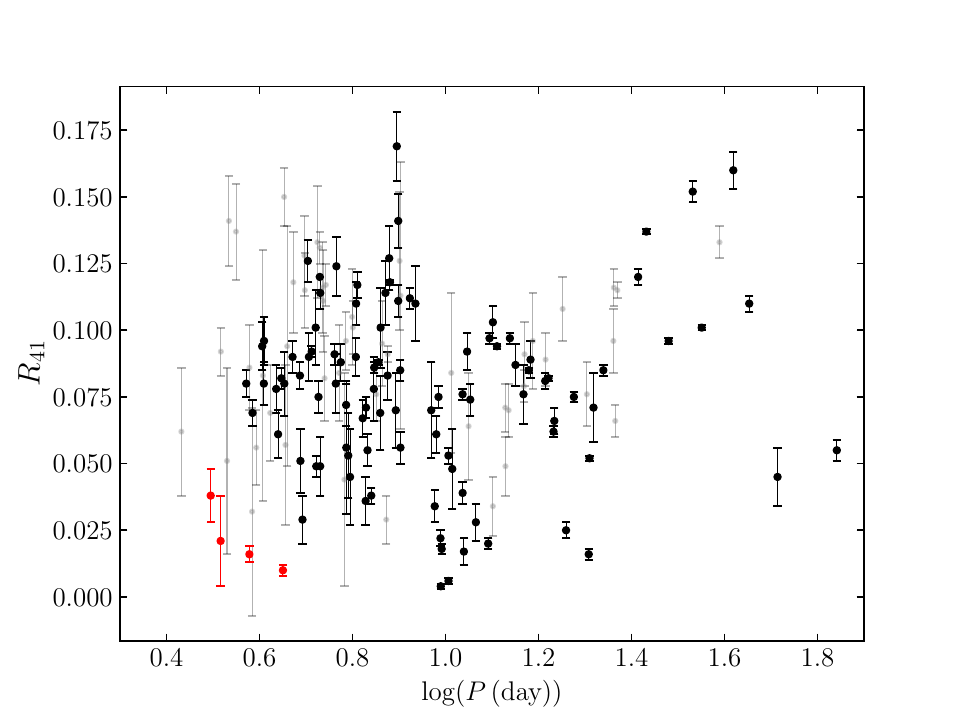}
\caption{} \label{fig:R41}
\end{subfigure}\hspace*{\fill}
\caption{Low-order Fourier parameters for RV curves of fundamental-mode and first-overtone Cepheids. For the fundamental mode black points and grey points are variables of Quality (1+1a) and 2 respectively (see Table \ref{Tab:quality}). Vertical dashed lined indicate a pulsation period of 4.68 and 10 day to guide the eye in the resonance vicinity of first-overtone and fundamental mode Cepheids respectively.\label{fig:fourier}}
\end{figure*}

\begin{figure*} 

\begin{subfigure}{0.5\textwidth}
\includegraphics[width=\linewidth]{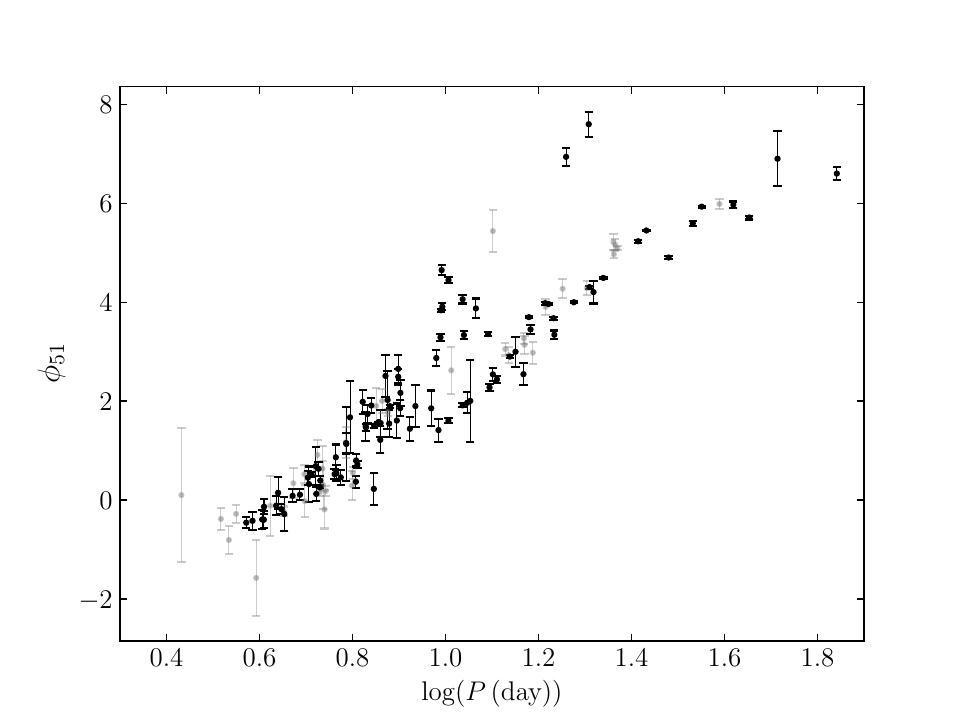}
\caption{} \label{fig:phi51}
\end{subfigure}\hspace*{\fill}
\begin{subfigure}{0.50\textwidth}
\includegraphics[width=\linewidth]{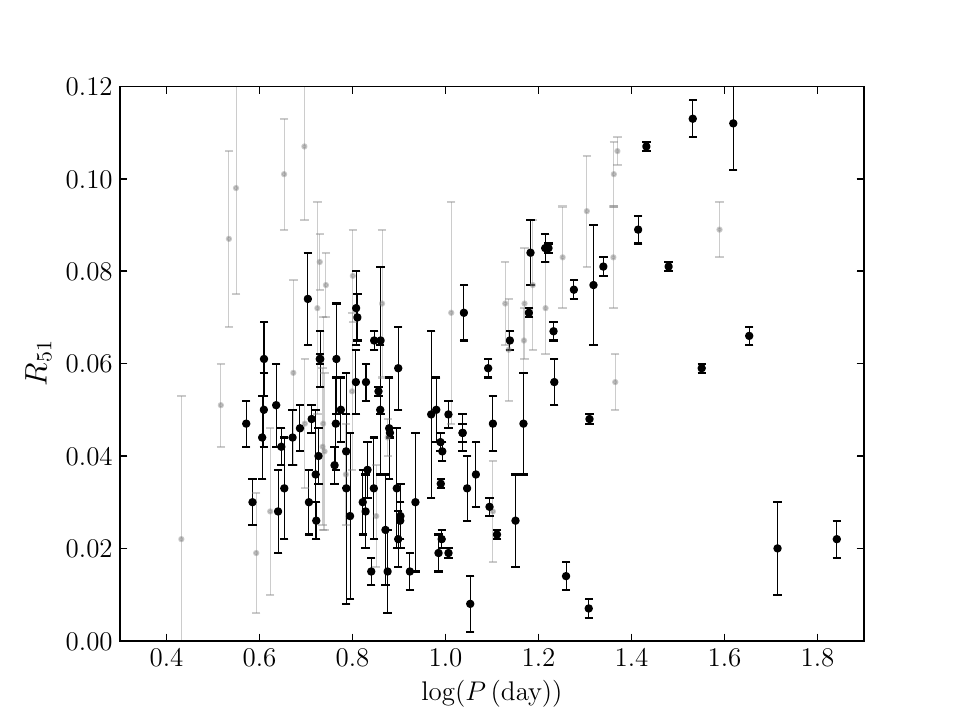}
\caption{} \label{fig:R51}
\end{subfigure}

\medskip
\begin{subfigure}{0.50\textwidth}
\includegraphics[width=\linewidth]{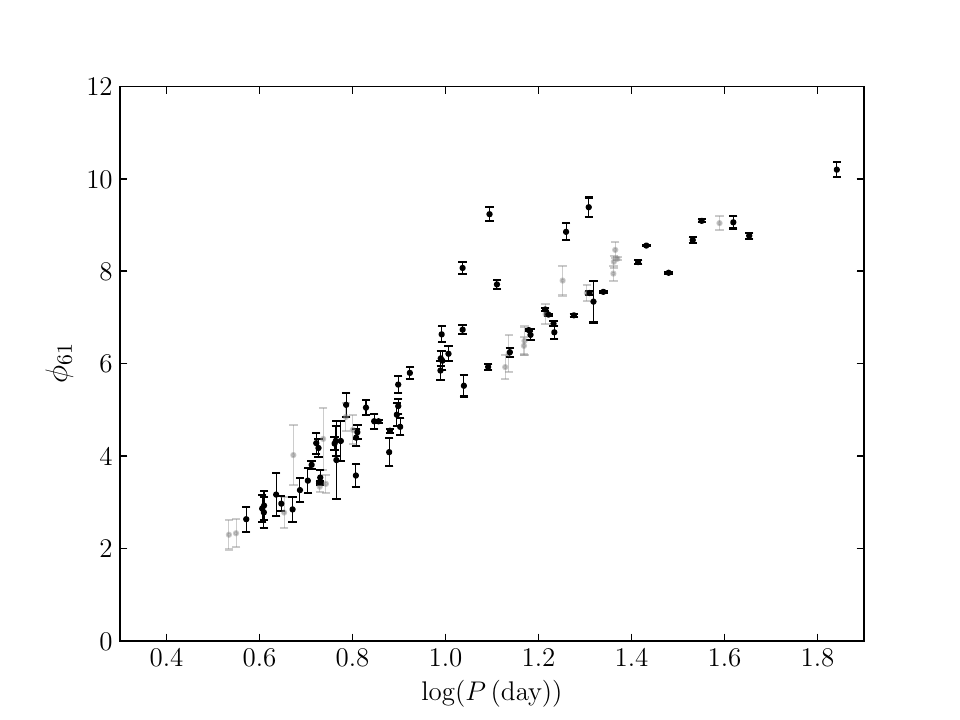}
\caption{} \label{fig:phi61}
\end{subfigure}\hspace*{\fill}
\begin{subfigure}{0.50\textwidth}
\includegraphics[width=\linewidth]{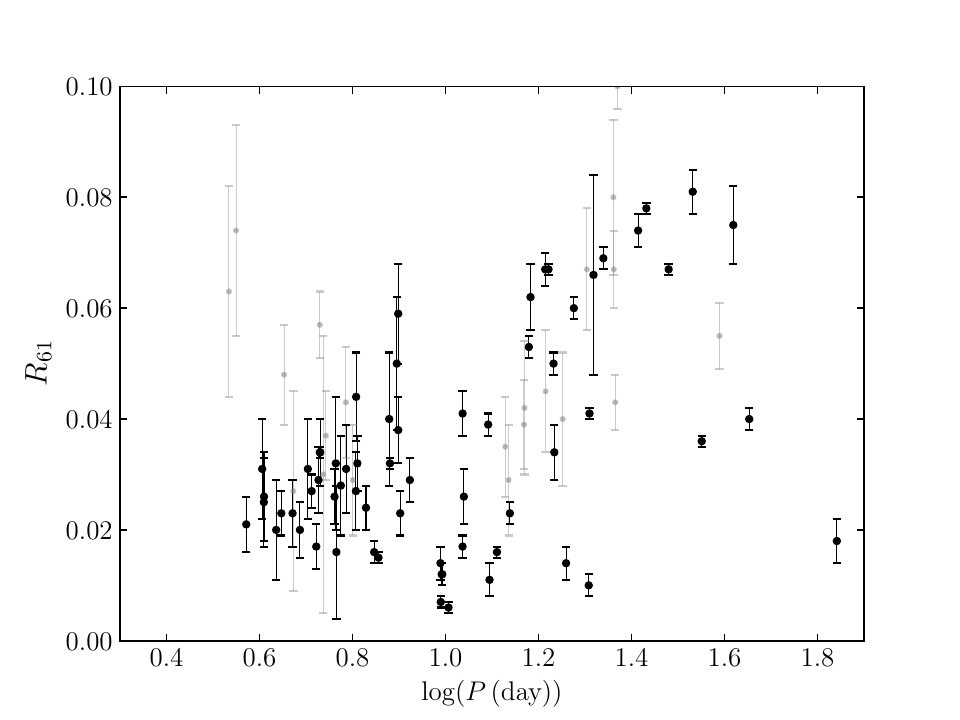}
\caption{} \label{fig:R61}
\end{subfigure}
\medskip
\begin{subfigure}{0.50\textwidth}
\includegraphics[width=\linewidth]{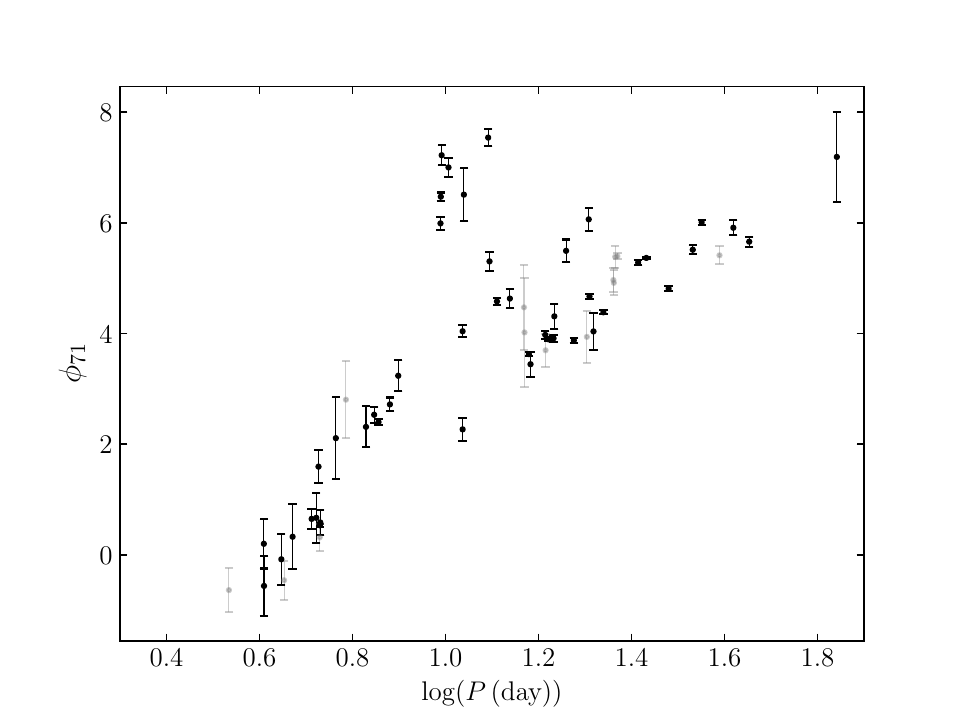}
\caption{} \label{fig:phi71}
\end{subfigure}\hspace*{\fill}
\begin{subfigure}{0.50\textwidth}
\includegraphics[width=\linewidth]{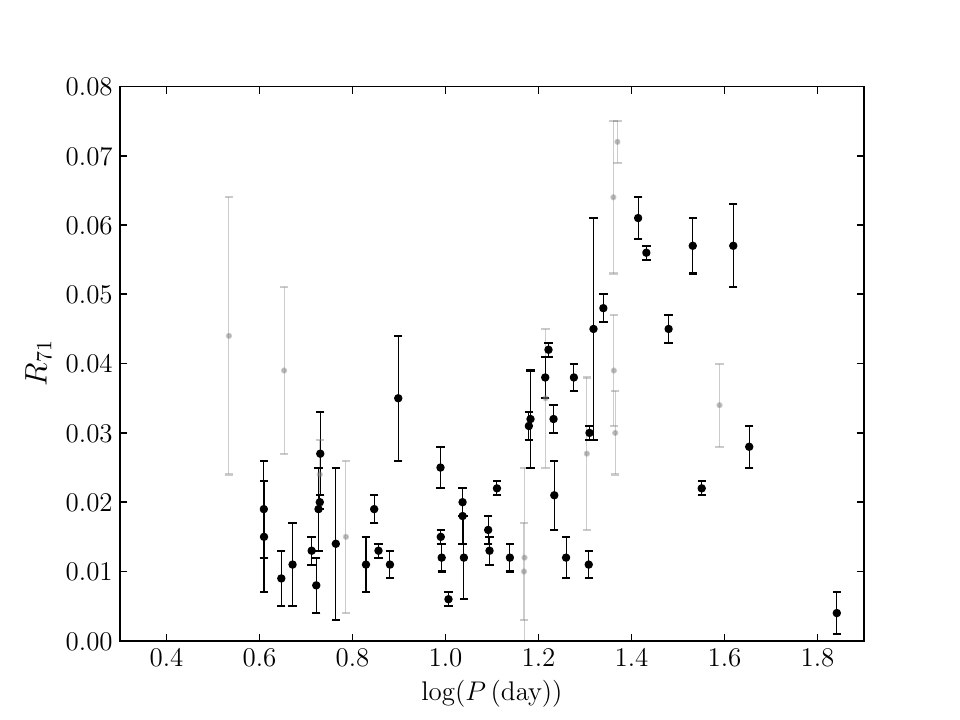}
\caption{} \label{fig:R71}
\end{subfigure}
\caption{Fourier parameters of 5th, 6th and 7th terms for RV curves of the fundamental-mode Cepheids. Black points and Grey points are variables of Quality (1+1a) and 2 respectively (see Table \ref{Tab:quality}).\label{fig:fourier_high}}
\end{figure*}

The plots of entire collection of Cepheid radial velocity curves are available as an on-line material. We displayed selected velocity curves in Fig.~\ref{fig:Progression}, which nicely exhibits the bump progression in the fundamental-mode Cepheids. We also present the amplitude of the first Fourier term $A_1$ (Fig.~\ref{fig:A1}), the low-order Fourier parameters (Fig.~\ref{fig:fourier}) and high-order Fourier parameters (Fig.~\ref{fig:fourier_high}). Fourier parameters for all the star are provided in Tables \ref{Tab:data_funda_quality1}, \ref{Tab:data_funda_quality1a}, \ref{Tab:data_fourier_quality2}, \ref{Tab:data_fourier_quality1_overtone} and \ref{Tab:unknown}. The different trends of Fourier parameters are qualitatively discussed in Sect.~\ref{sect:discussion}.

\begin{figure*}
\begin{subfigure}{0.5\textwidth}
\includegraphics[width=\linewidth]{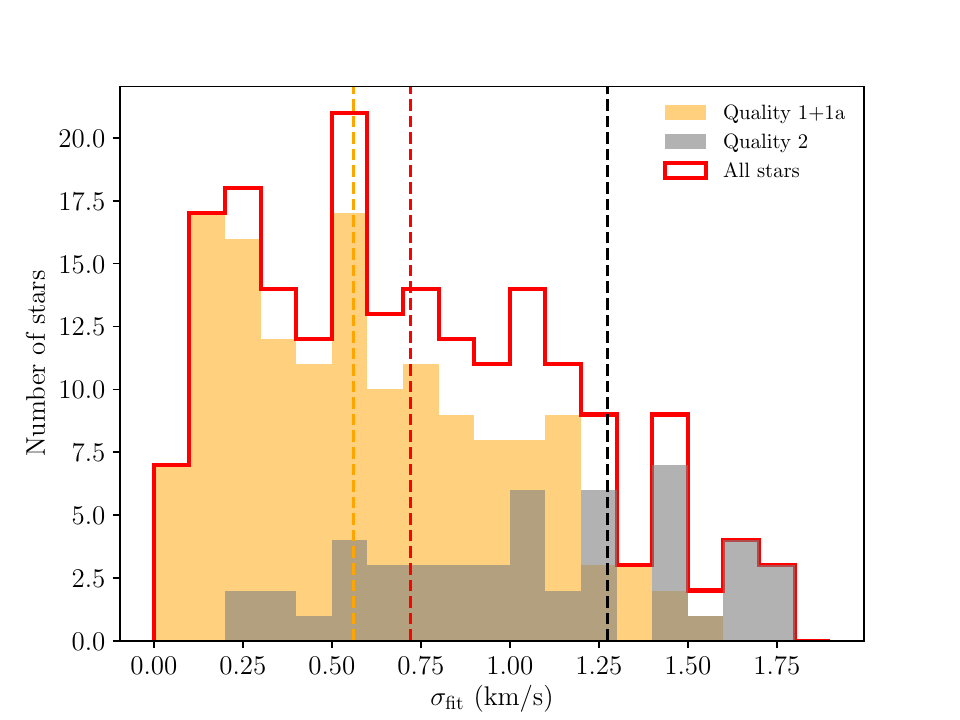}
\caption{}  \label{fig:sigma}
\end{subfigure}\hspace*{\fill}
\begin{subfigure}{0.50\textwidth}
\includegraphics[width=\linewidth]{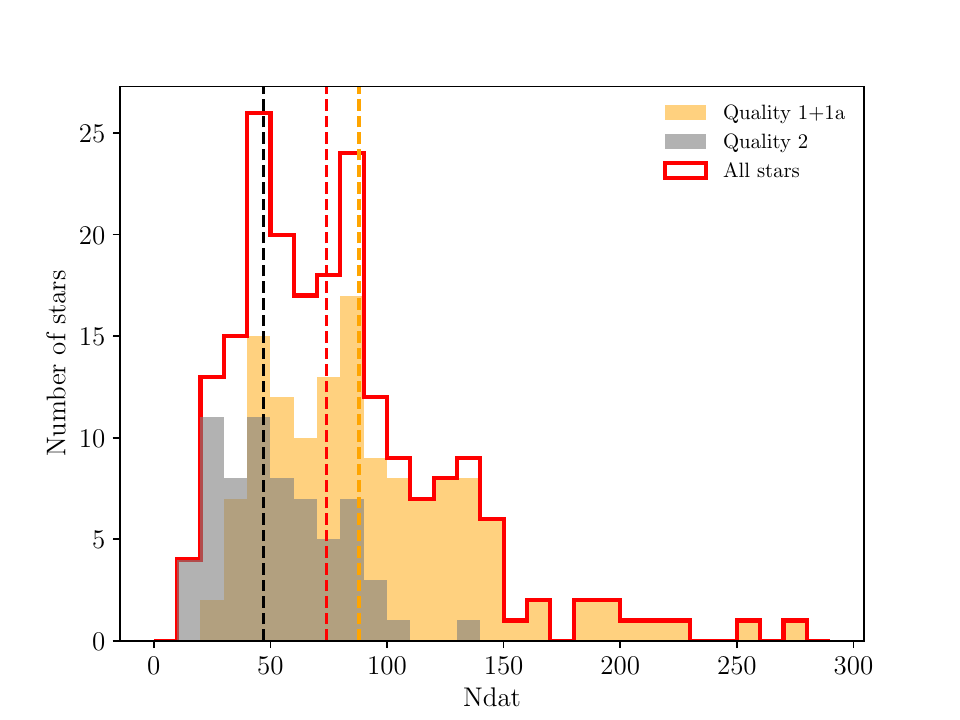}
\caption{}  \label{fig:ndat}
\end{subfigure}
\begin{subfigure}{0.5\textwidth}
\includegraphics[width=\linewidth]{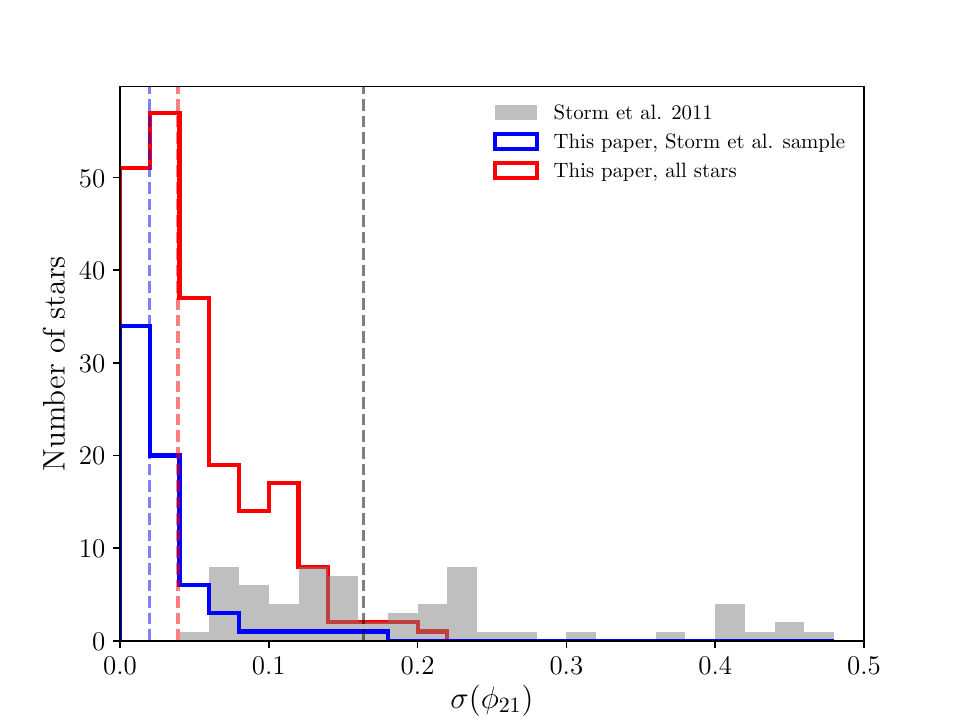}
\caption{} \label{fig:ephi21}
\end{subfigure}\hspace*{\fill}
\begin{subfigure}{0.50\textwidth}
\includegraphics[width=\linewidth]{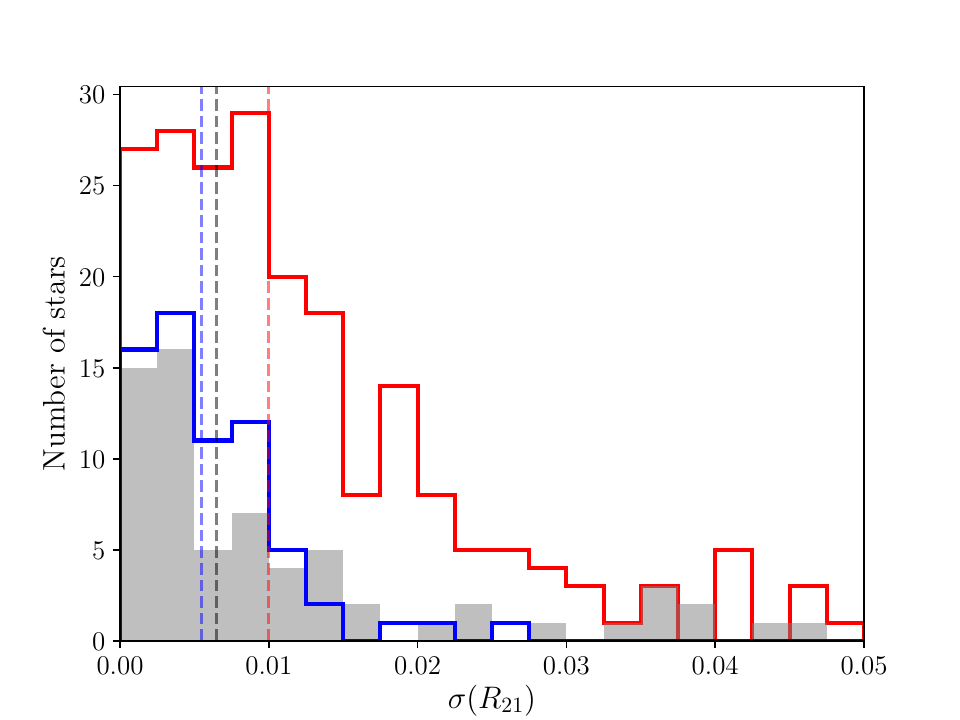}
\caption{} \label{fig:eR21}
\end{subfigure}
\medskip
\begin{subfigure}{0.50\textwidth}
\includegraphics[width=\linewidth]{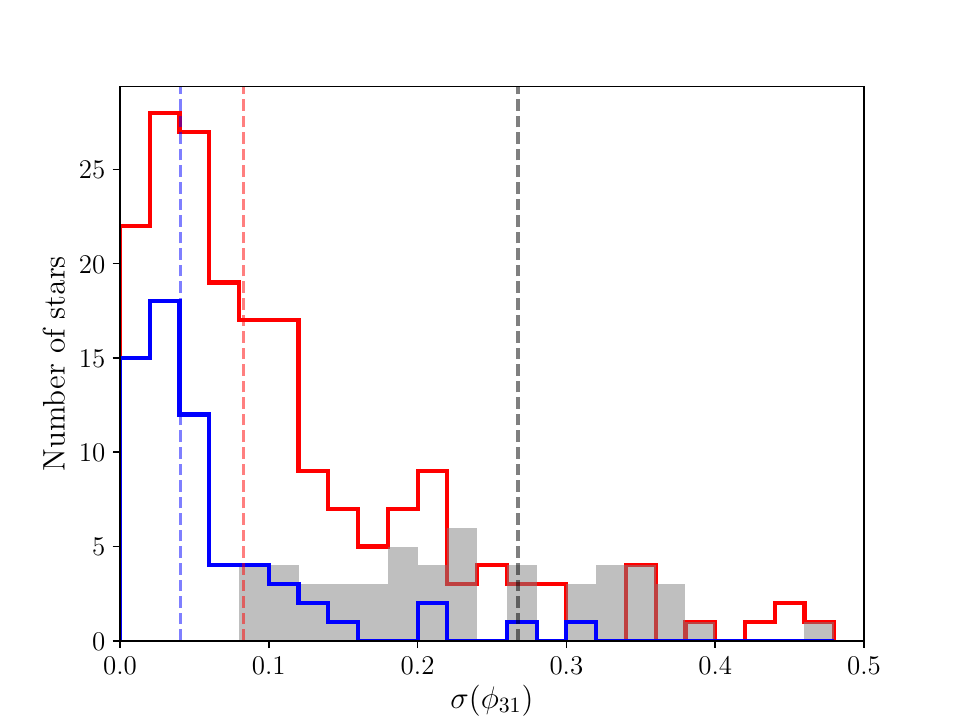}
\caption{} \label{fig:ephi31}
\end{subfigure}\hspace*{\fill}
\begin{subfigure}{0.50\textwidth}
\includegraphics[width=\linewidth]{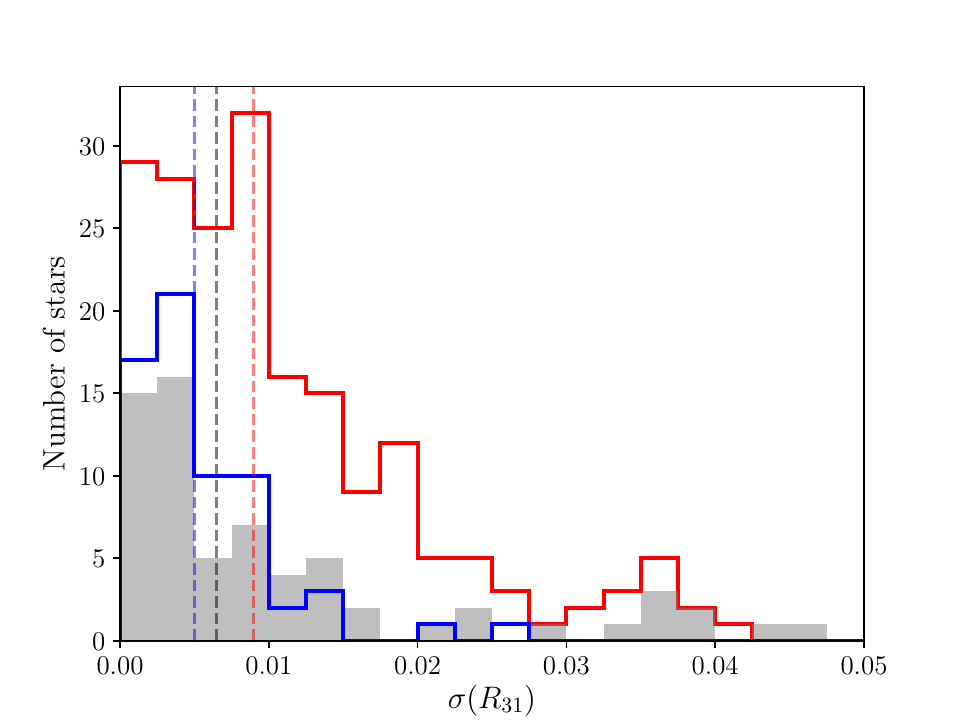}
\caption{} \label{fig:eR31}
\end{subfigure}
\caption{Histograms of (a) the standard deviation of the fits $\sigma_\mathrm{fit}$ and (b) the number of data points Ndat for our Cepheid sample. Histogram of uncertainties of low-order Fourier parameters are displayed for (c) $\sigma(\phi_{21})$, (d) $\sigma(R_{21})$, (e) $\sigma(\phi_{31})$  and (f) $\sigma(R_{31})$ for Cepheids of this paper and, for comparison, for Cepheids from \cite{storm11a}. In each plot, the median of each sample is represented by a vertical dashed line. See Sect.~\ref{sect:result}.}\label{fig:storm_comp}
\end{figure*}

Our sample consists of 218 variables. This is the largest collection of Cepheids for which Fourier parameters of their RV curves are determined. In Fig.~\ref{fig:storm_comp} we present the precision of the Fourier fits and of the derived Fourier parameters. The median number of data points per star and the median dispersion of the fit, together with quartiles, are (see Figs.~\ref{fig:sigma} and \ref{fig:ndat}):
\begin{align}
    \mathrm{Ndat}=74_{47}^{110}, \quad \sigma_\mathrm{fit}=0.72_{0.38}^{1.15}\,\mathrm{km/s}.
\end{align}
For the best defined RV curves (quality flags 1 and 1a) we obtain
\begin{align}
    \mathrm{Ndat}=88_{61}^{130}, \quad \sigma_\mathrm{fit}=0.56_{0.26}^{0.90}\,\mathrm{km/s}.
\end{align}
Although for many Cepheids we obtained fits of excellent quality (see RV curves plots online), the dispersion of the fit alone is a poor indicator of the Fourier parameters precision. Thus, we also display the derived uncertainties of low-order Fourier parameters (up to the third harmonics) in Figs~\ref{fig:ephi21}, \ref{fig:ephi31}, \ref{fig:eR21} and \ref{fig:eR31}. For Cepheids of our sample we find a median uncertainty of
\begin{align}
    \sigma(\phi_{21})=0.04_{0.02}^{0.07}, \quad \sigma(R_{21})=0.009_{0.005}^{0.018},  \\
    \sigma(\phi_{31})=0.08_{0.04}^{0.16}, \quad \sigma(R_{31})=0.009_{0.005}^{0.016}.
\end{align}
It is also possible to verify the obtained precision using the trends of the Fourier phases of low and high-orders (Figs.~\ref{fig:fourier} and \ref{fig:fourier_high}). When plotted vs. pulsation period, the Fourier phases $\phi_{n1}$ display remarkably tight progressions, which is in qualitative agreement with theoretical predictions \citep{Buchler1990}. More importantly, the dispersion of each Fourier phase progression is in excellent agreement with the derived uncertainties, which demonstrates that they the are realistic. Intrarelations of Fourier phases discussed in Sect.~\ref{sect:intrarelation} and displayed in Fig.~\ref{fig:intra_phi} provide further evidences of internal consistency on the obtained precisions, and, therefore, of the robustness of our fitting method.


We compare the precision achieved in our work with the precision of the similar Fourier parameter surveys published in the literature. First, we cross-matched our results with those of \cite{storm11a}. We display histograms of their errors in Figs~\ref{fig:ephi21}, \ref{fig:ephi31}, \ref{fig:eR21} and \ref{fig:eR31}. Comparing the median of each sample we conclude that our median precision {\it for the same set of stars} is 8 times better in case of $\sigma(\phi_{21})$, 6 times better in case of $\sigma(\phi_{31})$, 18\% in case of $\sigma(R_{21})$ and 22\% better in case of $\sigma(R_{31})$. For the first overtone Cepheids, we obtain a median uncertainty of
\begin{align}
    \sigma(\phi_{21})=0.06_{0.04}^{0.09}, \quad \sigma(R_{21})=0.011_{0.006}^{0.017},
\end{align}
\noindent as compared to previous results from \cite{Kienzle1999}
\begin{align}
    \sigma(\phi_{21})=0.09_{0.06}^{0.16}, \quad \sigma(R_{21})=0.0155_{0.013}^{0.018},
\end{align}
which yield an improved precision of about 33 and 29\% respectively.

As a conclusion, we not only obtained the largest sample of available Fourier parameters of RV curves (218 stars, determined from a same method), but we also significantly improved the precision of Fourier parameters. This improved precision is valuable for fine-tuning hydrodynamical pulsation models \citep[see, e.g.][]{Paxton2019,Kovacs2023}, a problem that will be addressed in forthcoming papers in preparation in our group. As an immediate by-product, however, we are able to identify first-overtone mode Cepheids based on the Fourier phase $\phi_{21}$ in Sect.~\ref{sect:1O} and to identify some binary Cepheids with their orbital period in Sect.~\ref{sect:binary_results}.











\begin{table}[]
\caption{\small Quality classification of the Cepheids sample.
}\label{Tab:quality}
\begin{tabular}{l|r|r|r}
\hline
\hline
  Quality      & F & 1O & U\\
\hline
1         &    85  &   30  & 4\\
1a         &    27  &   1 &\\
2         &    66   &    2  &3\\
\hline
Total    &   178      &   33   &7\\

\hline
\hline
\end{tabular}
\normalsize
    \begin{tablenotes}
    \item \textbf{Notes :} Consecutive rows list numbers of stars with quality classification 1, 1a and 2, as defined in Sect.~\ref{sect:quality}, for fundamental (F), first-overtone (1O) and undetermined mode (U) Cepheids.
    \end{tablenotes}

\end{table}


\subsection{Identification of First-overtone Cepheids}\label{sect:1O}
Many known overtone Cepheids in the literature have been identified with the help of the Fourier phases of their light curves \citep{AntonelloPoretti1986,AntonelloPoretti1990,Mantegazza1992,Poretti1994}. Unfortunately, light-curve Fourier parameters \textit{cannot} differentiate the pulsation modes at longer periods ($P>5.0-5.5\,$day) because $\phi_{21}$ and $\phi_{31}$ sequences of fundamental and first-overtone merge with each other \citep{AntonelloPoretti1990,Soszynski2008}. We stress that inability to identify the pulsation mode from the light-curve in this period range is not a matter of precision of the observation; it is rather an intrinsic physical property of classical Cepheids.

Alternatively, other photometric survey of Galactic Cepheids such as OGLE \citep{Soszynski2008,Soszynski2010} or \textit{Gaia} used a criterion based on the $R_{21}$-$P$ plane \citep{Clementini2016DR1}, which distinguishes the first-overtone and the fundamental-mode pulsators. This criterion is robust for a large samples of Cepheids, but can fail in individual cases. Indeed, we still encounter the same problem as in the case of $\phi_{21}$, namely, that light-curve $R_{21}$ sequences of the fundamental mode and the first-overtone merge for periods $P>5.5\,$day. Moreover, the light-curve $R_{21}$ of the fundamental-mode Cepheids depends on the metallicity \citep{Klagyivik2007,Szabados2012,Majaess2013,Hocde2023} and the location of the star within the instability strip \citep{Sandage2004,Sandage2009}. In particular, fundamental-mode Cepheids located close to the edge of the instability strip might have a low amplitude and, consequently, low $R_{21}$, in the range of those observed in the overtone pulsators. The fundamental-mode Cepheid Y~Oph is a good example of such a behaviour (this paper; Hocd\'e et al. 2023b, submitted). In the framework of specific processing and validation of all sky Cepheids from \textit{Gaia} DR3, \cite{Ripepi2022DR3Reclass_catalogue,Ripepi2023DR3} performed a re-classification of \textit{Gaia} DR2 and DR3 pulsation modes obtained from The Specific Objects Study (SOS) Cep\&RRL pipeline, after in-depth analysis of light-curves, period-luminosity relation and results from literature. For Cepheids of pulsation period above 5$\,$days, most of their re-classification change fundamental into first-overtone mode as shown in Table \ref{Tab:reclass}.

In the light of these difficulties, the mode identification based on
   $\phi_{21}$ of the RV curves is a more robust, and thus preferable, approach. As
   shown already by \cite{Kienzle1999}, the progression of Fourier phases
   for radial velocity curves is different than for the light curves and
   allows to discriminate between the overtone and the fundamental-mode at all
   periods (except a narrow range around $P = 5\,$day, where the two
   progressions cross). In fact, for Cepheids with $P > 5.5$ day analysis of
   the RV curves is the {\it only} unambiguous way to identify the pulsation
   mode. Using this property, we identify certainty V495~Cyg as a new overtone
   Cepheid. This star has been classified by \textit{Gaia} as
   fundamental-mode pulsators, but this cannot be considered reliable, because of its pulsation period above $5.5\,$day. We confirm overtone
   nature of several other Cepheids, classified as such by \textit{Gaia}: VZ~CMa,
   V351~Cep, DT~Cyg, V532~Cyg, V411~Lac and EV~Sct. We also confirm overtone
   nature of GH~Car, originally classified by \cite{Kienzle1999}. This last
   star has later been re-classified by \textit{Gaia} as a fundamental-mode pulsator.
   However, for $P = 5.7\,$day Gaia's classification based on the light curve
   is not reliable. Finally, we have examined the pulsation mode of Polaris, which was debated in the last decade \citep[see, e.g.][]{Turner2013,Neilson2014,Anderson2018}. Although a consensus emerged for Polaris to be a first-overtone pulsator based on different level of evidences (see \cite{Torres2023} and references therein), we are able to firmly conclude, on the basis of Fourier parameters of its radial velocity curve, that Polaris is an overtone pulsator.

\begin{figure} 
\begin{subfigure}{0.52\textwidth}
\includegraphics[width=\linewidth]{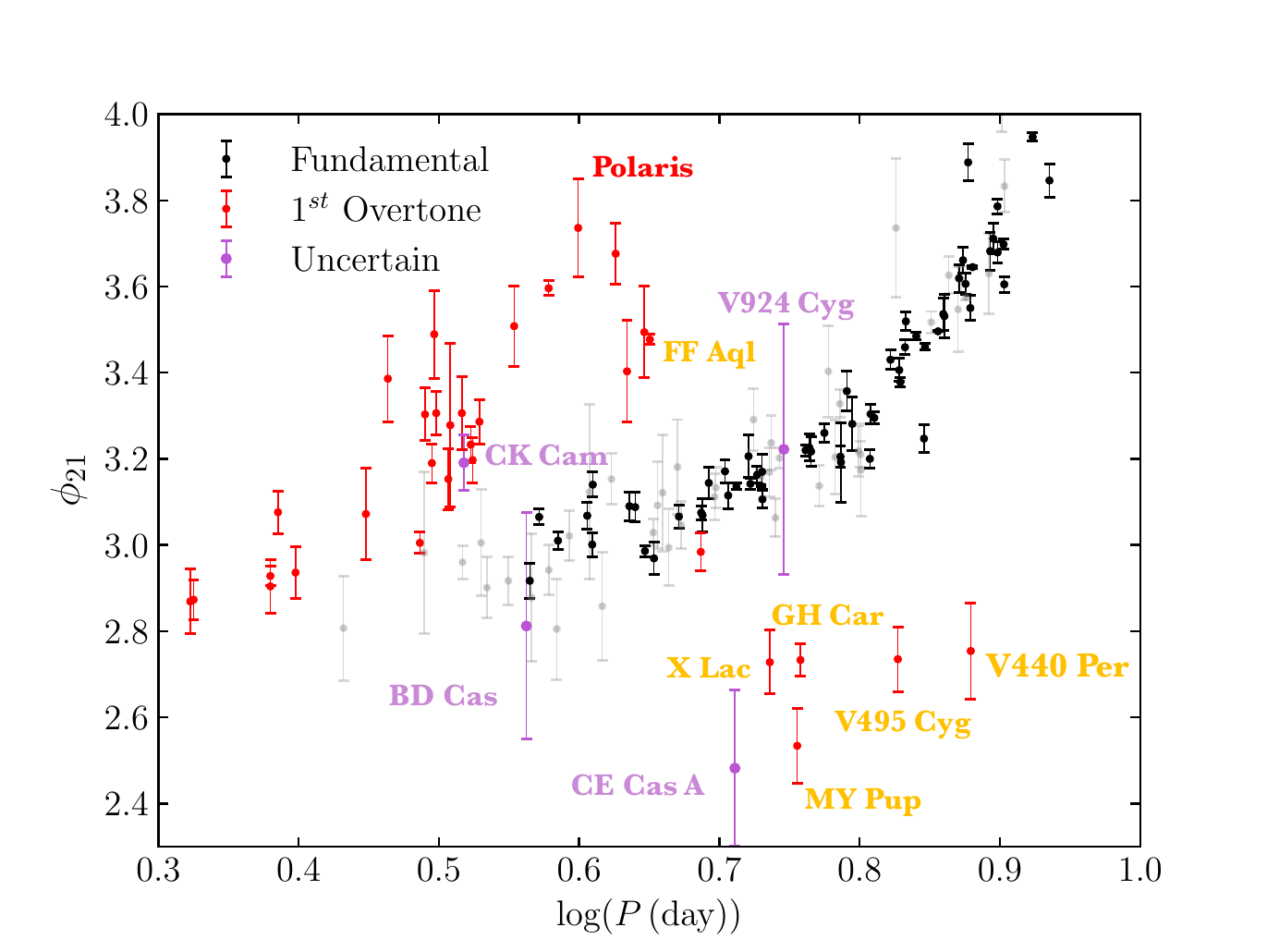}
\caption{}  \label{fig:1O}
\end{subfigure}\hspace*{\fill}

\begin{subfigure}{0.52\textwidth}
\includegraphics[width=\linewidth]{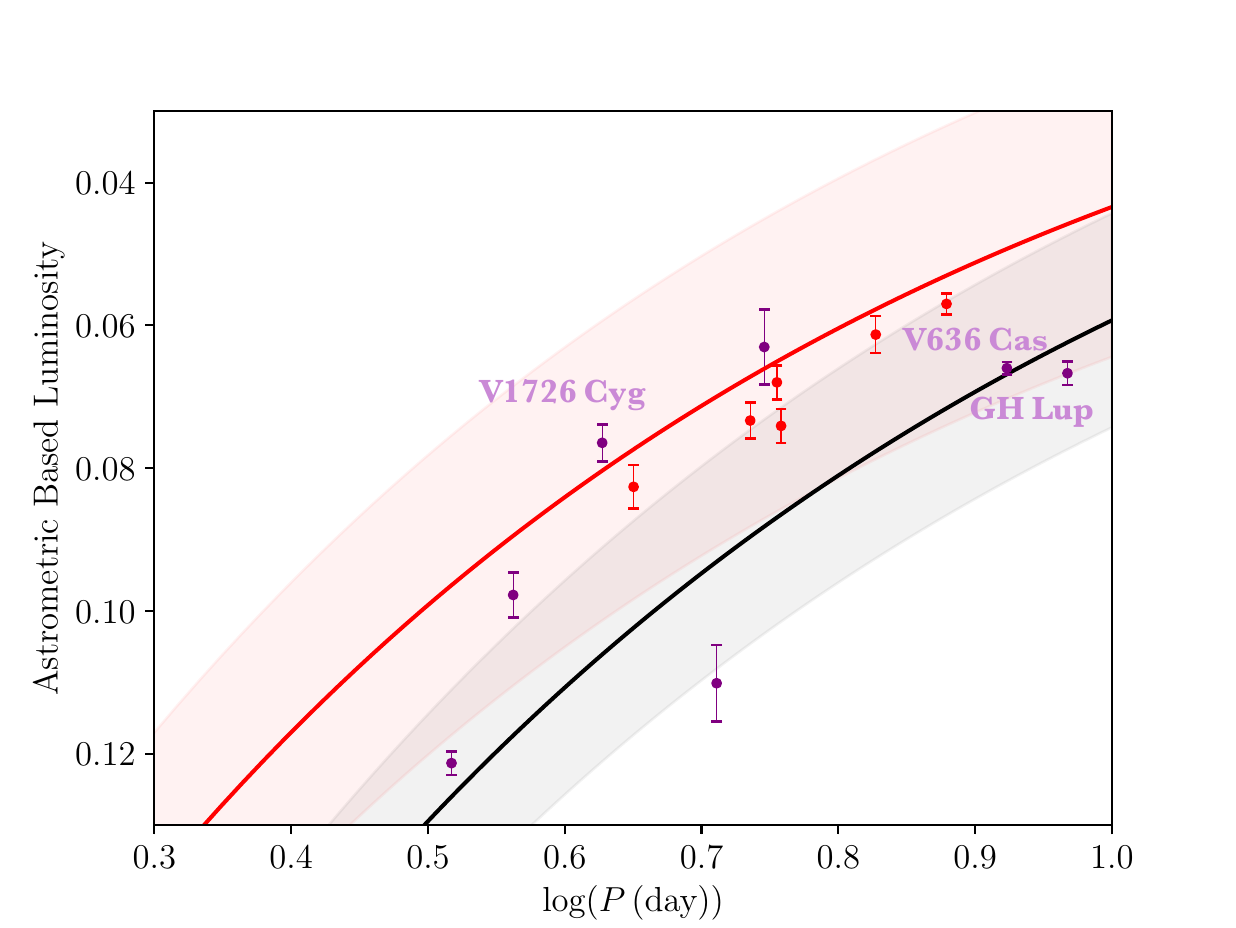}
\caption{}  \label{fig:ABL}
\end{subfigure}
\caption{(a) Velocity Fourier phases $\phi_{21}$ for first-overtone and short-period fundamental-mode Cepheids. Orange labels correspond to overtone Cepheids for which our results disagree either by {\it Gaia} SOS Cep\&RRL pipeline or reclassification from \citep{Ripepi2023DR3} (see Table \ref{Tab:reclass}). Purple symbols correspond to stars with pulsation mode uncertain or unknown according to our criteria based on $\phi_{21}$. V1726~Cyg and GH~Lup cannot be plotted since the 2nd Fourier term has not been detected in these stars, and V636~Cas has very high $\phi_{21} \simeq 4.5$ rad. All three variables are presented in the bottom plot. (b) Astrometric Based Luminosities in the Wesenheit G-band. Relations for the first-overtone and for the fundamental-mode Cepheids are also plotted as red and black curves, respectively.}
\end{figure}

\subsubsection{Comparison with mode identification of {\it Gaia} survey}

As discussed above, with the mode identification method used by the SOS Cep\&RRL pipeline from
   {\it Gaia}, some Cepheids might be misclassified. For example, the
   two well established first-overtones Cepheids MY~Pup \citep{Kienzle1999} and V440~Per \citep{Kienzle1999,Baranowski2009} are classified as Fundamental mode from \textit{Gaia} DR3 catalogue. In
   order to demonstrate the robustness of our criteria based on
   velocity $\phi_{21}$, we compared systematically our classification
   with that of {\it Gaia} DR3 catalogue. In total, we found 5 first-overtones
   Cepheids that are classified in {\it Gaia} DR3 as fundamental-mode
   pulsators : X~Lac, MY~Pup, GH~Car, V495~Cyg and V440~Per. We
   display these stars in Fig.~\ref{fig:1O} and mark them with orange labels. As expected, we find disagreements only for pulsation periods above
   5.4~days, since criteria based on the light-curves are ambiguous
   in this period range. Four of these stars have very recently been reclassified by \cite{Ripepi2023DR3} as first-overtone variables in agreement with our results. In Fig.~\ref{fig:1O} we also display several Cepheids, pulsation mode of which is uncertain or unknown according to our criteria, but can be identified according to {\it Gaia} DR3 (CK Cam, BD~Cas, CE~Cas~A, V924~Cyg). These variables are plotted as purple points. Finally, we also mark in the Figure the  Cepheid FF~Aql, that is the only Cepheid of our sample below 5$\,$days for which the results from \textit{Gaia} DR3 SOS Cep\&RRL pipeline and of \cite{Ripepi2023DR3} disagree. We summarize the mode
   identifications for all these stars in Table \ref{Tab:reclass}.


In Fig.~\ref{fig:ABL} we verify our mode identifications with the help of the astrometry-based luminosity (ABL) \citep{FeastCatchpole1997,ArenouLuri1999}, established from reddening-free Wesenheit magnitudes, $W_G$ as defined by \cite{Ripepi2019}, i.e. $W_G =
   G - 1.90(G_{Bp} - G_{Rp})$. The ABL vs. period relations displayed
   in Fig.~\ref{fig:ABL} for first-overtone and for fundamental-mode variables
   are calibrated with a clean sample of Cepheids (RUWE < 1.41,
   $\delta\omega/\omega < 0.2$) \citep{Ripepi2023DR3}. The relations are
   plotted together with their dispersion adopted from \cite{Ripepi2023DR3}. We have also derived observed ABL for our sample of stars
   (Table \ref{Tab:reclass}), using the apparent Wesenheit magnitudes and Gaia DR3 parallaxes \citep[corrected from zero-point,][]{Lindegren2021}. We have also applied residual parallax offset \citep{Riess2021parallax}.  As we can see from Fig.~\ref{fig:ABL}, all the stars that we have identified as long period first-overtone Cepheids (red symbols) are in excellent agreement with the ABL vs. period relation for the first overtone. Only for one star, GH~Car, the ABL plot yields ambiguous
   conclusion since the star falls between the ABL relations of the
   two modes, deviating from either of them by more than 3$\sigma$.
   Still, it is twice closer to the first overtone line. Among the
   stars with pulsation mode uncertain/unknown according to our
   criteria, BD~Cas, V1726~Cyg and V924~Cyg agree with the
   first-overtone ABL relation while CK~Cam, CE~Cas~A and GH~Lup are
   in better agreement with the fundamental-mode relation.

\subsubsection{Comments on individual overtone Cepheids}
Below we enclose comments on five individual stars : three identified long-period overtone Cepheids and on Polaris and FF~Aql. Comments on remaining overtone Cepheids and on Cepheids of uncertain/unknown mode identifications are deferred to Appendix~\ref{app:overtone} and \ref{app:unknown} respectively.

\textbf{X Lac, V495 Cyg}: The radial velocity curves of X Lac and V495 Cyg have already been discussed by \cite{Kienzle1999}, who referred to unpublished results of Krzyt, Moskalik, Gorynya et al. (1999). The Fourier amplitudes $A_1$ for both variables are above 10$\,$km/s and are rather typical for fundamental-mode Cepheids. On the other hand, the values of their RV Fourier phases $\phi_{21}$ are well below the fundamental-mode progression (by 5.2$\sigma$ and 8.9$\sigma$, respectively) and are similar to
those of long-period overtone Cepheids. Fourier phase $\phi_{31}$ is
available only for V495~Cyg and it deviates from the
fundamental-mode progression as well (by 7.6$\sigma$). \cite{Kienzle1999} have suggested that X~Lac and V495~Cyg might be overtone
pulsators, but they have stopped short of classifying them as such.
Following the approach of \cite{AntonelloPoretti1990}, who recommended
using for mode identification solely $\phi_{21}$ and $\phi_{31}$, we conclude
with certainty that both Cepheids are overtone pulsators. We also
provide their RV Fourier parameters for the first time. We note, that both stars are classified in {\it Gaia} DR3 catalog as fundamental-mode pulsators. Only X~Lac is re-classified in \cite{Ripepi2023DR3} as overtone pulsator. We have to keep in mind that at pulsation periods of 5.4 day and 6.7 day {\it Gaia's} mode identification, which is based on the light  curves, cannot be reliable \citep{AntonelloPoretti1990}.

\textbf{GH Car:} \cite{AntonelloPoretti1990} have pointed out that this Cepheid is
located in the light-curve Fourier plots at the period where the
first overtone and the fundamental-mode progressions merge.
Therefore, its mode of pulsation cannot be established on the basis
of the light-curve Fourier phases. With the radial velocity curve,
GH~Car has been identified as an overtone Cepheid by \cite{Kienzle2000}, who however have not published its Fourier parameters. We confirm overtone nature of GH~Car and provide its RV Fourier parameters for the first time. We note that GH~Car is classified in {\it Gaia} DR3 catalog as fundamental-mode pulsator, but in \cite{Ripepi2023DR3} as overtone pulsator. For the same reason as stated in the previous case, using the RV curves is the only way to unambiguously identify the pulsation mode in this star.

{\bf FF~Aql:} our first-overtone identification aligns well with
   {\it Gaia} DR3 catalog, but \cite{Ripepi2023DR3} re-classifies it
   as a fundamental-mode pulsator. The
   value of velocity Fourier phase of this Cepheid, $\phi_{21}=3.477
   \pm 0.012$, places it firmly among overtone pulsators and
   29$\sigma$ away from the fundamental-mode progression (see
   Fig.~\ref{fig:1O}). The light-curve Fourier phase $\phi_{21}$ is measured
   less accurately, but its value leads to the same mode
   identification \citep{AntonelloPoretti1986}. On this basis we can
   conclude with certainty that FF~Aql is an overtone pulsator.

\textbf{$\alpha$ UMi (Polaris):} pulsation mode of Polaris has been
vigorously debated in the last decade \citep[see, e.g.][]{vanLeeuwen2013,Turner2013,Neilson2014,Anderson2018,Evans2018}. This variable has
the lowest pulsation amplitude of all Cepheids and, consequently,
almost sinusoidal light and radial velocity curves. Using Fourier
phases to establishing its pulsation mode was so far unsuccessful. The primary objective of this study is to detect and characterize the first harmonic of the RV curve of Polaris, which is a notably low amplitude component (below 0.1km/s). This necessitates a large number of measurements to beat down the errors. A challenging aspect lies in the fact that Polaris exhibits fast variations in both its amplitude and period \citep{neilson12polarismdot,Usenko2018,Anderson2019,Usenko2020}. For this reason, phasing together long datasets yields excessive scatter of the RV curve. To address this problem, we require a short dataset to ensure that neither amplitude nor phase undergo significant variability
   during the observations. At the same time, it is imperative that
   this short dataset consists of a large number of measurements.
   Regrettably, this criterion is not met when examining any of the
   subsets within the very precise dataset of \cite{Anderson2019}.
   However, the conditions are met by a subset of data from \cite{Eaton2020}. The subset is 83-day long, from HJD=2454713 to HJD=2454796 and consists of 167 data points. This dataset yields Fourier fit with dispersion of 0.09$\,$km/s and allows to determine the RV Fourier phase of $\phi_{21} = 3.736 \pm 0.113$.

The measured value of $\phi_{21}$ is much more accurate, but within error it is the same as determined previously by \cite{MoskalikOgloza2000}. This is
so, despite an over 50\% increase of the pulsation amplitude between the two
epochs. The new value of $\phi_{21}$ fits very well the overtone progression and deviates from the fundamental-mode progression by 5.4$\sigma$. Thus, for the first time we are able to firmly demonstrate with the RV Fourier
parameters, that Polaris is an overtone Cepheid. Our mode
identification agrees with earlier result of \cite{MoskalikOgloza2000}, who reached the same conclusion with a very different
method, namely with measuring the phase lag between light and the
radial velocity curves. We note, that the mode identifications of
our paper and of \cite{MoskalikOgloza2000} depend neither on
distance to Polaris nor on its interstellar reddening.

\begin{table}[]
\begin{center}
\caption{\small Comparison of mode identification of our work and of {\it Gaia} survey.
}\label{Tab:reclass}
\begin{tabular}{l|c|c|c|c}
\hline
\hline
Star      & Period & GDR3 & R23 & This work\\
\hline
CK Cam     & 3.29  &    F  &   - & U/F\\
BD Cas     & 3.65  &    1O  &   -  & U/1O\\
V1726 Cyg     & 4.24  &    1O  &   -  & U/1O\\
FF Aql     & 4.47  &    1O  &   F  & 1O\\
CE Cas A     & 5.14  &    F  &   -  & U/F\\
X Lac     & 5.44  &    F  &   1O  & 1O\\
V924 Cyg     & 5.57  &    F  &   1O  & U/1O\\
MY Pup     & 5.69  &    F  &   1O  & 1O\\
GH Car     & 5.73  &    F  &   1O  & 1O\\
V495 Cyg     & 6.72  &    F  &   -  & 1O\\
V440 Per     & 7.57  &    F  &   1O  & 1O\\
V636 Cas     & 8.38  &    F  &   -  & U/F\\
GH Lup     & 9.28  &    F  &   -  & U/F\\
\hline
\hline
\end{tabular}
\normalsize
    \begin{tablenotes}
    \item \textbf{Notes :} For each star we indicate the fundamental or first-overtone mode identification from The Specific Objects Study Cep\&RRL pipeline \citep{gaia2023} (GDR3), re-classification  from \cite{Ripepi2023DR3} (R23) when applicable, and our identification based on radial velocity $\phi_{21}$. For uncertain or undetermined mode (U), we also indicate the mode identified from PL-ABL relation (see Fig.~\ref{fig:ABL}).
    \end{tablenotes}
\end{center}
\end{table}

\subsection{Detected spectroscopic Cepheid binaries}\label{sect:binary_results}
In five Cepheids of our sample we have found statistically very significant drift of the $\gamma$-velocity. Such $\gamma$-velocity variation might be caused by orbital motion in a long period binary system. For two of these variables, BP Cir and LR~TrA, the binary nature has been already recognized
\citep{Petterson2005,Szabados2013a}. To our knowledge the spectroscopic binarity of the other three Cepheids is identified here for the first time.

\textbf{VY Per:} we have found a slow $\gamma$-velocity drift, amounting to $\approx$7.0$\,$km/s in 500$\,$days. The observed change of $\gamma$-velocity indicates a binary orbital motion, with the period much longer than the duration of the observations.

\textbf{AQ Pup:}  \cite{Anderson2016a} noticed a small fluctuation of the $\gamma$-velocity in older archival data, but they attributed this value to problem of phase folding the data because of fast and irregular period changes. They also reported no significant $\gamma$-velocity variations within their precise data set. We reached a different conclusion by analyzing in details the single very precise dataset of \cite{Anderson2016a}. Fitting simultaneously pulsation RV curve and a trend, we have found a slow, approximately parabolic $\gamma$-velocity drift, amounting to $\simeq 2.0$ km/s in 800 days. Considering an excellent quality of the data, this drift is highly significant. To refine the analysis, we have divided the measurements into four separate subsets and determined the $\gamma$-velocity for each subset independently. This method of analysis is not sensitive to a possible variations of the pulsation period which might affect the phasing of the RV curve. We have found that the $\gamma$-velocity of AQ~Pup has changed between JD=2456652 and JD=2457459 by $\Delta V_{\gamma} = 1.81 \pm 0.045\,$km/s. The observed change of $\gamma$-velocity indicates a binary orbital motion, with the period much longer than the duration of the observations.

\textbf{QZ Nor:} the analysis of this star has been performed with a single very accurate dataset of \cite{anderson14}. Fitting
   simultaneously pulsation RV curve and a trend, we have found a
   slow, linear $\gamma$-velocity drift, amounting to $\simeq 0.5$
   km/s in 1000 days. Considering an excellent quality of the data,
   this drift is highly significant. Proceeding in the same was as in
   case of AQ~Pup, we have divided the measurements into four
   separate subsets and determined the $\gamma$-velocity for each of
   them independently. We have found that the $\gamma$-velocity of
   QZ~Nor has changed between JD=2455708 and JD=2456689 by $\Delta
   V_{\gamma} = 0.48 \pm 0.046$ km/s. The observed change of
   $\gamma$-velocity also indicates a binary orbital motion, with the
   period much longer than the duration of the observations.
\section{Discussion on Fourier parameters trend}\label{sect:discussion}
\subsection{Amplitude of the first Fourier term}
In case of the fundamental-mode Cepheids of short periods ($P<10\,$day), we observe in Fig.~\ref{fig:A1}, a very sharply defined upper envelope limit of the pulsation amplitude. This limit defines empirically a maximum $A_1$ of Galactic Cepheids to be between 16.5 and 15$\,$km/s, depending on the pulsation period. Conversely, the amplitudes of Cepheids are distributed below this upper limit. In case of the light curve, the amplitude of fundamental-mode Cepheids is highest on the blue edge of the instability strip and decreases towards the red edge \citep{Sandage2004,Sandage2009,Hocde2023}. It has been suggested in the literature that the radial velocity amplitude might be correlated with the location within the instability strip as well \citep{Cogan1980,Pont2001}. 

The light curve amplitude is also sensitive to the metallicity \citep{Klagyivik2007,Szabados2012,Majaess2013,Hocde2023}. In order to test if the RV amplitudes of Cepheids with $P<10\,$day depend on metallicity, we cross-matched our data with metallicity determined with the HR spectroscopy \citep{Luck2018}. We found no significant correlation between the velocity amplitude $A_1$ and the spectroscopic metallicity of the Cepheid. This is different from result of \citep{Szabados2012}, who found that the RV amplitude decreases with [Fe/H]. Extending the Cepheid sample to metal-poor objects of the Magellanic Clouds seems necessary to clarify this issue.

For the fundamental-mode Cepheids with $P>10\,$day, we observe in Fig.~\ref{fig:A1} a sharp rise of the pulsation amplitude $A_1$ between 10$\,$ and 15$\,$day. The upper limit envelope of $A_1$ reaches a broad maximum of about 25$\,$km/s for periods of 15-21$\,$day. For example, SV Mon ($P=15.23\,$day) and RZ Vel ($P=20.40$day) have amplitudes of $A_1=24.87\pm0.18\,$km/s and $A_1=24.64\pm0.03\,$km/s respectively. Some stars pulsate with much lower amplitudes though, e.g. V340 Nor, SZ Car and Y Oph, all with $A_1<10\,$km/s. There is only a small number of Cepheids with pulsation period longer than 21$\,$day. Their amplitude generally declines with increasing period, but with a few noticeable exceptions (e.g. AQ Pup). 

Finally, in Fig.~\ref{fig:A1} we see an upper limit envelope of the pulsation amplitude
also for the overtone Cepheids with $P<5\,$day. This limit is not
as sharply defined as for the fundamental-mode. Nevertheless, it clearly defines a maximum amplitude $A_1$ of overtone Cepheids to be between 9.5km/s and 9.0km/s, depending on the pulsation period.


\subsection{Amplitude ratio $R_{21}$, $R_{31}$ and $R_{41}$}
For the fundamental-mode Cepheids, the amplitude ratio of the second over the first Fourier component $R_{21}$  displays an almost symetric Z-shape centered on $R_{21}=0.3$ and $P=10\,$day (see Fig.~\ref{fig:R21}). The maximum of $R_{21}$, which corresponds to the most asymmetrical RV curves, is located at $P\approx8\,$day. In our sample, the maximum value is reached by W~Gem with $R_{21}=0.61\pm0.01$ at $P=7.91\,$day. Given the smooth variation of $A_1$ for $P<10\,$day, we can show that this sharp rise of $R_{21}$ is caused by the amplitude of the second Fourier component $A_2$. Interestingly, $R_{21}$ of the light curves display a somewhat different trend, with more asymmetrical curves (highest $R_{21}$) at $P\approx2.5 \,$day followed by decrease of light curve asymmetry until $P\approx10 \,$day \citep[see, e.g.][]{Soszynski2008,Soszynski2010,Pietrukowicz2021}. Thus, the amplitude $A_2$ of the light-curves starts decreasing already at $P\simeq 2.5\,$day, while for the RV curves it continues to increase up to  $P \simeq 8\,$day. Moreover, $R_{21}$ of the $V$-band light curve is known to have a minimum value at about $P \simeq 10\,$day, i.e. close to the resonance period, while in case of the RV curves the minimum of $R_{21}$ occurs at a longer period of $\sim12.5\,$day. For example, we found $R_{21}=0.029\pm0.002$ for XY Car ($P$=12.44$\,$day).  Although a quantitative comparison is needed, these differences between light and RV curves seem to be reproduced by hydrodynamical models \citep{Moskalik1992,Paxton2019}.

The plots of $R_{31}$ and $R_{41}$ appear more scattered and both parameters have larger relative uncertainties than $R_{21}$. This is expected since the third and fourth Fourier component have lower amplitudes and, thus, are less accurately determined from the data. Nevertheless, our amplitude ratios are precise enough to observe a first maximum of $R_{31}$ at a period close to 6 day. Among quality 1 Cepheids in our sample, the highest value are reached by AW Per ($R_{31}=0.25\pm0.01$) and XX Sgr ($R_{31}=0.24\pm0.01$) at $P\approx6.5\,$day. For $P>6.5\,$day a sudden decline of $R_{31}$ is observed, however it is difficult to precisely locate the minimum, because of a deficiency of our Cepheids sample with periods between 8 and $10\,$day. This defiency could be due to the so-called \textit{zone of avoidance} \citep{Genderen1970,Buchler1997,Szabados2012,Udalski2018}. \cite{Moskalik1992} postulated that this dip which appears around 8$\,$day could be associated with the $P_4/P_0=1/3$ resonance between the fundamental and fourth overtone \citep{Moskalik1989}. $R_{31}$ has its second maximum at around $P=10\,$day, then it declines again, reaching minimum around 17$\,$day, i.e. at a period slightly longer than in case of $R_{21}$. Beyond 17$\,$day, $R_{31}$ increases again and, similarly to $R_{21}$, it reaches its next maximum at around $P\simeq 40\,$day. In the case of $R_{41}$ the dispersion of the data is too large to describe the behaviour of the fourth Fourier term. 
\subsection{Fourier phase $\phi_{21}$}

Low-order Fourier parameters of fundamental-mode Cepheids presented in Fig.~\ref{fig:fourier} display a characteristic variation shaped by the $P_2/P_0=0.5$ resonance around 10$\,$day.  At short-periods below about 8$\,$day, we are able to confirm the tight progression of Fourier phases with the pulsation period as observed by \cite{Moskalik2000}, although this time with a better precision of the parameters. This tight progression was also predicted by hydrodynamical models \citep{Buchler1990}. In that study, they show that the dispersion of $\phi_{21}$ can be used to derive the width of the instability strip. 
On the other hand, we observe a large scatter of $\phi_{21}$ for periods between 9 to 13$\,$day. This can be explained by the strong sensitivity of $\phi_{21}$ to the value of
resonant period ratio $P_2/P_0$, which at a given period depends on the
fundamental parameters of the star (mainly metallicity, but also on mass,
luminosity and temperature). This sensitivity of $\phi_{21}$ to stellar
parameters is strongest within the resonance region, as predicted by
hydrodynamical calculation \citep[see, e.g.,][]{Buchler1990,Moskalik1992,Buchler1997}. 
By comparing the light curves of Cepheids in the Milky Way with those in metal-poor galaxies, 
it appears that with decreasing metallicity the resonance center shifts towards longer periods \citep{Beaulieu1998,Bhardwaj2015}. 
That implies that any dispersion of [Fe/H] within the Cepheid sample will cause the dispersion $P_2/P_0$ at a given pulsation period \citep{Buchler1990}. We note, that for our Cepheids with periods of 9-13$\,$days, the spectroscopic metallicities of \cite{Luck2018} range from [Fe/H]=$-0.2$ to $+0.3\,$dex. This range of metallicity could be large enough to explain, at least in part, the large scatter of $\phi_{21}$ observed in the resonance region.


At longer periods we observe an almost constant value of radial
velocity $\phi_{21}$ from about $P=13\,$day to $P=70\,$day. In the case of
light curves this branch appears in general more scattered, even
for a small sample of Galactic Cepheids \citep{Pont2001,SimonMoffett1985,Hocde2023}. This scatter is attributed to differences in metallicity between the stars \citep{Hocde2023}. Since the velocity $\phi_{21}$ in long-period Cepheids is almost constant, the observed metallicity effect on light-curve $\phi_{21}$ is likely of photospheric origin. Additional RV curves for long-period Cepheids are necessary to analyze this issue in more detail.




\subsection{Fourier phases $\phi_{31}$ and $\phi_{41}$}\label{sect:phi41}
The Fourier phase $\phi_{31}$ undergoes a rather smooth, monotonic increase across the entire range of pulsation periods. Contrary to $\phi_{21}$, it does not display a sharp variation in the vicinity of the $P_2/P_0=0.5$ resonance, but only a somewhat increased
 scatter in the progression. The intrarelations of Fourier phases
presented in Sec.~\ref{sect:intrarelation} show in the resonance region a linear correlation between $\phi_{21}$ and $\phi_{31}$. The phase $\phi_{31}$ increases moderately fast with the pulsation period up to $P\approx15\,$day, then for $P=15-20\,$day it undergoes a very steep increase (possibly a jump) by ~2$\,$rad. At even longer periods, we observe a very slow increase of $\phi_{31}$ from about 5 to 6$\,$rad. Thus, in the long-period Cepheids, both lowest order Fourier phases display a slow variation with the pulsation period.

The behavior of $\phi_{41}$ is somewhat different. This Fourier phase continues its steep, almost linear increase up to the longest period of 70$\,$day. Contrary to $\phi_{31}$, it increases with nearly constant rate also at periods between 15 and 70$\,$day. Despite the amplitude ratio $R_{41}$
versus period diagram being very scattered, the Fourier phase $\phi_{41}$ shows a very tight progression with the pulsation period. We also confirm that there is no break of $\phi_{41}$ around 7$\,$day, contrary to what is observed for the light-curves \citep{SimonMoffett1985}. The difference between $\phi_{41}$ of RV curves and of light-curves was first noted by \cite{kovacs90} who attributed it to unknown photospheric phenomena. 




\subsection{High-order Fourier parameters}
The presence of resonance features affecting higher order Fourier components, due to $P_0/P_1=3/2$ at 24$\,$day and to $P_0/P_3=3$ at 27$\,$day, was suggested by \cite{Antonello1996,Antonello2009}. Theoretical high-order Fourier parameters were also computed with hydrodynamical model \citep{Aikawa2000} to investigate these effects. In Fig.~\ref{fig:fourier_high} we displayed high-order Fourier parameters (5th, 6th and 7th Fourier terms) for the RV curves of all Cepheids of our sample. Fourier phases $\phi_{51}$ and $\phi_{61}$ display a linear trend over the entire range of pulsation periods similarly to $\phi_{41}$. At a close inspection, we notice that additional structures seem to be present in the plots. In case of $\phi_{51}$, we observe an excessive scatter at periods close to 10$\,$day, i.e. close to the resonance center. In this period range, several Cepheids deviate significantly upwards from the general $\phi_{51}$ progression. On the other hand, the progression of $\phi_{61}$ vs. pulsation period seems to split into two parallel sequences, that coexist for periods between 10 and 20$\,$day. We have to note, however, that this observation is based on a rather limited number of long-period Cepheids. The existence of the two $\phi_{61}$ sequences has to be confirmed when more RV curves becomes available.

The amplitude ratios $R_{51}$ and $R_{61}$ appear to be more scattered, therefore it is difficult to distinguish any trends (see Figs.~\ref{fig:R51} and \ref{fig:R61}). 
However, a weak minimum around the resonance period at $P\sim$10$\,$day is visible, especially in case of $R_{61}$ (see in Fig.~\ref{fig:R61}).
From these data, we are not able to identify any resonance features suggested by \cite{Antonello1996,Antonello2009}. Interestingly, the amplitude ratios of low and high order reach a secondary maximum at long periods, before a sharp decrease that follows when periods becomes even longer. For $R_{21}$, $R_{31}$ and $R_{41}$ the maximum occurs at $P=45\,$day, and for $R_{51}$ and $R_{61}$ at $P=27\,$day. As we can see from Hertzsprung progression displayed in Fig.~\ref{fig:FU_progression}, the latter value roughly coincides with a period at which a secondary bump reaches the top of the radial velocity curve (see T Mon). 


\subsection{Intrarelation of Fourier parameters}\label{sect:intrarelation}
\begin{figure*} 
\begin{subfigure}{0.52\textwidth}
\includegraphics[width=\linewidth]{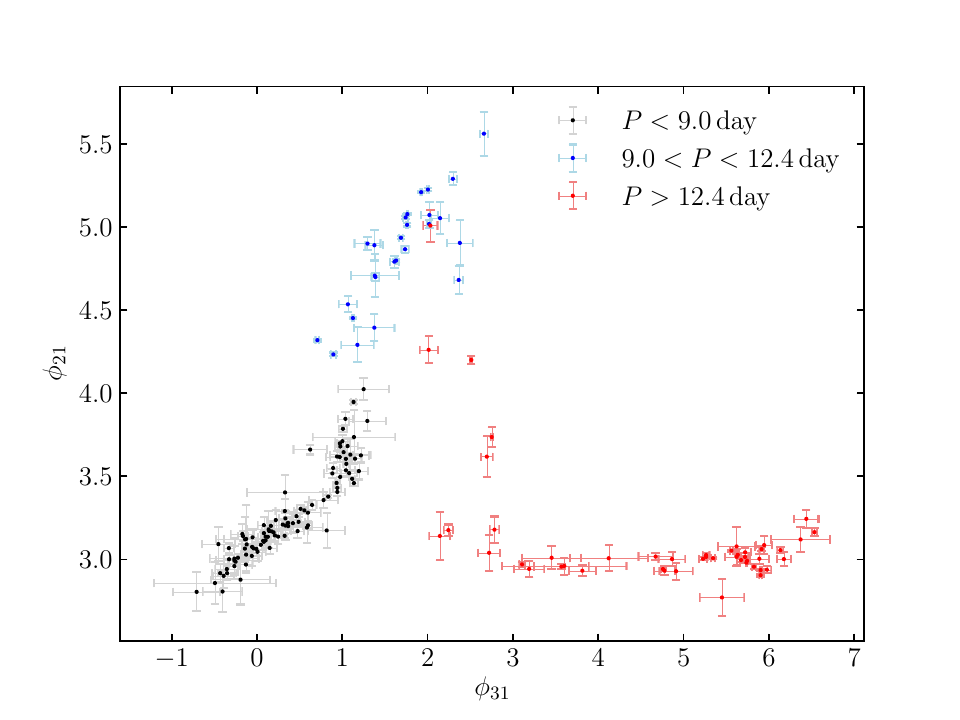}
\caption{} \label{fig:phi21_phi31}
\end{subfigure}\hspace*{\fill}
\begin{subfigure}{0.52\textwidth}
\includegraphics[width=\linewidth]{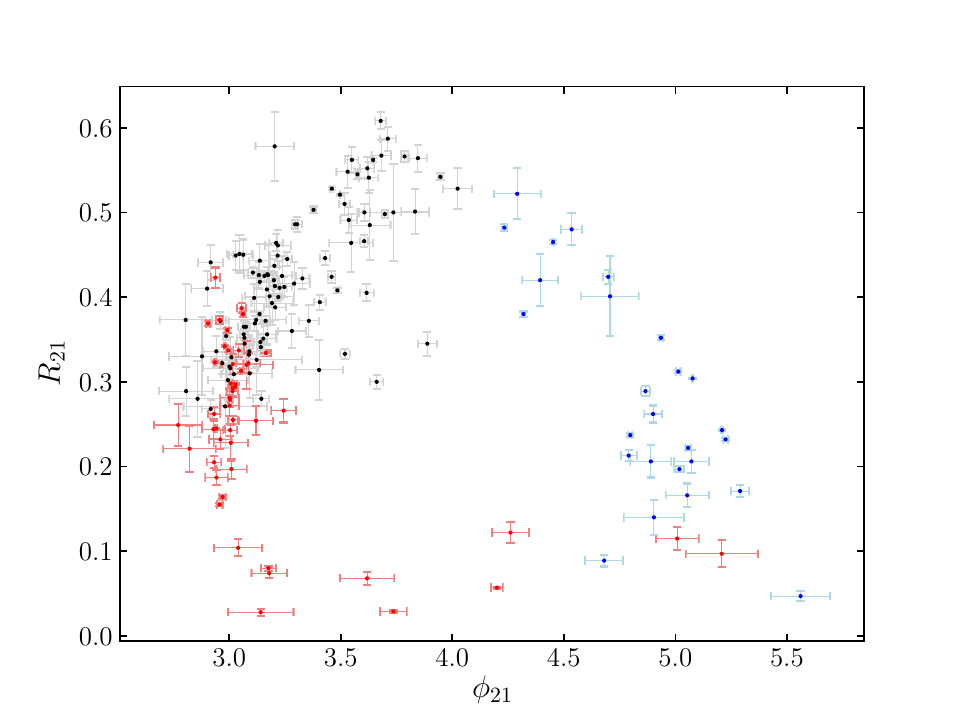}
\caption{} \label{fig:phi21_R21}
\end{subfigure}

\medskip
\begin{subfigure}{0.52\textwidth}
\includegraphics[width=\linewidth]{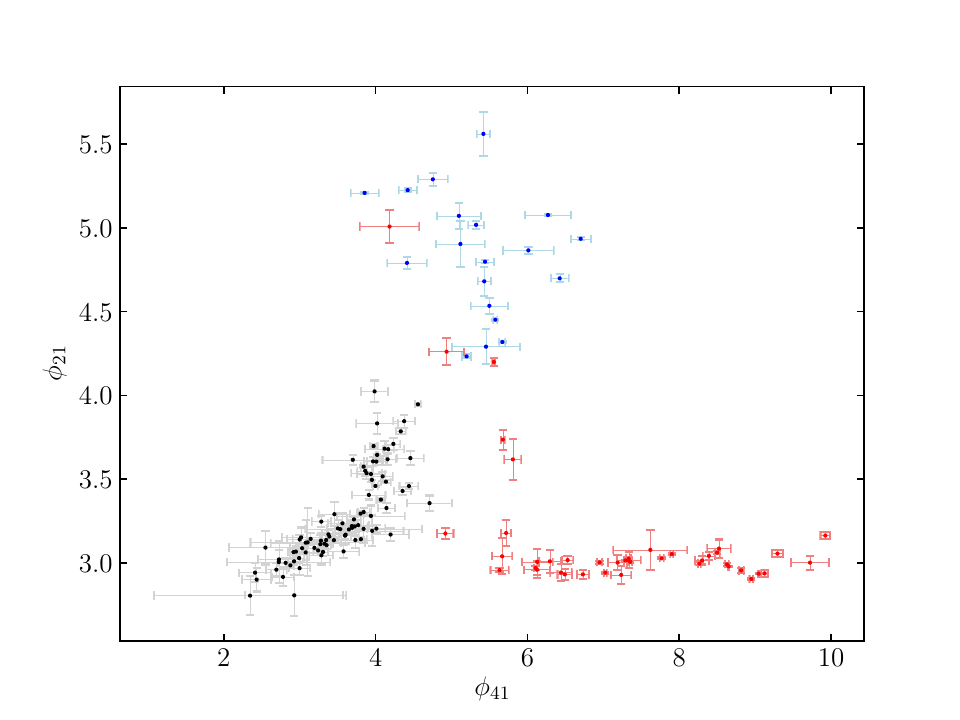}
\caption{} \label{fig:phi21_phi41}
\end{subfigure}\hspace*{\fill}
\begin{subfigure}{0.52\textwidth}
\includegraphics[width=\linewidth]{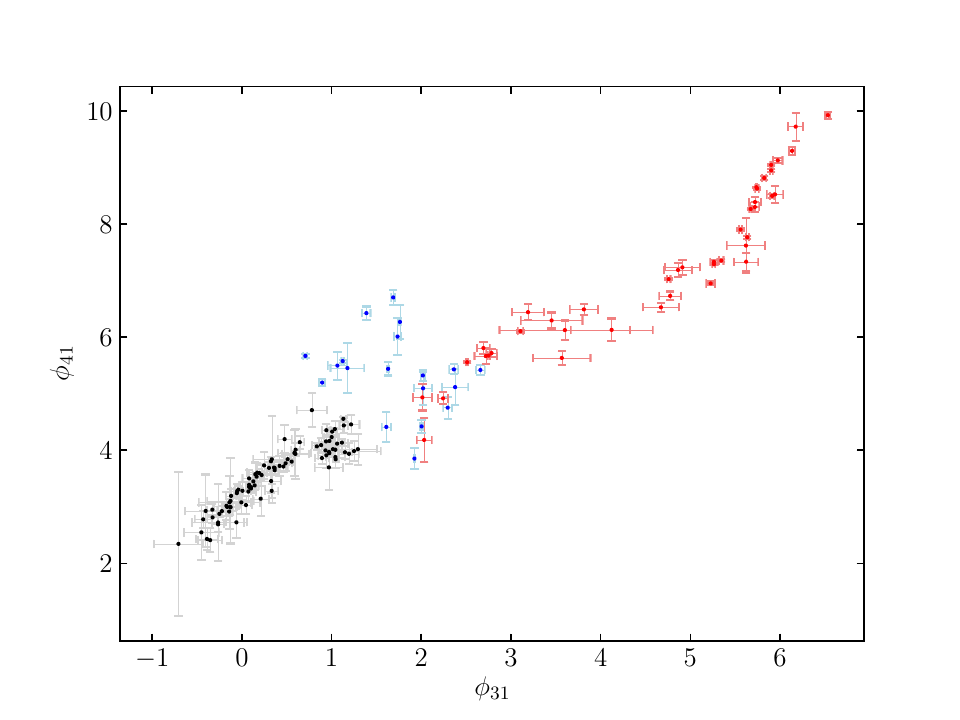}
\caption{} \label{fig:phi41_phi31}
\end{subfigure}
\caption{\small Intrarelation between low-order Fourier phases, and intrarelation between Fourier phase $\phi_{21}$ and amplitude ratio $R_{21}$. The meaning of the different symbols are indicated in the figure (a). \label{fig:intra_phi}}
\end{figure*}

We inspected intrarelation between Fourier parameters, in particular between the Fourier phases, similarly to \cite{kovacs90} and \cite{SimonMoffett1985} in the case of RV curves and light curves respectively. In the latter case, the authors found a discontinuity between the short-period ($P<10\,$day) and the long-period Cepheids and speculated that it might be caused by a different attractor or saturation mechanism, as originally proposed by \cite{Klapp1985}. From Fig.~\ref{fig:intra_phi} we see however that there is no apparent break at any phase-phase diagram. In particular, we observe a continuous variation of $\phi_{41}$ vs. $\phi_{31}$ which is not the case for the light curves \citep[see Fig. 17 in][]{SimonMoffett1985}, in agreement with our comments in Sect.~\ref{sect:phi41}. Moreover, $\phi_{31}$ and $\phi_{41}$ seem to be tightly correlated outside the resonance region. As expected, the intrarelations displayed in Fig.~\ref{fig:intra_phi} are less scattered than the ones presented by \cite{kovacs90} (see their Figure~5). We can notice for the first time a qualitative agreement between the shape of $\phi_{21}$ vs. $\phi_{31}$ relation determined observationally and that computed with hydrodynamical pulsation models of \cite{Buchler1990} presented in \cite{kovacs90}. Therefore, it is now possible to use Fourier parameters of RV curves to constrain hydrodynamical pulsation models. 

For Cepheids in the resonance region between 9 and 12.4$\,$day, we observe a
   linear relation between $\phi_{21}$ and $\phi_{31}$ (blue dots in Fig.~\ref{fig:intra_phi}). This suggests that in this period range the variations of these two Fourier phases are likely sensitive to the same stellar parameters. However, when $\phi_{41}$ is plotted vs. $\phi_{21}$ or vs. $\phi_{31}$ for the same stars, we see no
   trend but rather an unstructured cloud of points (Figs.~\ref{fig:phi21_phi41} and \ref{fig:phi41_phi31}). This
   suggests that the variation of $\phi_{41}$ in the resonance vicinity depends on different parameters than in case of $\phi_{21}$ or $\phi_{31}$. Last, the variation of $R_{21}$ with $\phi_{21}$ (Fig.~\ref{fig:phi21_R21}) displays an interesting loop, where $\phi_{21}$ variation lags behind the $R_{21}$ variation. In particular, Cepheids in the resonance region between 9 and 12.4$\,$day are characterized by a decrease of $R_{21}$ while $\phi_{21}$ continues to increase. In this period range, the relation between the two seems to be linear although with a large dispersion.

\section{Conclusions}\label{sect:conclusion}
In this paper, we took the advantage of the observational effort of the last decades which allows precise determination of the Fourier parameters of Cepheid velocity curves. These parameters are extremely valuable for constraining the dynamics of hydrodynamical pulsation models and for studying resonances between modes. The collected radial velocity curves will also be very useful for distance determination with parallax-of-pulsation method.
In this work we presented a careful analysis of each Cepheid independently. First of all, we selected for each star the highest quality RV measurements available in the literature. Since the datasets were often obtained with different instruments, we corrected for zero-point offsets between the datasets. We removed the contribution from orbital motion and residual trends. For four longest period variables in our sample, we also corrected for the rapid and irregular period change. We then used these data to perform a Fourier fit of the radial velocity curve of each Cepheid. The resulting Fourier fits were classified into two categories depending on their quality, taking into account the stability of the fit, the phase coverage and the dispersion of fit residuals.

Our final sample consists of 178 fundamental and 33 first overtone Cepheids with new precise radial velocity Fourier parameters. We also report 7 additional Cepheids whose pulsation mode is uncertain or undetermined.
We displayed Fourier parameters up to order six, we analyzed these parameters qualitatively and we compared some characteristic features to the literature. In 3 Cepheids, namely  VY~Per, AQ~Pup and QZ~Nor we have found previously unreported significant $\gamma$-velocity variations, indicative of a binary (or multiplicity) motion with long orbital period. We have also confirmed $\gamma$-velocity variations in LR~TrA and BP~Cir.
We identified V495 Cyg as a new first-overtone Cepheids and we confirmed the first-overtone nature of several others including Polaris. Obtaining more RV measurements for Cepheids located near the resonance center and for metal-poor Cepheids of the Magellanic Clouds, or of the outskirt of the Galaxy, will be essential to fully investigate the internal physics and pulsation dynamics of these variables. The pulsation period coverage and the precision obtained, in particular for Fourier phase $\phi_{21}$, will be useful for studying the dynamics of Cepheid pulsations with the help of hydrodynamical models. Further radial velocity measurements from modern high-resolution spectroscopic instruments will be valuable to improve these results.

\begin{acknowledgements}
We are grateful to astronomers who dedicated years of work to the measurement of radial velocity of Cepheids used in this paper. We thank Louis Balona, Jurek Krzesi{\'n}ski, Simon O'Toole and Frédéric Pont for making available to us their unpublished RV data. Natalia Gorynya thanks the Simeiz Observatory for generously granting the observing time on 1m telescope during 33 years of her Cepheid radial velocity program. VH, RS, OZ and RSR are supported by the National Science Center, Poland, Sonata BIS project 2018/30/E/ST9/00598. This research made use of the SIMBAD and VIZIER databases at CDS, Strasbourg (France) and the electronic bibliography maintained by the NASA/ADS system. This research also made use of Astropy, a community-developed core Python package for Astronomy \citep{astropy2018,astropy2022}.
\end{acknowledgements}
\bibliographystyle{aa}  
\bibliography{bibtex_vh} 

\begin{appendix}

\section{Table of references}
\onecolumn
\begin{center}
\begin{longtable}{lS[table-format=2.7]l|lS[table-format=2.7]l}
\caption{Sources of RV data used for Fourier decomposition in Sect.~\ref{sect:fourier}. }\\
\hline
\hline
\noalign{\smallskip}
\multicolumn{1}{l}{Star} &
\multicolumn{1}{r}{Period (day)} &
\multicolumn{1}{l|}{Reference} & 
\multicolumn{1}{l}{Star} &
\multicolumn{1}{r}{Period (day)} &
\multicolumn{1}{l}{Reference} \\ \hline 
\endfirsthead

\multicolumn{6}{c}%
{{\bfseries \tablename\ \thetable{} -- continued from previous page}} \\
\hline \multicolumn{1}{l}{Star} &
\multicolumn{1}{r}{Period (day)} &
\multicolumn{1}{l|}{Reference} &
\multicolumn{1}{l}{Star} &
\multicolumn{1}{r}{Period (day)} &
\multicolumn{1}{l}{Reference} \\ \hline 
\endhead

\hline \multicolumn{6}{r}{{Continued on next page}} \\ \hline
\endfoot

\hline \hline
\endlastfoot 
 $\eta$ Aql   & 7.176483   &  59& DL Cas    & 8.000616   &  12,14,32,41,55,57\\
 U Aql     & 7.0241494  &  11,14,22                &  FM Cas    & 5.809314   &  12,34,35                \\
 SZ Aql    &17.13956    &  9,11&  VW Cas    & 5.993868   &  36\\
 TT Aql    &13.754690   &  9,11,14,36,46,63,73&  V379 Cas  & 4.306696   &  34,35,36,47             \\
 FF Aql    & 4.4710413  &  14&  V636 Cas  & 8.375898   &  5,12,14,16,33,34,35\\
 FN Aql    & 9.481502   &  9,33,34,35,78&  V Cen     & 5.494326   &  11,28,49\\
 FM Aql    & 6.114113   &  9,33,34,35,36&  XX Cen    &10.95298    &  11,49\\
 KL Aql    & 6.107969   &  34,35,36,81&  AZ Cen    & 3.21175    &  47                     \\
 V336 Aql  & 7.304228   &  36&  KN Cen    &34.02126    &  4\\
 V496 Aql  & 6.807073   &  14,35,36,39,49,73&  V339 Cen  & 9.466516   &  19,70\\
 V600 Aql  & 7.238803   &  33,34,35,57,65&  $\delta$ Cep & 5.3663114  &  3\\
 V733 Aql  & 6.178632   &  33,34,35,40,81&  CP Cep    &17.86306    &  33,34,35,57,65\\
 V916 Aql  &13.443415   &  33,34,35,36&  CR Cep    & 6.23328    &  12,57\\
 V1162 Aql & 5.376202   &  33,34,35,36,39&  IR Cep    & 2.1142427  &  33,34,35,36,53\\
 V1344 Aql & 7.476736   &  49,78&  V351 Cep  & 2.805441   &  33,34,35,36\\
 V1496 Aql &65.924      &  36&  AV Cir    & 3.0652483  &  14,47\\
 V340 Ara  &20.80997    &  61,65&  AX Cir    & 5.273513   &  49,62\\
 RT Aur    & 3.7283085  &  23,33,34,35,36,48,64&  BP Cir    & 2.3981031  &  14,27,62\\
 RX Aur    &11.624021   &  46,64,69&  R Cru     & 5.825907   &  11,49,80\\
 SY Aur    &10.14552    &  4&  S Cru     & 4.689719   &  11,14\\
 YZ Aur    &18.19442    &  33,36,66,75&  T Cru     & 6.73336    &  11\\
 AN Aur    &10.289049   &  33,36&  AG Cru    & 3.837327   &  28,51,70\\
 BK Aur    & 8.002618   &  33,36&  BG Cru    & 3.342530   &  14\\
 RW Cam    &16.41383    &  14,36&  X Cyg     &16.385983   &  9,12,14,15,17,33,34,35,36,48,72\\
 RX Cam    & 7.912200   &  45&  SU Cyg    & 3.8455476  &  14,35,36,44,52\\
 CK Cam    & 3.294826   &  35,36,48                     &  SZ Cyg    &15.11073    &  4\\
 RY CMa    & 4.678352   &  8,33,34,35,36&  TX Cyg    &14.71162    &  36,64                        \\
 SS CMa    &12.35272    &  4&  VX Cyg    &20.13565    &  33,34,35,36                  \\
 VZ CMa    & 3.12633    &  47                &  VY Cyg    & 7.857234   &  36\\
 $\ell$ Car     &35.56103    &  2&  VZ Cyg    & 4.8643308  &  9,12,14,73                 \\
 U Car     &38.83867    &  11,14,82&  BZ Cyg    &10.14206    &  33,34,35,36\\
 V Car     & 6.69669    &  50,51,70&  CD Cyg    &17.074744   &  4,14,46\\
 UX Car    & 3.682284   &  50,51,70,71&  DT Cyg    & 2.4993188  &  35,36,48\\
 VY Car    &18.882747   &  4&  GH Cyg    & 7.817964   &  36\\
 WZ Car    &23.01552    &  11,19                        &  MW Cyg    & 5.954885   &  33,34,35,36,45\\
 XX Car    &15.71113    &  20,26&  V383 Cyg  & 4.61219    &  34,35,36,81\\
 XY Car    &12.436283   &  4,11&  V386 Cyg  & 5.257624   &  33,34,35,36\\
 XZ Car    &16.652244   &  4&  V402 Cyg  & 4.3649260  &  33,34,35,36,46,57\\
 YZ Car    &18.167617   &  4,11,14,27,49,62&  V438 Cyg  &11.21099    &  36,57\\
 AQ Car    & 9.769513   &  4&  V459 Cyg  & 7.251457   &  34,35,36,57,65\\
 FZ Car    & 3.57839    &  47                   &  V495 Cyg  & 6.718163   &  34,35,36,57\\
 GH Car    & 5.72523    &  76&  V520 Cyg  & 4.049071   &  33,34,35,36\\
 GI Car    & 4.43075    &  47                     &  V532 Cyg  & 3.283841   &  33,34,35,36,57\\
 HW Car    & 9.199134   &  4&  V538 Cyg  & 6.119064   &  34,35,36\\
 IT Car    & 7.5396     &  47                      &  V924 Cyg  & 5.571274   &  34,35,36\\
 RS Cas    & 6.295991   &  33,34,35,36                  &  V1154 Cyg & 4.9254682  &  21,33,34,35,36,46,79\\
 RW Cas    &14.789173   &  33,34,35,36&  V1334 Cyg & 3.332401   &  14\\
 RY Cas    &12.13966    &  34,35,36&  V1726 Cyg & 4.23745    &  34,35,36\\
 SU Cas    & 1.94933659 &  12,14,33,34,35,36,48,72&  V2340 Cyg & 7.96628    &  36\\
 SW Cas    & 5.440971   &  35,36&  $\beta$ Dor  & 9.8426613  &  14,63\\
 SY Cas    & 4.0711614  &  12,33,34,35,36,46&  $\zeta$ Gem &10.149357   &  14,22,73\\
 SZ Cas    &13.6401     &  18&  W Gem     & 7.913356   &  33,34,35,36,46,64\\
 XY Cas    & 4.50170    &  42&  AA Gem    &11.303215   &  46,75\\
 BD Cas    & 3.650692   &  33,34,35,36,47,81&  AD Gem    & 3.78778    &  43\\
 BY Cas    & 3.221953   &  33,34,35,36&  DX Gem    & 3.138357   &  11,12\\
 CD Cas    & 7.800801   &  36,65&  BB Her    & 7.507939   &  33,34,35,46,81\\
 CE Cas A  & 5.141061   &  8,33,35&  V Lac     & 4.982979   &  33,34,35,36\\
 CF Cas    & 4.875114   &  12,33,35,54,57,65      &  X Lac     & 5.444560   &  9,12\\
 DD Cas    & 9.8120375  &  4,12&  Y Lac     & 4.3237751  &  9,14,46\\
 Z Lac     &10.885804   &  9,33,34,35,36,45,74&  V335 Pup  & 4.861300   &  1\\
 RR Lac    & 6.416294   &  12,46&   U Sgr     & 6.7453096  &  14,33,34,35,56,72   \\
 BG Lac    & 5.331941   &  9,46&   W Sgr     & 7.5950154  &  14,22\\
 V411 Lac  & 2.908265   &  36&   X Sgr     & 7.012724   &  14,60,73\\
 GH Lup    & 9.2802     &  19&   Y Sgr     & 5.7733802  &  11,14,73\\
 T Mon     &27.03531    &  22&   WZ Sgr    &21.85040    &  4\\
 SV Mon    &15.23420    &  46&   XX Sgr    & 6.424252   &  36,39,73\\
 TW Mon    & 7.097067   &  66,67                   &   YZ Sgr    & 9.55400    &  36,73\\
 TZ Mon    & 7.428134   &  67                      &   AP Sgr    & 5.058064   &  36\\
 UY Mon    & 2.3982528  &  46,47&   AV Sgr    &15.41110    &  58,61,65\\
 XX Mon    & 5.456542   &  67                      &   BB Sgr    & 6.637116   &  34,35,73\\
 BE Mon    & 2.7037     &  12&   V350 Sgr  & 5.1542393  &  14,22,39,62\\
 CU Mon    & 4.707564   &  66,67                   &   S Sge     & 8.3821181  &  9,14,22,25,48\\
 CV Mon    & 5.378662   &  56,58,72&   GY Sge    &51.767      &  36\\
 FG Mon    & 4.4966203  &  67                      &   RV Sco    & 6.06109    &  6,28,51\\
 FI Mon    & 3.2878237  &  67                      &   RY Sco    &20.32508    &  4\\
 FT Mon    & 3.421772   &  66,67             &   V482 Sco  & 4.527826   &  10,28,51\\
 V508 Mon  & 4.1349     &  12&   V500 Sco  & 9.31660    &  36,49\\
 V510 Mon  & 7.457495   &  66,67                  &   V636 Sco  & 6.7969588  &  13,14,62\\
 R Mus     & 7.510302   &  14&   V950 Sco  & 3.380119   &  47\\
 S Mus     & 9.659889   &  14,27,62,82&   X Sct     & 4.198056   &  36,39,57,58,68\\
 RT Mus    & 3.086126   &  70,71&   Y Sct     &10.34139    &  11,36                        \\
 S Nor     & 9.7544350  &  14,27,56,82&   Z Sct     &12.90207    &  4\\
 U Nor     &12.64438    &  11,65&   RU Sct    &19.70510    &  14,34,35,36,56\\
 SY Nor    &12.645439   &  11,12,49,58&   SS Sct    & 3.671338   &  36,38,39                     \\
 QZ Nor    & 3.786834   &  1&   TY Sct    &11.05418    &  36,57,58\\
 V340 Nor  &11.288642   &  39,56,58&   UZ Sct    &14.747749   &  57,61,65\\
 Y Oph     &17.12575    &  14,22,63&   CK Sct    & 7.416227   &  36,57,58\\
 BF Oph    & 4.067537   &  36&   CM Sct    & 3.916994   &  49,57,58,65\\
 RS Ori    & 7.566967   &  46&   EV Sct    & 3.091102   &  56,57,72\\
 GQ Ori    & 8.616400   &  46&   CR Ser    & 5.301266   &  33,34,35,36\\
 SV Per    &11.129454   &  30,35,36,64,66             &   ST Tau    & 4.0342386  &  12,35,36,46\\
 UX Per    & 4.56542    &  75&   SZ Tau    & 3.1488148  &  12,29,33,34,35,36,48\\
 UY Per    & 5.36500    &  18,57&   EU Tau    & 2.1024881  &  12,31,33,34,35,36\\
 VX Per    &10.882532   &  4&   R TrA     & 3.389248   &  14,28\\
 VY Per    & 5.53185    &  18&   S TrA     & 6.32321    &  6,28,50,51\\
 AW Per    & 6.4636344  &  37&   LR TrA    & 2.428405   &  77\\
 BM Per    &22.96194    &  33,36,64,66&   $\alpha$ UMi & 3.97463    &  22\\
 V440 Per  & 7.57214    &  7                     &   RY Vel    &28.1201     &  11,19,65\\
 X Pup     &25.95909    &  4&   RZ Vel    &20.40291    &  77\\
 RS Pup    &41.5849     &  1&   SW Vel    &23.427444   &  11,63,65\\
 VZ Pup    &23.17518    &  14,73&   BG Vel    & 6.92389    &  77\\
 AP Pup    & 5.084534   &  47                    &   AH Vel    & 4.227412   &  14\\
 AQ Pup    &30.19016    &  4&   CS Vel    & 5.904798   &  56,58\\
 BC Pup    & 3.544217   &  66,67                   &   DR Vel    &11.1954     &  19,65\\
 BN Pup    &13.672484   &  65,73&   S Vul     &69.43       &  4,14\\
 HW Pup    &13.45711    &  67                       &   T Vul     & 4.4354084  &  9,12,14,24,48\\
 LS Pup    &14.1432     &  39,73&   U Vul     & 7.9907009  &  9,12,14,33,34,35,36,45\\
 MY Pup    & 5.694712   &  14,47&   X Vul     & 6.319652   &  12,33,34,35\\
           &            &       &  SV Vul    & 44.995      &  12,14,72  \\
\end{longtable}
\end{center}
    \begin{tablenotes}
    \item \textbf{Notes :} We list the pulsation period that we have derived for each star and which is used in the Fourier fit, see Sect.~\ref{sect:fourier} for details.
    \item \textbf{References :}  1: \cite{anderson14}, 2: \cite{anderson2016c}, 3: \cite{anderson15a}
    , 4: \cite{Anderson2016a}
    , 5: \cite{ArellanoFerro1984}, 6: Balona, L. A. (private communication), 7: \cite{Baranowski2009}, 8: \cite{Barnes1988}, 9: \cite{Barnes2005RV}, 10: \cite{Berdnikov2010}, 11: \cite{Bersier2002}, 12: \cite{Bersier1994}, 13: \cite{BohmVitense1998}, 14: \cite{borgniet2019}, 15: \cite{breitfellner93c}, 16: \cite{BurkiBenz82}, 17: \cite{butler93}, 18: \cite{Coker1989}, 19: \cite{Coulson1985a}, 20: \cite{Coulson1985b}, 21: \cite{Derekas2017},  22: \cite{Eaton2020}, 23: \cite{Evans76}, 24:  \cite{Evans1986}, 25: \cite{evans1993}, 26: \cite{Feast1967}, 27: \cite{Gallenne2019}, 28: \cite{gieren1981}, 29: \cite{Gieren1985RTAurSZtaur}, 30: \cite{Gieren1992}, 31: \cite{Gieren1989}, 32: \cite{Gieren1994}, 33: \cite{Gorynya1992}, 34: \cite{Gorynya1996Samus}, 35: \cite{Gorynya1998}, 36: Gorynya, N. A. et al. (in preparation), 37: \cite{Griffin2016}, 38: \cite{Groenewegen2008}, 39: \cite{Groenewegen2013}, 40: \cite{HarrisWallerstein84}, 41: \cite{Harris1987}, 42: \cite{Imbert1981}, 43: \cite{Imbert1983}, 44: \cite{Imbert1984}, 45: \cite{Imbert1996}, 46: \cite{Imbert1999},  47: \cite{Kienzle1999}, 48: \cite{KissVinko2000}, 49: Krzesi{\'n}ski, J. (private communication), 50: \cite{LloydEvans1968}, 51: \cite{LloydEvans1980}, 52: Madore, B. F., from \cite{Evans1988}, 53: \cite{Marschall1993},  54: \cite{Matthews1995}, 55: \cite{Mermilliod1987}, 56: \cite{mermilliod2008},  57: \cite{Metzger1991}, 58: \cite{Metzger1992},  59: \cite{Nardetto2023}, 60: O'Toole, S. (private communication), 61: \cite{Pedicelli2010}, 62: \cite{Petterson2004}, 63: \cite{Petterson2005}, 64: Pont, F. (private communication), 65: \cite{pont1994burki}, 66: \cite{Pont1997}, 67: \cite{Pont2001}, 68: \cite{Ripepi2021}, 69: \cite{Schmidt1974}, 70: \cite{Stibbs1955}, 71: \cite{Stobie1979}, 72: \cite{storm04}, 73:  \cite{storm11a}, 74: \cite{Sugars1996}, 75: \cite{Szabados1998}, 76: \cite{Szabados2013a},
    77: \cite{Szabados2013b}, 78: \cite{Szabados2014}, 79: \cite{Szabo2011}, 80: \cite{Usenko2014}, 81: \cite{Vinko1998}, 82: \cite{Wallerstein1992}.  \label{Tab:data_ref}
    \end{tablenotes}

\newpage
\section{Binary Cepheids}
\begin{table*}[h!]
\caption{Orbital periods of binary Cepheids adopted for the Fourier fit.}\label{Tab:binary}
\begin{center}
\begin{tabular}{l|l|l}
\hline
\hline
Star  & {$P_\mathrm{orb}$ (day)} & Source \\
\hline
U Aql    &    1862.8  $\pm$1.1    & This paper\\
FF Aql   &    1428.5  $\pm$1.6    & This paper\\
V496 Aql &    1351.31 $\pm$0.60   & This paper\\
V916 Aql &    2434    $\pm$26     & Gorynya et al. (in prep.)\\
V1344 Aql&    6294    $\pm$30     & Gorynya et al. (in prep.)\\
RX Cam   &    1113.8  $\pm$0.5    & \cite{Groenewegen2013}\\
YZ Car   &     829.68 $\pm$0.18   & This paper\\
BY Cas   &    4584    $\pm$33     & Gorynya et al. (in prep.)\\
DL Cas   &     684.27 $\pm$0.16   & \cite{Groenewegen2013}\\
XX Cen   &     711.7  $\pm$3.5    & This paper\\
AX Cir   &    6532    $\pm$25     & \cite{Petterson2004}\\
SU Cyg   &     549.085$\pm$0.026  & This paper\\
VZ Cyg   &    2183    $\pm$10     & \cite{Groenewegen2013}\\
MW Cyg   &     439.61 $\pm$0.18   & \cite{Groenewegen2013}\\
V1334 Cyg&    1932.8  $\pm$1.8    & \cite{Gallenne2018}\\
DX Gem   &    1043.1  $\pm$4.8 :  & This paper\\
Z Lac    &     382.63 $\pm$0.10   & \cite{Groenewegen2008}\\
S Mus    &     505.335$\pm$0.037  & This paper\\
SY Nor   &     551.86 $\pm$0.27   & This paper\\
AW Per   &   13590    $\pm$120    & This paper\\
S Sge    &     675.389$\pm$0.096  & This paper\\
W Sgr    &    1630.4  $\pm$3.3    & This paper\\
V350 Sgr &    1466.99 $\pm$0.78   & This paper\\
V636 Sco &    1320.6  $\pm$1.3    &  \cite{Gallenne2019}\\
U Vul    &    2510    $\pm$2.8    & \cite{Groenewegen2008}\\
\hline
\end{tabular}
\normalsize
\end{center}
\begin{tablenotes}
    \item \textbf{Notes :} The orbital period $P_\mathrm{orb}$ is a period of additional Fourier sum representing orbital motion, see Sect.~\ref{sect:binary} for details. 
\end{tablenotes}
\end{table*}
\subsection{Comments on binary Cepheids}
\noindent \textbf{V1344 Aql:} the recent study of \cite{V1344Aql2023} derived for
   this star orbital period of about $P_{orb} \simeq 12000$ day,
   which is about twice longer than the period derived by Gorynya et
   al. (in prep). We scrutinized both datasets to understand the
   origin of this discrepancy. The dataset of Gorynya et al. spans more than 23 years, with nearly 200
   measurements uniformly distributed in time. This dataset yields
   unambiguously the orbital period close to 6300 days and is not
   compatible with the period of $\sim 12000$ days. Thus, the orbital
   period found by \cite{V1344Aql2023} can be excluded.
\newline
\newline
\noindent \textbf{XX Cen:} \cite{Szabados1990} determined orbital period of this binary Cepheid to be
$P_\mathrm{orb}=909.4\,$day. Later, \cite{Groenewegen2008} has refined this value to
$P_\mathrm{orb}=924.1\,$day. Using only the most recent high
quality RV data and assuming circular orbit \citep[as done by][]{Groenewegen2008}, we
find somewhat shorter period of $P_\mathrm{orb}=711.7\,$day. The resulting fit has a dispersion of only 0.34$\,$km/s, which is more than 4 times lower than for the fit in which $P_\mathrm{orb}$ from \cite{Groenewegen2008} is assumed.
\newline
\newline
\noindent \textbf{DX Gem:} this overtone Cepheid is a spectroscopic binary \citep{Bersier2002}. The total range of orbital velocity variations is $\sim20\,$km/s, but the currently available data are not sufficient to uniquely determine the orbital period. Among several possibilities, we have chosen $P_\mathrm{orb}=1043.1\,$day, which yields the lowest dispersion of the fit. The Fourier parameters of the pulsation RV curve are almost insensitive to the choice of $P_\mathrm{orb}$.
\newline
\newline
\noindent \textbf{SY Nor:} this is another spectroscopic binary \citep{Bersier2002}. The
total range of orbital velocity variations is $\sim30\,$km/s. To the best
of our knowledge, the full orbital solution was never published, but
the unique determination of orbital period is already possible. Our
value of $P_\mathrm{orb}=551.86\,$day is almost the same as 551.7$\,$days, the value found by
\cite{Bersier2002}.
\newline
\newline
\noindent \textbf{BP Cir:} the presence of hot companion of BP~Cir was detected for the first time by International Ultraviolet Explorer satellite \citep{ArellanoMadore1985,Evans1992} and later revealed by long-baseline interferometric observations \citep{Gallenne2019}. From data of \cite{Petterson2004} alone, we have found a slow $\gamma$-velocity drift, amounting to $\approx$7.0$\,$km/s in 2300$\,$days. Variability of $\gamma$-velocity of BP Cir has already been noticed
by Petterson et al., who wrote that "observations indicate
continuing decrease in $\gamma$-velocity". The observed drift
indicates a binary orbital motion, with the period much longer than the duration of the observations.
\newline
\newline
\noindent \textbf{LR TrA:} a slow $\gamma$-velocity drift has been noticed in this
Cepheid already by \cite{Szabados2013b}. We have confirmed this
finding, we have found a $\gamma$-velocity change of ~1.5$\,$km/s in 450$\,$days. Considering an excellent quality of analyzed data, a drift of this size is highly significant. The observed change of
$\gamma$-velocity indicates a binary orbital motion, with the period much longer than the duration of the observations.

\onecolumn
\begin{center}
\begin{longtable}{lll|lll}
\caption{New orbital periods of binary Cepheid adopted in this paper compared with the previously published values.}\label{Tab:binary_comp}\\
\hline
\hline
\noalign{\smallskip}
\multicolumn{1}{l}{Star} &
\multicolumn{1}{l}{ {$P_\mathrm{orb}$ (day)}} &
\multicolumn{1}{l|}{Reference} &
\multicolumn{1}{l}{Star} &
\multicolumn{1}{l}{ {$P_\mathrm{orb}$ (day)}} &
\multicolumn{1}{l}{Reference} \\ \hline 
\endfirsthead

\multicolumn{6}{c}%
{{\bfseries \tablename\ \thetable{} -- continued from previous page}} \\
\hline \multicolumn{1}{l}{Star} &
\multicolumn{1}{l}{ {$P_\mathrm{orb}$ (day)}} &
\multicolumn{1}{l|}{Reference}&
\multicolumn{1}{l}{Star} &
\multicolumn{1}{l}{ {$P_\mathrm{orb}$ (day)}} &
\multicolumn{1}{l}{Reference}  \\ \hline 
\endhead

\hline \multicolumn{6}{r}{{Continued on next page}} \\ \hline
\endfoot

\hline \hline
\endlastfoot 
\hline
U Aql &      1862.8  $\pm$1.1     &  This paper                   & S Mus     &       505.335$\pm$0.037&  This paper\\
      &     1831.4  $\pm$6.5      &   \cite{Gallenne2019}         &           &        506.3  $\pm$0.5 &   \cite{Gallenne2019}\\
      &     1853.6  $\pm$3.0      &  \cite{Groenewegen2013}       &           &     505.15 $\pm$0.08&   \cite{Groenewegen2008}\\
      &       1856.4  $\pm$4.3    &  \cite{Welch1987}             &           &      504.9  $\pm$0.1 &  \cite{Petterson2004}\\
      &                           &                               &          &        504.58          & \cite{Petterson1998}\\
FF Aql &     1428.5  $\pm$1.6     &  This paper                   &           &       505.44 $\pm$0.35& \cite{evans1990}\\
      &      1430.3  $\pm$2.6     &  \cite{Gallenne2019}          &           &    506.4  $\pm$2     & \cite{EvansLloyd1982}\\
      &     1432.7  $\pm$0.9      &  \cite{Groenewegen2013}       &           &                        & \\
      &      1432.4  $\pm$1.1     &  \cite{Groenewegen2008}       & AW Per    &     13590    $\pm$120  &  This paper\\
      &      1434    $\pm$1       &  \cite{benedict07}            &           &     13954    $\pm$181  &   \cite{Griffin2016}\\
      &      1434.2               & \cite{GOrynya2000PROCEEDING}  &           &      14293.70 $\pm$283 &  \cite{Groenewegen2013}\\
      &      1433                 & \cite{Gorynya1996Rastorguev}  &           &      14594    $\pm$324 &  \cite{Evans2000AWPer}\\
      &       1429.7  $\pm$1.1    & \cite{evans1990}              &           &      13100    $\pm$1000 & \cite{Welch1989}\\
      &         1435.0.           &    \cite{Abt1959}             &           &                         &                  \\
      &                           &                               & S Sge     &       675.389$\pm$0.096&  This paper\\
V496 Aql &    1351.31 $\pm$0.60   &  This paper                   &           & 675.72 $\pm$0.04&  \cite{Groenewegen2008}\\
      &     1351.5  $\pm$5.8      &   \cite{borgniet2019}         &           &    675.75          & \cite{GOrynya2000PROCEEDING}\\   
      &     1066.2  $\pm$1.9      &   \cite{Groenewegen2013}      &           &    675.77          & \cite{Gorynya1996Rastorguev}\\   
      &       1331    $\pm$6.5    & \cite{Groenewegen2008}        &           &    675.99 $\pm$0.14&  \cite{BreifellnerGillet93b}\\    
      &       1447                &\cite{GOrynya2000PROCEEDING}   &           &     675.79 $\pm$0.18&  \cite{evans1993}\\    
      &                           &                               &           &    676.2           & \cite{Slovak1987}\\    
YZ Car&       829.68 $\pm$0.18    &  This paper                   &           &    682             & \cite{Herbig1952}\\    
      &        831.6  $\pm$0.9    & \cite{Gallenne2019}           &           &                     & \\
      &          830.22 $\pm$0.34 &  \cite{Anderson2016a}         &   W Sgr   &      1630.4  $\pm$3.3  &  This paper\\
      &        657.3  $\pm$0.3    &   \cite{Petterson2004}        &           &      1615.5  $\pm$11.0&   \cite{Gallenne2019}\\
      &        850    $\pm$11     &   \cite{Coulson1983YZCar}     &           &      1651    $\pm$8.8 &  \cite{Groenewegen2008}\\
      &                           &                               &           &      1582    $\pm$3   &  \cite{benedict07}\\
XX Cen&       711.7  $\pm$3.5     &  This paper                   &           &      1582    $\pm$9   &   \cite{Petterson2004}\\
      &     924.1  $\pm$1.1    &  \cite{Groenewegen2008}       &           &        1603             & \cite{AlbrowCottrell1996}\\
      &        909.4  $\pm$29.0   &  \cite{Szabados1990}          &           &      1740    $\pm$43  &  \cite{Bersier1994}\\
      &       ~800                &  \cite{Coulson1986}           &           &      1780    $\pm$5   &  \cite{Babel1989}\\
      &                           &                               &           &                           &                  \\
SU Cyg&       549.085$\pm$0.026   &  This paper                   & V350 Sgr  &   1466.99 $\pm$0.78 &  This paper\\
      &     538.5  $\pm$4.2    &  \cite{borgniet2019}          &           &  1465.3  $\pm$0.4 & \cite{Gallenne2019}\\
      &     549.24 $\pm$0.02   &   \cite{Groenewegen2008}      &           &  1468.9  $\pm$0.9 &  \cite{Groenewegen2013}\\
      &        549.25             & \cite{GOrynya2000PROCEEDING}  &           &    1472.91 $\pm$1.33&   \cite{Evans2011V350Sgr}\\
      &        549.16 $\pm$0.08   &  \cite{Evans1988}             &           &     1482    $\pm$2.4 &  \cite{Groenewegen2008}\\
      &        549.20 $\pm$0.03   & \cite{Imbert1984}             &           &  1464.9  $\pm$0.4 &  \cite{Petterson2004}\\
      &                           &                               &           & 1481.8           & \cite{GOrynya2000PROCEEDING}\\
SY Nor&       551.88 $\pm$0.27    &  This paper                   &           & 1477.0  $\pm$3.6 &   \cite{evanssugars1997}\\
      &        551.7              & \cite{Bersier2002}            &           &  1467             & \cite{Gorynya1996Rastorguev}\\
      &                            &                              &           & 1129    $\pm$24  &  \cite{Szabados1990}\\
      &                            &                              &           &                  & \\
\hline
\end{longtable}
\normalsize
\end{center}

\newpage
\section{Fourier parameters}
\subsection{Comments on individual Cepheids of Tables C.1-C.3}\label{sect:individual}
\noindent \textbf{KL Aql, V733 Aql:} in older literature these two variables are considered to be Pop.$\,$II Cepheids \citep[see, e.g.,][]{Balog1997,Vinko1998}. Recently, they have been both reclassified as Classical Cepheids, i.e. Cepheids of Pop.$\,$I \citep{Groenewegen2017}. Our analysis of the RV curves confirms this new classification. All Fourier parameters of KL~Aql and V733~Aql firmly place these two variables among fundamental-mode Classical Cepheids.
\newline
\newline
\noindent \textbf{TT Aql:} individual zero-point corrections are applied separately to each season of Gorynya in prep. measurements. These offsets change from year to year in a systematic way, suggesting a binary motion with a very small peak-to-peak amplitude of $\Delta V_r \sim1.0 \,$km/s.
\newline
\newline
\noindent \textbf{SV Vul, GY Sge, V1496 Aql, S Vul:} these four long-period Cepheids all display very fast, irregular $O-C$ variations. To phase their velocity curves properly, we have taken into account the $O-C$ shifts, calculated for each season separately. This procedure is described in detail in Sect.~\ref{sect:OC_correction}.
\newline
\newline
\noindent \textbf{V Cen:} New data of \cite{Bersier2002} and Krzesi{\'n}ski (private comm.)
do not provide a good coverage of the RV maximum. Therefore, we
have included into the fit also older data of \cite{gieren1981}. Because of a very long timebase (6614$\,$d), phasing of all these data with a common period does not work too well and yields RV curve with excessive scatter. In order to improve the phasing, we have applied individual $O-C$ shifts to each season separately. This reduces the scatter of the RV curve by $\sim$10\% and improves overall stability of the fit.
\newline
\newline
\noindent \textbf{$\ell$ Car:} because of amplitude instability, only data collected after
HJD=2457000 are used in the analysis. The lack of amplitude stability might be linked to cycle-to-cycle spectral line and atmospheric velocity gradient variability observed in this star \citep{anderson16lcar,anderson2016c}.
\newline
\newline
\noindent  \textbf{YZ Car:} because of lower quality, older data of \cite{Petterson2004} collected before HJD=2450500 are not used in the analysis.
\newline
\newline
\noindent  \textbf{SV Per:} because of lower quality, older data of \cite{Gieren1992} collected before HJD=2445000 are not used in the analysis.
\newline
\newline
\noindent  \textbf{RS Pup:} because of period instability, only data collected after
HJD=2456200 are used in the analysis. 
\newline
\newline
\noindent \textbf{DR Vel:} first harmonic of the radial velocity curve has not been detected in this Cepheid, perhaps because of low quality of available data. The second harmonic has been marginally detected. The measured value of $\phi_{31}$ is consistent with fundamental-mode pulsations. Therefore, we classify DR~Vel as a fundamental-mode Cepheid.



\onecolumn
\begin{center}
\begin{longtable}{lSS[table-format=2.0]lcS[table-format=2.2]cccccc}
\caption{Low-order Fourier parameters of fundamental-mode Cepheids, stars of Quality 1. \label{Tab:data_funda_quality1}} \\
\hline
\hline
\hline
\noalign{\smallskip}
\multicolumn{1}{l}{Star} &
\multicolumn{1}{S}{{Period}} &
\multicolumn{1}{r}{Ndat} &
\multicolumn{1}{l}{$n$} &
\multicolumn{1}{c}{$\sigma_\mathrm{fit}$} &
\multicolumn{1}{c}{$A_1$} &
\multicolumn{1}{c}{$\phi_{21}$} &
\multicolumn{1}{c}{$R_{21}$} &
\multicolumn{1}{c}{$\phi_{31}$} &
\multicolumn{1}{c}{$R_{31}$} &
\multicolumn{1}{c}{$\phi_{41}$} &
\multicolumn{1}{c}{$R_{41}$} \\
\hline
\endfirsthead

\multicolumn{12}{c}%
{{\bfseries \tablename\ \thetable{} -- continued from previous page}} \\
\noalign{\medskip}
\hline
\hline
\noalign{\smallskip}
\multicolumn{1}{l}{Star} &
\multicolumn{1}{S}{{Period}} &
\multicolumn{1}{r}{Ndat} &
\multicolumn{1}{l}{$n$} &
\multicolumn{1}{c}{$\sigma_\mathrm{fit}$} &
\multicolumn{1}{c}{$A_1$} &
\multicolumn{1}{c}{$\phi_{21}$} &
\multicolumn{1}{c}{$R_{21}$} &
\multicolumn{1}{c}{$\phi_{31}$} &
\multicolumn{1}{c}{$R_{31}$} &
\multicolumn{1}{c}{$\phi_{41}$} &
\multicolumn{1}{c}{$R_{41}$} \\
\hline
\endhead

\hline
\noalign{\vskip 0.065truecm}
\multicolumn{2}{r}{{Continued on next page}} \\
\hline
\endfoot

\hline
\hline
\endlastfoot
\noalign{\medskip}
SS Sct       &  3.67 &  67 &  3   & 0.78 & 14.18 & 2.917 & 0.268 & 5.933 & 0.095 &       &       \\
             &       &     &      &      &  0.14 & 0.041 & 0.010 & 0.108 & 0.010 &       &       \\
\noalign{\medskip}
RT Aur       &  3.73 & 123 &  6   & 0.62 & 15.53 & 3.065 & 0.356 & 6.141 & 0.166 & 2.918 & 0.080 \\
             &       &     &      &      &  0.08 & 0.018 & 0.006 & 0.037 & 0.005 & 0.073 & 0.005 \\
\noalign{\medskip}
SU Cyg       &  3.85 & 331 &  5+5 & 1.13 & 16.49 & 3.010 & 0.329 & 6.060 & 0.145 & 2.925 & 0.069 \\
             &       &     &      &      &  0.09 & 0.020 & 0.006 & 0.041 & 0.005 & 0.083 & 0.005 \\
\noalign{\medskip}
SY Cas       &  4.07 & 119 &  7   & 0.95 & 15.57 & 3.140 & 0.347 & 6.119 & 0.163 & 2.999 & 0.096 \\
             &       &     &      &      &  0.14 & 0.029 & 0.009 & 0.061 & 0.008 & 0.092 & 0.009 \\
\noalign{\medskip}
T Vul        &  4.44 & 197 &  7   & 0.52 & 14.50 & 2.986 & 0.354 & 6.032 & 0.154 & 2.877 & 0.082 \\
             &       &     &      &      &  0.06 & 0.013 & 0.005 & 0.029 & 0.004 & 0.051 & 0.004 \\
\noalign{\medskip}
XY Cas       &  4.50 &  47 &  5+t & 0.70 & 14.78 & 2.969 & 0.322 & 6.154 & 0.164 & 2.998 & 0.080 \\
             &       &     &      &      &  0.16 & 0.038 & 0.013 & 0.075 & 0.011 & 0.142 & 0.012 \\
\noalign{\medskip}
S Cru        &  4.69 &  53 &  7   & 0.39 & 15.19 & 3.066 & 0.365 & 6.244 & 0.191 & 3.307 & 0.090 \\
             &       &     &      &      &  0.09 & 0.027 & 0.007 & 0.033 & 0.006 & 0.065 & 0.006 \\
\noalign{\medskip}
VZ Cyg       &  4.86 &  89 &  6+2 & 0.46 & 15.00 & 3.075 & 0.365 & 6.226 & 0.169 & 3.242 & 0.083 \\
             &       &     &      &      &  0.07 & 0.016 & 0.005 & 0.030 & 0.005 & 0.061 & 0.005 \\
\noalign{\medskip}
CF Cas       &  4.88 &  87 &  4   & 0.94 & 13.73 & 3.069 & 0.345 & 0.149 & 0.164 & 3.577 & 0.051 \\
             &       &     &      &      &  0.16 & 0.038 & 0.012 & 0.083 & 0.011 & 0.208 & 0.012 \\
\noalign{\medskip}
V1154 Cyg    &  4.93 & 181 &  4   & 0.93 & 11.40 & 3.144 & 0.280 & 0.210 & 0.106 & 3.144 & 0.029 \\
             &       &     &      &      &  0.10 & 0.036 & 0.009 & 0.090 & 0.009 & 0.310 & 0.009 \\
\noalign{\medskip}
AP Pup       &  5.08 &  45 &  5   & 0.51 & 14.02 & 3.115 & 0.369 & 0.101 & 0.183 & 3.325 & 0.090 \\
             &       &     &      &      &  0.12 & 0.030 & 0.009 & 0.054 & 0.009 & 0.087 & 0.009 \\
\noalign{\medskip}
V350 Sgr     &  5.15 & 162 &  7+4 & 0.30 & 15.09 & 3.136 & 0.380 & 0.127 & 0.194 & 3.451 & 0.092 \\
             &       &     &      &      &  0.03 & 0.008 & 0.002 & 0.014 & 0.002 & 0.028 & 0.002 \\
\noalign{\medskip}
V386 Cyg     &  5.26 &  79 &  5   & 1.19 & 14.41 & 3.206 & 0.388 & 0.080 & 0.195 & 3.503 & 0.101 \\
             &       &     &      &      &  0.19 & 0.049 & 0.015 & 0.073 & 0.014 & 0.146 & 0.014 \\
\noalign{\medskip}
BG Lac       &  5.33 &  61 &  7   & 0.41 & 14.24 & 3.163 & 0.372 & 0.194 & 0.178 & 3.597 & 0.075 \\
             &       &     &      &      &  0.08 & 0.019 & 0.006 & 0.033 & 0.006 & 0.077 & 0.006 \\
\noalign{\medskip}
$\delta$ Cep &  5.37 & 135 & 11+t & 0.06 & 15.57 & 3.137 & 0.418 & 0.100 & 0.227 & 3.349 & 0.120 \\
             &       &     &      &      &  0.01 & 0.002 & 0.000 & 0.002 & 0.000 & 0.004 & 0.000 \\
\noalign{\medskip}
V1162 Aql    &  5.38 & 109 &  4+t & 1.04 & 12.76 & 3.170 & 0.356 & 0.476 & 0.154 & 4.197 & 0.049 \\
             &       &     &      &      &  0.15 & 0.041 & 0.012 & 0.078 & 0.012 & 0.243 & 0.011 \\
\noalign{\medskip}
FM Cas       &  5.81 &  66 &  7   & 0.75 & 13.55 & 3.226 & 0.411 & 0.487 & 0.188 & 3.771 & 0.080 \\
             &       &     &      &      &  0.16 & 0.031 & 0.013 & 0.073 & 0.011 & 0.139 & 0.011 \\
\noalign{\medskip}
MW Cyg       &  5.95 & 202 &  6+3 & 1.09 & 15.04 & 3.260 & 0.445 & 0.463 & 0.224 & 3.714 & 0.088 \\
             &       &     &      &      &  0.12 & 0.021 & 0.008 & 0.042 & 0.008 & 0.088 & 0.007 \\
\noalign{\medskip}
FM Aql       &  6.11 & 139 &  6+t & 1.03 & 15.79 & 3.205 & 0.413 & 0.598 & 0.201 & 4.010 & 0.072 \\
             &       &     &      &      &  0.13 & 0.024 & 0.009 & 0.046 & 0.008 & 0.118 & 0.008 \\
\noalign{\medskip}
V538 Cyg     &  6.12 &  52 &  5+t & 1.34 & 12.37 & 3.192 & 0.393 & 0.585 & 0.147 & 3.955 & 0.056 \\
             &       &     &      &      &  0.28 & 0.092 & 0.026 & 0.185 & 0.023 & 0.412 & 0.025 \\
\noalign{\medskip}
CR Cep       &  6.23 &  41 &  5+t & 0.81 & 10.50 & 3.281 & 0.360 & 0.597 & 0.121 & 3.937 & 0.045 \\
             &       &     &      &      &  0.20 & 0.062 & 0.020 & 0.149 & 0.021 & 0.445 & 0.018 \\
\noalign{\medskip}
AW Per       &  6.46 & 355 &  6+5 & 0.92 & 15.36 & 3.295 & 0.486 & 0.555 & 0.248 & 3.801 & 0.117 \\
             &       &     &      &      &  0.08 & 0.014 & 0.005 & 0.025 & 0.005 & 0.044 & 0.005 \\
\noalign{\medskip}
BB Sgr       &  6.64 &  88 &  5   & 0.59 & 12.90 & 3.430 & 0.446 & 0.941 & 0.167 & 4.355 & 0.067 \\
             &       &     &      &      &  0.09 & 0.022 & 0.008 & 0.047 & 0.007 & 0.114 & 0.007 \\
\noalign{\smallskip}
\noalign{\eject}
\noalign{\medskip}
T Cru        &  6.73 &  39 &  5   & 0.40 & 11.52 & 3.406 & 0.394 & 0.941 & 0.099 & 3.911 & 0.036 \\
             &       &     &      &      &  0.09 & 0.027 & 0.009 & 0.089 & 0.008 & 0.219 & 0.009 \\
\noalign{\medskip}
U Sgr        &  6.75 & 320 &  7   & 0.72 & 14.88 & 3.378 & 0.503 & 0.835 & 0.207 & 4.069 & 0.071 \\
             &       &     &      &      &  0.06 & 0.011 & 0.004 & 0.022 & 0.004 & 0.058 & 0.004 \\
\noalign{\medskip}
V636 Sco     &  6.80 & 133 &  5+3 & 0.56 & 12.49 & 3.459 & 0.424 & 1.136 & 0.136 & 4.439 & 0.055 \\
             &       &     &      &      &  0.08 & 0.017 & 0.007 & 0.048 & 0.007 & 0.125 & 0.006 \\
\noalign{\medskip}
V496 Aql     &  6.81 & 314 &  3+5 & 0.60 &  8.89 & 3.519 & 0.333 & 1.080 & 0.062 &       &       \\
             &       &     &      &      &  0.05 & 0.021 & 0.006 & 0.093 & 0.005 &       &       \\
\noalign{\medskip}
BG Vel       &  6.92 &  33 &  5   & 0.10 & 11.15 & 3.485 & 0.408 & 1.115 & 0.090 & 4.136 & 0.038 \\
             &       &     &      &      &  0.03 & 0.008 & 0.003 & 0.029 & 0.002 & 0.062 & 0.003 \\
\noalign{\medskip}
U Aql        &  7.02 & 106 & 10+3 & 0.20 & 14.71 & 3.460 & 0.528 & 0.932 & 0.208 & 3.998 & 0.086 \\
             &       &     &      &      &  0.03 & 0.007 & 0.002 & 0.012 & 0.002 & 0.028 & 0.002 \\
\noalign{\medskip}
$\eta$ Aql   &  7.18 &  98 & 10+t & 0.05 & 15.51 & 3.496 & 0.521 & 0.975 & 0.200 & 3.951 & 0.088 \\
             &       &     &      &      &  0.01 & 0.001 & 0.001 & 0.003 & 0.001 & 0.006 & 0.001 \\
\noalign{\medskip}
V600 Aql     &  7.24 &  93 &  5   & 1.18 & 13.58 & 3.536 & 0.491 & 1.042 & 0.142 & 3.881 & 0.069 \\
             &       &     &      &      &  0.18 & 0.037 & 0.015 & 0.105 & 0.013 & 0.201 & 0.014 \\
\noalign{\medskip}
V459 Cyg     &  7.25 &  68 &  5   & 1.20 & 14.20 & 3.531 & 0.548 & 1.194 & 0.179 & 3.938 & 0.101 \\
             &       &     &      &      &  0.21 & 0.050 & 0.019 & 0.108 & 0.017 & 0.184 & 0.015 \\
\noalign{\medskip}
TZ Mon       &  7.43 &  36 &  5   & 0.54 & 14.92 & 3.619 & 0.552 & 0.938 & 0.157 & 4.158 & 0.114 \\
             &       &     &      &      &  0.16 & 0.032 & 0.013 & 0.079 & 0.011 & 0.105 & 0.012 \\
\noalign{\medskip}
V1344 Aql    &  7.48 &  48 &  3   & 0.24 &  6.83 & 3.661 & 0.300 & 0.623 & 0.039 &       &       \\
             &       &     &      &      &  0.05 & 0.030 & 0.008 & 0.195 & 0.007 &       &       \\
\noalign{\medskip}
BB Her       &  7.51 &  85 &  5   & 0.72 & 12.96 & 3.606 & 0.500 & 1.149 & 0.130 & 3.966 & 0.083 \\
             &       &     &      &      &  0.12 & 0.025 & 0.010 & 0.076 & 0.009 & 0.116 & 0.009 \\
\noalign{\medskip}
IT Car       &  7.54 &  40 &  2   & 0.44 &  8.03 & 3.888 & 0.345 &       &       &       &       \\
             &       &     &      &      &  0.10 & 0.043 & 0.014 &       &       &       &       \\
\noalign{\medskip}
W Sgr        &  7.60 & 135 &  9+3 & 0.15 & 14.49 & 3.645 & 0.562 & 1.015 & 0.181 & 4.020 & 0.118 \\
             &       &     &      &      &  0.02 & 0.004 & 0.002 & 0.008 & 0.001 & 0.012 & 0.001 \\
\noalign{\medskip}
GH Cyg       &  7.82 &  93 &  4+t & 1.37 & 13.50 & 3.682 & 0.567 & 1.062 & 0.142 & 4.118 & 0.070 \\
             &       &     &      &      &  0.20 & 0.044 & 0.018 & 0.118 & 0.016 & 0.261 & 0.014 \\
\noalign{\medskip}
RX Cam       &  7.91 &  85 &  6+5 & 0.52 & 14.42 & 3.786 & 0.566 & 1.007 & 0.142 & 4.331 & 0.111 \\
             &       &     &      &      &  0.09 & 0.017 & 0.007 & 0.051 & 0.006 & 0.064 & 0.006 \\
\noalign{\medskip}
U Vul        &  7.99 & 440 &  6+4 & 0.74 & 12.00 & 3.698 & 0.498 & 0.971 & 0.132 & 3.973 & 0.085 \\
             &       &     &      &      &  0.05 & 0.012 & 0.005 & 0.035 & 0.004 & 0.052 & 0.004 \\
\noalign{\medskip}
DL Cas       &  8.00 & 191 &  5+4 & 0.76 & 13.25 & 3.605 & 0.466 & 1.042 & 0.091 & 4.010 & 0.056 \\
             &       &     &      &      &  0.08 & 0.018 & 0.007 & 0.074 & 0.007 & 0.126 & 0.006 \\
\noalign{\medskip}
S Sge        &  8.38 & 115 &  6+4 & 0.37 & 14.15 & 3.947 & 0.542 & 1.131 & 0.131 & 4.557 & 0.112 \\
             &       &     &      &      &  0.05 & 0.010 & 0.004 & 0.030 & 0.004 & 0.035 & 0.004 \\
\noalign{\medskip}
GQ Ori       &  8.62 &  29 &  5   & 0.70 & 14.42 & 3.846 & 0.564 & 1.036 & 0.156 & 4.376 & 0.110 \\
             &       &     &      &      &  0.19 & 0.039 & 0.016 & 0.090 & 0.014 & 0.142 & 0.014 \\
\noalign{\medskip}
HW Car       &  9.20 &  66 &  3   & 0.11 &  9.05 & 5.057 & 0.222 & 1.742 & 0.078 &       &       \\
             &       &     &      &      &  0.02 & 0.011 & 0.002 & 0.029 & 0.002 &       &       \\
\noalign{\medskip}
V500 Sco     &  9.32 &  75 &  5   & 1.33 & 13.20 & 4.535 & 0.480 & 1.065 & 0.159 & 5.498 & 0.070 \\
             &       &     &      &      &  0.23 & 0.047 & 0.019 & 0.109 & 0.018 & 0.247 & 0.018 \\
\noalign{\medskip}
FN Aql       &  9.48 & 140 &  4   & 0.77 & 14.49 & 5.290 & 0.171 & 2.295 & 0.151 & 4.754 & 0.034 \\
             &       &     &      &      &  0.10 & 0.040 & 0.007 & 0.048 & 0.007 & 0.195 & 0.006 \\
\noalign{\smallskip}
\noalign{\eject}
\noalign{\medskip}
YZ Sgr       &  9.55 &  76 &  5   & 0.58 & 13.91 & 4.699 & 0.424 & 1.388 & 0.176 & 0.142 & 0.061 \\
             &       &     &      &      &  0.09 & 0.024 & 0.008 & 0.046 & 0.007 & 0.117 & 0.007 \\
\noalign{\medskip}
S Mus        &  9.66 & 211 &  5+3 & 0.48 & 12.28 & 4.233 & 0.482 & 0.895 & 0.142 & 5.198 & 0.075 \\
             &       &     &      &      &  0.05 & 0.012 & 0.004 & 0.033 & 0.004 & 0.062 & 0.004 \\
\noalign{\medskip}
AQ Car       &  9.77 &  59 & 10+t & 0.08 & 13.95 & 5.077 & 0.304 & 1.762 & 0.168 & 6.270 & 0.004 \\
             &       &     &      &      &  0.02 & 0.004 & 0.001 & 0.008 & 0.001 & 0.300 & 0.001 \\
\noalign{\medskip}
DD Cas       &  9.81 & 123 &  7   & 0.21 & 14.49 & 5.225 & 0.232 & 2.003 & 0.165 & 4.423 & 0.018 \\
             &       &     &      &      &  0.03 & 0.010 & 0.002 & 0.015 & 0.002 & 0.119 & 0.002 \\
\noalign{\medskip}
$\beta$ Dor  &  9.84 & 139 &  6   & 0.18 & 13.91 & 5.013 & 0.312 & 1.759 & 0.177 &  ---  &  ---  \\
             &       &     &      &      &  0.03 & 0.010 & 0.002 & 0.013 & 0.002 &  ---  &  ---  \\
\noalign{\medskip}
SY Aur       & 10.15 & 109 &  5+t & 0.16 & 10.51 & 4.319 & 0.380 & 0.708 & 0.106 & 5.670 & 0.053 \\
             &       &     &      &      &  0.02 & 0.008 & 0.003 & 0.022 & 0.002 & 0.040 & 0.003 \\
\noalign{\medskip}
$\zeta$ Gem  & 10.15 & 312 &  7   & 0.15 & 11.85 & 5.209 & 0.243 & 1.924 & 0.127 & 3.855 & 0.006 \\
             &       &     &      &      &  0.01 & 0.005 & 0.001 & 0.009 & 0.001 & 0.185 & 0.001 \\
\noalign{\medskip}
Y Sct        & 10.34 &  72 &  4   & 1.45 & 18.09 & 5.072 & 0.206 & 2.020 & 0.194 & 5.097 & 0.048 \\
             &       &     &      &      &  0.25 & 0.079 & 0.014 & 0.100 & 0.014 & 0.290 & 0.015 \\
\noalign{\medskip}
VX Per       & 10.88 & 118 &  9   & 0.17 & 13.49 & 4.452 & 0.465 & 1.124 & 0.148 & 5.577 & 0.076 \\
             &       &     &      &      &  0.02 & 0.005 & 0.002 & 0.014 & 0.002 & 0.026 & 0.002 \\
\noalign{\medskip}
Z Lac        & 10.89 & 276 &  7+1 & 0.81 & 18.35 & 5.018 & 0.197 & 2.017 & 0.191 & 5.323 & 0.039 \\
             &       &     &      &      &  0.07 & 0.022 & 0.004 & 0.024 & 0.004 & 0.102 & 0.004 \\
\noalign{\medskip}
XX Cen       & 10.95 &  40 &  7+1 & 0.34 & 16.42 & 4.866 & 0.289 & 1.735 & 0.179 & 6.012 & 0.017 \\
             &       &     &      &      &  0.09 & 0.021 & 0.006 & 0.042 & 0.006 & 0.334 & 0.005 \\
\noalign{\medskip}
TY Sct       & 11.05 &  54 &  3   & 1.26 & 17.38 & 5.053 & 0.166 & 2.145 & 0.140 &       &       \\
             &       &     &      &      &  0.24 & 0.097 & 0.014 & 0.106 & 0.015 &       &       \\
\noalign{\medskip}
SV Per       & 11.13 &  80 &  5   & 0.93 & 21.01 & 4.681 & 0.089 & 2.364 & 0.161 & 5.431 & 0.092 \\
             &       &     &      &      &  0.16 & 0.085 & 0.007 & 0.050 & 0.008 & 0.087 & 0.007 \\
\noalign{\medskip}
V340 Nor     & 11.29 &  50 &  3   & 0.41 &  8.58 & 4.900 & 0.262 & 1.295 & 0.067 &       &       \\
             &       &     &      &      &  0.09 & 0.039 & 0.010 & 0.152 & 0.010 &       &       \\
\noalign{\medskip}
AA Gem       & 11.30 &  52 &  5+t & 0.53 & 17.53 & 5.561 & 0.047 & 2.659 & 0.142 & 5.420 & 0.074 \\
             &       &     &      &      &  0.11 & 0.133 & 0.006 & 0.047 & 0.006 & 0.089 & 0.006 \\
\noalign{\medskip}
RX Aur       & 11.62 &  51 &  5   & 0.47 & 14.30 & 4.791 & 0.213 & 1.610 & 0.143 & 4.413 & 0.028 \\
             &       &     &      &      &  0.09 & 0.036 & 0.007 & 0.053 & 0.007 & 0.259 & 0.007 \\
\noalign{\medskip}
SS CMa       & 12.35 &  80 &  8   & 0.18 & 16.86 & 4.798 & 0.237 & 1.630 & 0.167 & 5.441 & 0.020 \\
             &       &     &      &      &  0.03 & 0.009 & 0.002 & 0.012 & 0.002 & 0.118 & 0.002 \\
\noalign{\medskip}
XY Car       & 12.44 &  74 &  8   & 0.21 & 20.95 & 3.736 & 0.029 & 2.751 & 0.133 & 5.675 & 0.097 \\
             &       &     &      &      &  0.03 & 0.059 & 0.002 & 0.013 & 0.002 & 0.023 & 0.002 \\
\noalign{\medskip}
SY Nor       & 12.65 &  42 &  5+1 & 0.58 & 22.73 & 3.179 & 0.074 & 2.783 & 0.116 & 5.719 & 0.103 \\
             &       &     &      &      &  0.16 & 0.079 & 0.006 & 0.057 & 0.006 & 0.066 & 0.006 \\
\noalign{\medskip}
SZ Cas       & 13.64 &  70 &  2   & 0.76 &  8.30 & 5.208 & 0.097 &       &       &       &       \\
             &       &     &      &      &  0.14 & 0.162 & 0.016 &       &       &       &       \\
\noalign{\medskip}
TT Aql       & 13.75 & 253 & 11   & 0.53 & 23.20 & 2.970 & 0.164 & 3.106 & 0.066 & 6.107 & 0.097 \\
             &       &     &      &      &  0.05 & 0.014 & 0.002 & 0.032 & 0.002 & 0.024 & 0.002 \\
\noalign{\medskip}
LS Pup       & 14.14 &  27 &  5   & 0.60 & 22.14 & 3.618 & 0.068 & 2.693 & 0.129 & 5.808 & 0.087 \\
             &       &     &      &      &  0.19 & 0.122 & 0.008 & 0.069 & 0.008 & 0.111 & 0.008 \\
\noalign{\medskip}
SZ Cyg       & 15.11 & 107 & 12   & 0.14 & 21.94 & 2.942 & 0.245 & 4.758 & 0.061 & 0.742 & 0.085 \\
             &       &     &      &      &  0.03 & 0.005 & 0.001 & 0.021 & 0.001 & 0.017 & 0.001 \\
\noalign{\smallskip}
\noalign{\eject}
\noalign{\medskip}
SV Mon       & 15.23 &  40 &  8   & 0.69 & 24.87 & 2.932 & 0.262 & 4.775 & 0.056 & 0.447 & 0.089 \\
             &       &     &      &      &  0.18 & 0.028 & 0.008 & 0.121 & 0.007 & 0.077 & 0.007 \\
\noalign{\medskip}
X Cyg        & 16.39 & 467 & 10   & 0.90 & 23.45 & 3.026 & 0.294 & 5.263 & 0.067 & 1.049 & 0.081 \\
             &       &     &      &      &  0.06 & 0.010 & 0.003 & 0.040 & 0.003 & 0.033 & 0.003 \\
\noalign{\medskip}
XZ Car       & 16.65 & 116 & 13+t & 0.21 & 23.79 & 3.015 & 0.293 & 5.263 & 0.071 & 1.006 & 0.082 \\
             &       &     &      &      &  0.03 & 0.005 & 0.001 & 0.017 & 0.001 & 0.019 & 0.001 \\
\noalign{\medskip}
CD Cyg       & 17.07 & 149 & 10   & 0.39 & 24.84 & 3.004 & 0.279 & 5.227 & 0.044 & 0.667 & 0.062 \\
             &       &     &      &      &  0.05 & 0.009 & 0.002 & 0.049 & 0.002 & 0.034 & 0.002 \\
\noalign{\medskip}
Y Oph        & 17.13 &  94 &  3   & 0.23 &  8.47 & 3.141 & 0.028 & 2.143 & 0.034 &       &       \\
             &       &     &      &      &  0.03 & 0.147 & 0.004 & 0.122 & 0.004 &       &       \\
\noalign{\medskip}
YZ Car       & 18.17 &  92 &  7+2 & 0.23 & 13.97 & 3.176 & 0.080 & 2.243 & 0.049 & 4.919 & 0.025 \\
             &       &     &      &      &  0.04 & 0.034 & 0.003 & 0.055 & 0.003 & 0.107 & 0.003 \\
\noalign{\medskip}
VY Car       & 18.88 &  80 & 10   & 0.23 & 24.06 & 3.007 & 0.298 & 5.345 & 0.076 & 1.071 & 0.075 \\
             &       &     &      &      &  0.04 & 0.007 & 0.002 & 0.025 & 0.002 & 0.023 & 0.002 \\
\noalign{\medskip}
RU Sct       & 19.71 & 116 &  3   & 1.03 & 21.87 & 3.004 & 0.243 & 5.887 & 0.059 &       &       \\
             &       &     &      &      &  0.15 & 0.030 & 0.007 & 0.112 & 0.007 &       &       \\
\noalign{\medskip}
RY Sco       & 20.33 &  52 &  8+t & 0.15 & 16.26 & 2.957 & 0.155 & 3.567 & 0.006 & 5.633 & 0.016 \\
             &       &     &      &      &  0.03 & 0.013 & 0.002 & 0.320 & 0.002 & 0.120 & 0.002 \\
\noalign{\medskip}
RZ Vel       & 20.40 &  65 &  9   & 0.16 & 24.64 & 3.028 & 0.297 & 5.634 & 0.083 & 1.484 & 0.052 \\
             &       &     &      &      &  0.03 & 0.005 & 0.001 & 0.018 & 0.001 & 0.026 & 0.001 \\
\noalign{\medskip}
WZ Sgr       & 21.85 &  81 & 13   & 0.21 & 23.62 & 3.053 & 0.313 & 5.560 & 0.104 & 1.619 & 0.085 \\
             &       &     &      &      &  0.04 & 0.006 & 0.002 & 0.019 & 0.002 & 0.020 & 0.002 \\
\noalign{\medskip}
T Mon        & 27.04 & 590 & 16+t & 0.35 & 20.77 & 2.980 & 0.342 & 5.736 & 0.182 & 2.368 & 0.137 \\
             &       &     &      &      &  0.02 & 0.004 & 0.001 & 0.006 & 0.001 & 0.008 & 0.001 \\
\noalign{\medskip}
AQ Pup       & 30.19 &  98 & 11+t & 0.18 & 25.58 & 2.996 & 0.337 & 5.673 & 0.141 & 1.983 & 0.096 \\
             &       &     &      &      &  0.04 & 0.009 & 0.002 & 0.013 & 0.002 & 0.016 & 0.001 \\
\noalign{\medskip}
$\ell$ Car   & 35.56 & 591 & 12   & 0.26 & 15.89 & 2.936 & 0.323 & 5.902 & 0.172 & 2.763 & 0.101 \\
             &       &     &      &      &  0.02 & 0.004 & 0.001 & 0.008 & 0.001 & 0.012 & 0.001 \\
\noalign{\medskip}
SV Vul       & 45.00 & 124 &  8   & 0.36 & 19.84 & 2.905 & 0.369 & 5.902 & 0.191 & 2.666 & 0.110 \\
             &       &     &      &      &  0.05 & 0.008 & 0.003 & 0.015 & 0.002 & 0.024 & 0.003 \\
\noalign{\medskip}
S Vul        & 69.43 &  88 &  7   & 0.21 & 12.31 & 3.164 & 0.334 & 0.250 & 0.120 & 3.641 & 0.055 \\
             &       &     &      &      &  0.03 & 0.023 & 0.003 & 0.026 & 0.004 & 0.064 & 0.004 \\
\noalign{\smallskip}
\hline

\end{longtable}
\end{center}
\begin{tablenotes}
    \item \textbf{Notes :} In this table we provided only the low-order parameters up to fourth Fourier order and the standard deviation of the fit. The full table with all derived Fourier parameters and their uncertainties is only available in electronic form. This table displays the name of the star, the pulsation period in days, the number of data points used in the fit Ndat, the order of the fit $n$ which can be additionally supplemented with an orbital motion correction or a trend t (see Sect~\ref{sect:binary} and \ref{sect:trend} respectively).
\end{tablenotes}

\subsection{Low-order Fourier parameters of fundamental-mode Cepheids, stars of Quality 1a.}
\onecolumn
\begin{center}
\begin{longtable}{lSS[table-format=2.0]lcS[table-format=2.2]cccccc}
\caption{Low-order Fourier parameters of fundamental-mode Cepheids, stars of Quality 1a. \label{Tab:data_funda_quality1a}} \\
\hline
\hline
\hline
\noalign{\smallskip}
\multicolumn{1}{l}{Star} &
\multicolumn{1}{S}{{Period}} &
\multicolumn{1}{r}{Ndat} &
\multicolumn{1}{l}{$n$} &
\multicolumn{1}{c}{$\sigma_\mathrm{fit}$} &
\multicolumn{1}{c}{$A_1$} &
\multicolumn{1}{c}{$\phi_{21}$} &
\multicolumn{1}{c}{$R_{21}$} &
\multicolumn{1}{c}{$\phi_{31}$} &
\multicolumn{1}{c}{$R_{31}$} &
\multicolumn{1}{c}{$\phi_{41}$} &
\multicolumn{1}{c}{$R_{41}$} \\
\hline
\endfirsthead

\multicolumn{12}{c}%
{{\bfseries \tablename\ \thetable{} -- continued from previous page}} \\
\noalign{\medskip}
\hline
\hline
\noalign{\smallskip}
\multicolumn{1}{l}{Star} &
\multicolumn{1}{S}{{Period}} &
\multicolumn{1}{r}{Ndat} &
\multicolumn{1}{l}{$n$} &
\multicolumn{1}{c}{$\sigma_\mathrm{fit}$} &
\multicolumn{1}{c}{$A_1$} &
\multicolumn{1}{c}{$\phi_{21}$} &
\multicolumn{1}{c}{$R_{21}$} &
\multicolumn{1}{c}{$\phi_{31}$} &
\multicolumn{1}{c}{$R_{31}$} &
\multicolumn{1}{c}{$\phi_{41}$} &
\multicolumn{1}{c}{$R_{41}$} \\
\hline
\endhead

\hline
\noalign{\vskip 0.065truecm}
\multicolumn{2}{r}{{Continued on next page}} \\
\hline
\endfoot

\hline
\hline
\endlastfoot
\noalign{\medskip}
ST Tau       &  4.03 &  89 &  6   & 0.87 & 15.67 & 3.068 & 0.352 & 5.953 & 0.167 & 2.950 & 0.094 \\
             &       &     &      &      &  0.14 & 0.032 & 0.009 & 0.058 & 0.009 & 0.102 & 0.009 \\
\noalign{\medskip}
BF Oph       &  4.07 &  64 &  8   & 0.58 & 15.17 & 3.001 & 0.318 & 5.956 & 0.173 & 2.814 & 0.080 \\
             &       &     &      &      &  0.13 & 0.028 & 0.010 & 0.058 & 0.008 & 0.107 & 0.008 \\
\noalign{\medskip}
Y Lac        &  4.32 &  82 &  6   & 0.91 & 16.50 & 3.090 & 0.336 & 6.163 & 0.168 & 3.194 & 0.078 \\
             &       &     &      &      &  0.15 & 0.033 & 0.009 & 0.060 & 0.010 & 0.125 & 0.009 \\
\noalign{\medskip}
V402 Cyg     &  4.36 & 105 &  5   & 0.87 & 13.46 & 3.088 & 0.332 & 0.045 & 0.136 & 3.032 & 0.061 \\
             &       &     &      &      &  0.13 & 0.034 & 0.010 & 0.070 & 0.010 & 0.157 & 0.009 \\
\noalign{\medskip}
AP Sgr       &  5.06 &  72 &  6+t & 0.70 & 15.65 & 3.171 & 0.427 & 0.142 & 0.228 & 3.379 & 0.126 \\
             &       &     &      &      &  0.13 & 0.026 & 0.009 & 0.043 & 0.009 & 0.085 & 0.008 \\
\noalign{\medskip}
AX Cir       &  5.27 & 100 &  7+2 & 0.33 & 13.17 & 3.142 & 0.341 & 0.324 & 0.150 & 3.806 & 0.049 \\
             &       &     &      &      &  0.05 & 0.012 & 0.004 & 0.027 & 0.004 & 0.075 & 0.004 \\
\noalign{\medskip}
CV Mon       &  5.38 & 106 &  8   & 0.59 & 15.09 & 3.106 & 0.429 & 0.080 & 0.222 & 3.354 & 0.114 \\
             &       &     &      &      &  0.10 & 0.020 & 0.006 & 0.033 & 0.006 & 0.054 & 0.006 \\
\noalign{\medskip}
Y Sgr        &  5.77 &  66 &  6   & 0.33 & 15.07 & 3.220 & 0.400 & 0.365 & 0.201 & 3.687 & 0.091 \\
             &       &     &      &      &  0.06 & 0.013 & 0.004 & 0.025 & 0.004 & 0.046 & 0.004 \\
\noalign{\medskip}
R Cru        &  5.83 &  44 &  6+t & 0.68 & 15.23 & 3.217 & 0.449 & 0.358 & 0.243 & 3.696 & 0.124 \\
             &       &     &      &      &  0.17 & 0.035 & 0.013 & 0.062 & 0.009 & 0.107 & 0.011 \\
\noalign{\medskip}
V733 Aql     &  6.18 &  77 &  4   & 0.95 & 10.59 & 3.357 & 0.372 & 0.780 & 0.118 & 4.710 & 0.053 \\
             &       &     &      &      &  0.17 & 0.046 & 0.019 & 0.169 & 0.014 & 0.296 & 0.016 \\
\noalign{\medskip}
RR Lac       &  6.42 &  56 &  6   & 0.47 & 14.10 & 3.200 & 0.420 & 0.366 & 0.203 & 3.648 & 0.090 \\
             &       &     &      &      &  0.10 & 0.022 & 0.007 & 0.041 & 0.007 & 0.084 & 0.007 \\
\noalign{\medskip}
XX Sgr       &  6.42 & 116 &  6   & 0.90 & 15.24 & 3.304 & 0.486 & 0.510 & 0.242 & 3.843 & 0.110 \\
             &       &     &      &      &  0.12 & 0.023 & 0.009 & 0.041 & 0.008 & 0.079 & 0.008 \\
\noalign{\medskip}
X Sgr        &  7.01 &  77 &  5   & 0.71 & 12.29 & 3.247 & 0.412 & 0.332 & 0.159 & 3.284 & 0.078 \\
             &       &     &      &      &  0.12 & 0.033 & 0.012 & 0.075 & 0.011 & 0.128 & 0.012 \\
\noalign{\medskip}
RS Ori       &  7.57 &  38 &  6   & 0.72 & 14.90 & 3.550 & 0.562 & 0.892 & 0.193 & 3.863 & 0.127 \\
             &       &     &      &      &  0.19 & 0.029 & 0.015 & 0.077 & 0.012 & 0.104 & 0.012 \\
\noalign{\medskip}
VY Cyg       &  7.86 &  84 &  6   & 1.02 & 14.51 & 3.711 & 0.587 & 1.001 & 0.177 & 4.234 & 0.169 \\
             &       &     &      &      &  0.18 & 0.035 & 0.014 & 0.082 & 0.012 & 0.093 & 0.013 \\
\noalign{\medskip}
W Gem        &  7.91 &  96 &  9   & 0.84 & 14.88 & 3.679 & 0.608 & 0.976 & 0.188 & 4.166 & 0.141 \\
             &       &     &      &      &  0.14 & 0.024 & 0.010 & 0.062 & 0.009 & 0.071 & 0.010 \\
\noalign{\medskip}
S Nor        &  9.75 & 142 &  7   & 0.30 & 14.08 & 4.935 & 0.352 & 1.687 & 0.184 & 0.419 & 0.022 \\
             &       &     &      &      &  0.04 & 0.010 & 0.003 & 0.020 & 0.003 & 0.132 & 0.003 \\
\noalign{\medskip}
RY Cas       & 12.14 &  40 &  3   & 1.18 & 15.89 & 4.890 & 0.206 & 1.376 & 0.173 &       &       \\
             &       &     &      &      &  0.27 & 0.092 & 0.019 & 0.104 & 0.019 &       &       \\
\noalign{\medskip}
Z Sct        & 12.90 & 130 & 12+t & 0.23 & 22.37 & 4.200 & 0.057 & 2.511 & 0.140 & 5.557 & 0.094 \\
             &       &     &      &      &  0.03 & 0.026 & 0.001 & 0.012 & 0.002 & 0.017 & 0.001 \\
\noalign{\medskip}
TX Cyg       & 14.71 &  61 &  5   & 1.21 & 22.22 & 3.040 & 0.104 & 2.720 & 0.084 & 5.667 & 0.076 \\
             &       &     &      &      &  0.24 & 0.108 & 0.010 & 0.127 & 0.010 & 0.134 & 0.011 \\
\noalign{\medskip}
SZ Aql       & 17.14 &  64 & 11   & 0.61 & 25.13 & 3.017 & 0.255 & 4.672 & 0.024 & 0.246 & 0.066 \\
             &       &     &      &      &  0.12 & 0.022 & 0.005 & 0.201 & 0.005 & 0.076 & 0.005 \\
\noalign{\medskip}
V340 Ara     & 20.81 &  34 &  8   & 1.16 & 24.38 & 3.078 & 0.320 & 5.620 & 0.094 & 1.336 & 0.071 \\
             &       &     &      &      &  0.42 & 0.118 & 0.028 & 0.215 & 0.019 & 0.487 & 0.013 \\
\noalign{\medskip}
X Pup        & 25.96 &  84 & 12   & 0.39 & 25.30 & 2.992 & 0.361 & 5.744 & 0.175 & 2.344 & 0.120 \\
             &       &     &      &      &  0.08 & 0.009 & 0.003 & 0.018 & 0.003 & 0.023 & 0.003 \\
\noalign{\smallskip}
\noalign{\eject}
\noalign{\medskip}
KN Cen       & 34.02 &  70 & 14   & 0.48 & 22.07 & 2.956 & 0.373 & 5.824 & 0.212 & 2.532 & 0.152 \\
             &       &     &      &      &  0.09 & 0.014 & 0.005 & 0.024 & 0.004 & 0.032 & 0.004 \\
\noalign{\medskip}
RS Pup       & 41.58 & 410 & 16   & 0.84 & 18.83 & 2.938 & 0.423 & 5.975 & 0.255 & 2.842 & 0.160 \\
             &       &     &      &      &  0.17 & 0.021 & 0.012 & 0.053 & 0.007 & 0.044 & 0.007 \\
\noalign{\medskip}
GY Sge       & 51.77 & 128 &  5   & 1.10 & 13.30 & 3.002 & 0.281 & 6.175 & 0.135 & 3.439 & 0.045 \\
             &       &     &      &      &  0.14 & 0.042 & 0.011 & 0.084 & 0.011 & 0.248 & 0.011 \\
\noalign{\medskip}
V1496 Aql    & 65.92 & 123 &  3   & 1.03 & 10.49 & 3.244 & 0.266 & 0.154 & 0.094 &       &       \\
             &       &     &      &      &  0.14 & 0.055 & 0.014 & 0.143 & 0.013 &       &       \\
\noalign{\smallskip}
\hline

\end{longtable}
\end{center}
\begin{tablenotes}
    \item \textbf{Notes :} Same as Table \ref{Tab:data_funda_quality1}.
\end{tablenotes}

\subsection{Low-order Fourier parameters of fundamental-mode Cepheids, stars of Quality 2.}
\onecolumn
\begin{center}
\begin{longtable}{lSS[table-format=2.0]lcS[table-format=2.2]ccccccl}
\caption{Low-order Fourier parameters of fundamental-mode Cepheids, stars of Quality 2. \label{Tab:data_fourier_quality2}} \\
\hline
\hline
\hline
\noalign{\smallskip}
\multicolumn{1}{l}{Star} &
\multicolumn{1}{S}{{Period}} &
\multicolumn{1}{r}{Ndat} &
\multicolumn{1}{l}{$n$} &
\multicolumn{1}{c}{$\sigma_\mathrm{fit}$} &
\multicolumn{1}{c}{$A_1$} &
\multicolumn{1}{c}{$\phi_{21}$} &
\multicolumn{1}{c}{$R_{21}$} &
\multicolumn{1}{c}{$\phi_{31}$} &
\multicolumn{1}{c}{$R_{31}$} &
\multicolumn{1}{c}{$\phi_{41}$} &
\multicolumn{1}{c}{$R_{41}$} &
\multicolumn{1}{l}{flag} \\
\hline
\endfirsthead

\multicolumn{13}{c}%
{{\bfseries \tablename\ \thetable{} -- continued from previous page}} \\
\noalign{\medskip}
\hline
\hline
\noalign{\smallskip}
\multicolumn{1}{l}{Star} &
\multicolumn{1}{S}{{Period}} &
\multicolumn{1}{r}{Ndat} &
\multicolumn{1}{l}{$n$} &
\multicolumn{1}{c}{$\sigma_\mathrm{fit}$} &
\multicolumn{1}{c}{$A_1$} &
\multicolumn{1}{c}{$\phi_{21}$} &
\multicolumn{1}{c}{$R_{21}$} &
\multicolumn{1}{c}{$\phi_{31}$} &
\multicolumn{1}{c}{$R_{31}$} &
\multicolumn{1}{c}{$\phi_{41}$} &
\multicolumn{1}{c}{$R_{41}$} &
\multicolumn{1}{l}{flag} \\
\hline
\endhead

\hline
\noalign{\vskip 0.065truecm}
\multicolumn{2}{r}{{Continued on next page}} \\
\hline
\endfoot

\hline
\hline
\endlastfoot
\noalign{\medskip}
BE Mon       &  2.70 &  17 &  5   & 1.04 & 14.48 & 2.807 & 0.289 & 5.879 & 0.138 & 2.928 & 0.062 & 2bd  \\
             &       &     &      &      &  0.42 & 0.121 & 0.029 & 0.233 & 0.034 & 0.645 & 0.024 &      \\
\noalign{\medskip}
RT Mus       &  3.09 &  32 &  2   & 2.69 & 15.61 & 2.982 & 0.271 &       &       &       &       & 2bce \\
             &       &     &      &      &  0.66 & 0.187 & 0.049 &       &       &       &       &      \\
\noalign{\medskip}
FI Mon       &  3.29 &  25 &  5   & 0.55 & 18.68 & 2.960 & 0.372 & 6.018 & 0.171 & 2.692 & 0.092 & 2b   \\
             &       &     &      &      &  0.17 & 0.039 & 0.010 & 0.066 & 0.009 & 0.135 & 0.009 &      \\
\noalign{\medskip}
R TrA        &  3.39 &  42 &  4   & 1.80 & 12.29 & 3.005 & 0.316 & 6.018 & 0.125 & 2.724 & 0.051 & 2be  \\
             &       &     &      &      &  0.45 & 0.123 & 0.037 & 0.290 & 0.037 & 0.679 & 0.035 &      \\
\noalign{\medskip}
FT Mon       &  3.42 &  28 &  8   & 1.07 & 19.40 & 2.901 & 0.410 & 5.893 & 0.200 & 2.433 & 0.141 & 2b   \\
             &       &     &      &      &  0.36 & 0.071 & 0.021 & 0.123 & 0.018 & 0.190 & 0.017 &      \\
\noalign{\medskip}
BC Pup       &  3.54 &  26 &  6   & 1.07 & 18.72 & 2.917 & 0.441 & 5.852 & 0.233 & 2.780 & 0.137 & 2b   \\
             &       &     &      &      &  0.34 & 0.056 & 0.021 & 0.096 & 0.019 & 0.147 & 0.018 &      \\
\noalign{\medskip}
UX Car       &  3.68 &  50 &  3   & 3.01 & 15.29 & 2.878 & 0.330 & 6.090 & 0.141 &       &       & 2be  \\
             &       &     &      &      &  0.68 & 0.149 & 0.046 & 0.344 & 0.039 &       &       &      \\
\noalign{\medskip}
AD Gem       &  3.79 &  38 &  4   & 1.15 & 16.56 & 2.942 & 0.336 & 5.929 & 0.144 & 2.411 & 0.086 & 2b   \\
             &       &     &      &      &  0.30 & 0.058 & 0.018 & 0.134 & 0.017 & 0.211 & 0.016 &      \\
\noalign{\medskip}
AG Cru       &  3.84 &  45 &  4   & 2.56 & 15.45 & 2.805 & 0.373 & 5.576 & 0.155 & 2.346 & 0.032 & 2bce \\
             &       &     &      &      &  0.57 & 0.117 & 0.042 & 0.275 & 0.038 & 1.268 & 0.039 &      \\
\noalign{\medskip}
CM Sct       &  3.92 &  23 &  5   & 0.61 & 14.51 & 3.021 & 0.309 & 6.222 & 0.115 & 2.727 & 0.056 & 2bd  \\
             &       &     &      &      &  0.18 & 0.058 & 0.015 & 0.117 & 0.015 & 0.276 & 0.014 &      \\
\noalign{\medskip}
V520 Cyg     &  4.05 &  50 &  4+t & 2.59 & 13.23 & 3.123 & 0.326 & 6.156 & 0.176 & 3.105 & 0.083 & 2bce \\
             &       &     &      &      &  0.76 & 0.203 & 0.042 & 0.258 & 0.059 & 0.753 & 0.047 &      \\
\noalign{\medskip}
V508 Mon     &  4.13 &  17 &  3   & 1.01 & 12.26 & 2.858 & 0.280 & 5.790 & 0.061 &       &       & 2d   \\
             &       &     &      &      &  0.47 & 0.126 & 0.045 & 0.716 & 0.030 &       &       &      \\
\noalign{\medskip}
X Sct        &  4.20 &  89 &  5   & 1.95 & 17.39 & 3.153 & 0.351 & 6.111 & 0.138 & 3.018 & 0.069 & 2be  \\
             &       &     &      &      &  0.31 & 0.059 & 0.019 & 0.137 & 0.018 & 0.251 & 0.018 &      \\
\noalign{\medskip}
FG Mon       &  4.50 &  24 &  8   & 0.49 & 17.07 & 3.029 & 0.449 & 6.156 & 0.250 & 2.992 & 0.150 & 2b   \\
             &       &     &      &      &  0.20 & 0.031 & 0.017 & 0.064 & 0.009 & 0.070 & 0.011 &      \\
\noalign{\medskip}
V482 Sco     &  4.53 &  33 &  4   & 1.68 & 15.28 & 3.092 & 0.310 & 5.831 & 0.149 & 2.549 & 0.057 & 2be  \\
             &       &     &      &      &  0.43 & 0.101 & 0.029 & 0.192 & 0.031 & 0.480 & 0.030 &      \\
\noalign{\medskip}
UX Per       &  4.57 &  18 &  4   & 1.69 & 16.62 & 3.121 & 0.373 & 6.144 & 0.192 & 3.079 & 0.094 & 2bd  \\
             &       &     &      &      &  0.73 & 0.135 & 0.056 & 0.343 & 0.035 & 0.462 & 0.045 &      \\
\noalign{\medskip}
V383 Cyg     &  4.61 &  51 &  3+t & 1.49 & 13.98 & 2.994 & 0.302 & 6.018 & 0.117 &       &       & 2b   \\
             &       &     &      &      &  0.34 & 0.089 & 0.024 & 0.214 & 0.023 &       &       &      \\
\noalign{\medskip}
RY CMa       &  4.68 &  46 &  3   & 2.23 & 15.51 & 3.181 & 0.401 & 0.135 & 0.118 &       &       & 2be  \\
             &       &     &      &      &  0.49 & 0.110 & 0.030 & 0.252 & 0.035 &       &       &      \\
\noalign{\medskip}
CU Mon       &  4.71 &  20 &  6   & 0.77 & 16.85 & 3.046 & 0.451 & 0.006 & 0.205 & 3.285 & 0.118 & 2bd  \\
             &       &     &      &      &  0.29 & 0.055 & 0.022 & 0.096 & 0.019 & 0.157 & 0.019 &      \\
\noalign{\medskip}
AS Per       &  4.97 &  85 &  5   & 1.50 & 16.13 & 3.112 & 0.399 & 0.072 & 0.231 & 3.272 & 0.128 & 2b   \\
             &       &     &      &      & 0.25  & 0.053 & 0.016 & 0.086 & 0.016 & 0.166 & 0.015 &      \\
\noalign{\medskip}
V Lac        &  4.98 &  84 &  5+t & 1.24 & 16.39 & 3.133 & 0.426 & 6.231 & 0.221 & 3.280 & 0.115 & 2b   \\
             &       &     &      &      &  0.20 & 0.047 & 0.015 & 0.065 & 0.013 & 0.121 & 0.014 &      \\
\noalign{\medskip}
CR Ser       &  5.30 &  65 &  5+t & 1.77 & 15.31 & 3.291 & 0.416 & 0.326 & 0.235 & 3.457 & 0.133 & 2be  \\
             &       &     &      &      &  0.35 & 0.072 & 0.025 & 0.109 & 0.023 & 0.207 & 0.021 &      \\
\noalign{\medskip}
UY Per       &  5.37 &  84 &  7+t & 0.52 & 16.30 & 3.158 & 0.425 & 0.081 & 0.231 & 3.391 & 0.131 & 2b   \\
             &       &     &      &      &  0.09 & 0.016 & 0.006 & 0.029 & 0.006 & 0.044 & 0.006 &      \\
\noalign{\smallskip}
\noalign{\eject}
\noalign{\medskip}
SW Cas       &  5.44 &  55 &  5   & 1.28 & 14.56 & 3.169 & 0.409 & 0.171 & 0.183 & 3.605 & 0.116 & 2b   \\
             &       &     &      &      &  0.25 & 0.057 & 0.020 & 0.112 & 0.017 & 0.172 & 0.017 &      \\
\noalign{\medskip}
XX Mon       &  5.46 &  22 &  6   & 0.88 & 15.66 & 3.237 & 0.425 & 0.220 & 0.245 & 3.563 & 0.111 & 2bcd \\
             &       &     &      &      &  0.32 & 0.063 & 0.024 & 0.100 & 0.021 & 0.189 & 0.019 &      \\
\noalign{\medskip}
V Cen        &  5.49 &  70 &  5   & 1.27 & 14.79 & 3.063 & 0.450 & 6.277 & 0.195 & 3.079 & 0.082 & 2b   \\
             &       &     &      &      &  0.25 & 0.044 & 0.018 & 0.105 & 0.015 & 0.203 & 0.016 &      \\
\noalign{\medskip}
VY Per       &  5.53 &  81 &  6+t & 0.75 & 16.22 & 3.202 & 0.437 & 0.163 & 0.243 & 3.534 & 0.117 & 2b   \\
             &       &     &      &      &  0.13 & 0.023 & 0.008 & 0.036 & 0.008 & 0.071 & 0.008 &      \\
\noalign{\medskip}
CS Vel       &  5.90 &  31 &  4   & 0.96 & 15.26 & 3.137 & 0.443 & 0.247 & 0.234 & 3.733 & 0.084 & 2c   \\
             &       &     &      &      &  0.26 & 0.047 & 0.020 & 0.088 & 0.021 & 0.232 & 0.018 &      \\
\noalign{\medskip}
VW Cas       &  5.99 &  17 &  3   & 1.22 & 13.20 & 3.403 & 0.314 & 0.329 & 0.088 &       &       & 2cd  \\
             &       &     &      &      &  0.44 & 0.106 & 0.035 & 0.444 & 0.034 &       &       &      \\
\noalign{\medskip}
RV Sco       &  6.06 &  53 &  4   & 2.20 & 13.49 & 3.204 & 0.578 & 0.333 & 0.238 & 3.842 & 0.044 & 2e   \\
             &       &     &      &      &  0.46 & 0.086 & 0.041 & 0.206 & 0.031 & 0.766 & 0.040 &      \\
\noalign{\medskip}
KL Aql       &  6.11 &  73 &  7   & 0.82 & 13.91 & 3.328 & 0.422 & 0.645 & 0.243 & 4.143 & 0.096 & 2b   \\
             &       &     &      &      &  0.15 & 0.032 & 0.012 & 0.052 & 0.012 & 0.113 & 0.011 &      \\
\noalign{\medskip}
RS Cas       &  6.30 &  76 &  5   & 1.47 & 15.37 & 3.218 & 0.461 & 0.421 & 0.235 & 3.725 & 0.105 & 2b   \\
             &       &     &      &      &  0.26 & 0.058 & 0.018 & 0.089 & 0.017 & 0.174 & 0.018 &      \\
\noalign{\medskip}
X Vul        &  6.32 &  97 &  6   & 0.89 & 14.05 & 3.210 & 0.464 & 0.302 & 0.240 & 3.690 & 0.101 & 2b   \\
             &       &     &      &      &  0.13 & 0.030 & 0.010 & 0.049 & 0.010 & 0.103 & 0.010 &      \\
\noalign{\medskip}
S TrA        &  6.32 &  65 &  3   & 2.67 & 13.79 & 3.174 & 0.426 & 0.817 & 0.175 &       &       & 2e   \\
             &       &     &      &      &  0.46 & 0.107 & 0.037 & 0.217 & 0.036 &       &       &      \\
\noalign{\medskip}
V Car        &  6.70 &  43 &  3   & 3.01 & 11.77 & 3.736 & 0.500 & 1.136 & 0.126 &       &       & 2e   \\
             &       &     &      &      &  0.70 & 0.161 & 0.057 & 0.485 & 0.057 &       &       &      \\
\noalign{\medskip}
TW Mon       &  7.10 &  24 &  5   & 0.37 & 12.89 & 3.517 & 0.510 & 0.883 & 0.121 & 4.092 & 0.076 & 2cd  \\
             &       &     &      &      &  0.12 & 0.025 & 0.013 & 0.099 & 0.009 & 0.133 & 0.010 &      \\
\noalign{\medskip}
V336 Aql     &  7.30 &  48 &  5   & 1.08 & 14.43 & 3.626 & 0.541 & 1.217 & 0.195 & 4.458 & 0.095 & 2b   \\
             &       &     &      &      &  0.23 & 0.043 & 0.018 & 0.094 & 0.017 & 0.174 & 0.016 &      \\
\noalign{\medskip}
CK Sct       &  7.42 &  60 &  2+t & 1.68 & 10.08 & 3.547 & 0.464 &       &       &       &       & 2e   \\
             &       &     &      &      &  0.31 & 0.098 & 0.034 &       &       &       &       &      \\
\noalign{\medskip}
V510 Mon     &  7.46 &  20 &  4   & 0.27 &  9.94 & 3.616 & 0.405 & 0.970 & 0.054 & 3.699 & 0.029 & 2cd  \\
             &       &     &      &      &  0.09 & 0.031 & 0.010 & 0.159 & 0.011 & 0.399 & 0.009 &      \\
\noalign{\medskip}
R Mus        &  7.51 &  70 &  5   & 0.21 & 14.15 & 3.575 & 0.545 & 1.046 & 0.191 & 3.842 & 0.091 & 2d   \\
             &       &     &      &      &  0.05 & 0.007 & 0.006 & 0.032 & 0.003 & 0.039 & 0.003 &      \\
\noalign{\medskip}
CD Cas       &  7.80 &  20 &  3   & 1.48 & 14.28 & 3.630 & 0.485 & 1.094 & 0.163 &       &       & 2d   \\
             &       &     &      &      &  0.52 & 0.093 & 0.041 & 0.236 & 0.036 &       &       &      \\
\noalign{\medskip}
V2340 Cyg    &  7.97 &  55 &  4   & 1.12 & 11.31 & 4.024 & 0.528 & 1.250 & 0.072 & 3.986 & 0.126 & 2c   \\
             &       &     &      &      &  0.21 & 0.065 & 0.024 & 0.296 & 0.021 & 0.176 & 0.026 &      \\
\noalign{\medskip}
BK Aur       &  8.00 &  44 &  4+t & 1.48 & 14.07 & 3.833 & 0.501 & 1.293 & 0.127 & 4.020 & 0.113 & 2c   \\
             &       &     &      &      &  0.42 & 0.061 & 0.027 & 0.216 & 0.027 & 0.273 & 0.050 &      \\
\noalign{\medskip}
V339 Cen     &  9.47 &  54 &  3+t & 2.54 & 12.09 & 4.707 & 0.401 & 1.380 & 0.169 &       &       & 2e   \\
             &       &     &      &      &  0.53 & 0.128 & 0.047 & 0.281 & 0.042 &       &       &      \\
\noalign{\medskip}
BZ Cyg       & 10.14 &  83 &  3+t & 2.03 & 12.09 & 4.394 & 0.420 & 1.375 & 0.125 &       &       & 2e   \\
             &       &     &      &      &  0.35 & 0.080 & 0.031 & 0.236 & 0.027 &       &       &      \\
\noalign{\medskip}
AN Aur       & 10.29 &  44 &  5   & 1.46 & 14.09 & 4.291 & 0.522 & 1.177 & 0.185 & 5.454 & 0.084 & 2c   \\
             &       &     &      &      &  0.38 & 0.105 & 0.030 & 0.189 & 0.024 & 0.446 & 0.030 &      \\
\noalign{\smallskip}
\noalign{\eject}
\noalign{\medskip}
DR Vel       & 11.20 &  38 &  3   & 2.54 & 17.51 &  ---  &  ---  & 2.685 & 0.099 &       &       & 2e   \\
             &       &     &      &      &  0.59 &  ---  &  ---  & 0.381 & 0.034 &       &       &      \\
\noalign{\medskip}
V438 Cyg     & 11.21 &  57 &  4   & 2.10 & 20.77 & 4.904 & 0.140 & 2.377 & 0.141 & 5.117 & 0.064 & 2e   \\
             &       &     &      &      &  0.43 & 0.136 & 0.021 & 0.149 & 0.019 & 0.320 & 0.020 &      \\
\noalign{\medskip}
U Nor        & 12.64 &  30 &  5   & 0.66 & 17.84 & 5.008 & 0.115 & 2.032 & 0.128 & 4.183 & 0.034 & 2c   \\
             &       &     &      &      &  0.19 & 0.096 & 0.014 & 0.083 & 0.014 & 0.393 & 0.011 &      \\
\noalign{\medskip}
V916 Aql     & 13.44 & 134 &  6+1 & 1.63 & 23.07 & 2.943 & 0.187 & 3.190 & 0.051 & 0.159 & 0.071 & 2be  \\
             &       &     &      &      &  0.22 & 0.050 & 0.009 & 0.177 & 0.009 & 0.139 & 0.009 &      \\
\noalign{\medskip}
HW Pup       & 13.46 &  28 &  4   & 0.73 & 19.74 & 4.261 & 0.122 & 2.012 & 0.125 & 4.935 & 0.049 & 2c   \\
             &       &     &      &      &  0.21 & 0.082 & 0.012 & 0.106 & 0.010 & 0.231 & 0.011 &      \\
\noalign{\medskip}
BN Pup       & 13.67 &  34 &  6   & 0.95 & 25.49 & 2.960 & 0.232 & 3.601 & 0.015 & 6.126 & 0.070 & 2bc  \\
             &       &     &      &      &  0.25 & 0.051 & 0.011 & 0.729 & 0.010 & 0.169 & 0.010 &      \\
\noalign{\medskip}
UZ Sct       & 14.75 &  43 &  7   & 0.58 & 21.94 & 2.932 & 0.205 & 3.813 & 0.047 & 0.209 & 0.079 & 2bc  \\
             &       &     &      &      &  0.15 & 0.033 & 0.007 & 0.159 & 0.006 & 0.096 & 0.006 &      \\
\noalign{\medskip}
RW Cas       & 14.79 &  90 &  7   & 1.91 & 24.69 & 3.010 & 0.197 & 3.453 & 0.037 & 0.012 & 0.091 & 2e   \\
             &       &     &      &      &  0.30 & 0.068 & 0.012 & 0.345 & 0.012 & 0.141 & 0.012 &      \\
\noalign{\medskip}
AV Sgr       & 15.41 &  36 &  5   & 1.30 & 24.84 & 3.007 & 0.228 & 4.122 & 0.041 & 6.129 & 0.096 & 2b   \\
             &       &     &      &      &  0.42 & 0.077 & 0.021 & 0.457 & 0.016 & 0.202 & 0.018 &      \\
\noalign{\medskip}
XX Car       & 15.71 &  60 &  2   & 2.77 & 20.37 & 2.822 & 0.221 &       &       &       &       & 2e   \\
             &       &     &      &      &  0.54 & 0.117 & 0.027 &       &       &       &       &      \\
\noalign{\medskip}
RW Cam       & 16.41 &  72 &  7+t & 1.27 & 22.48 & 3.002 & 0.272 & 4.863 & 0.063 & 0.905 & 0.089 & 2b   \\
             &       &     &      &      &  0.26 & 0.043 & 0.012 & 0.158 & 0.011 & 0.124 & 0.010 &      \\
\noalign{\medskip}
CP Cep       & 17.86 &  43 &  6   & 1.02 & 20.68 & 2.929 & 0.244 & 4.911 & 0.062 & 0.952 & 0.108 & 2bc  \\
             &       &     &      &      &  0.25 & 0.052 & 0.012 & 0.194 & 0.012 & 0.134 & 0.012 &      \\
\noalign{\medskip}
YZ Aur       & 18.19 &  46 &  3   & 1.80 & 24.10 & 3.120 & 0.254 & 0.087 & 0.047 &       &       & 2ce  \\
             &       &     &      &      &  0.37 & 0.077 & 0.017 & 0.347 & 0.018 &       &       &      \\
\noalign{\medskip}
VX Cyg       & 20.14 &  66 &  7   & 1.45 & 23.07 & 3.016 & 0.321 & 5.622 & 0.088 & 1.051 & 0.076 & 2b   \\
             &       &     &      &      &  0.28 & 0.048 & 0.012 & 0.137 & 0.012 & 0.156 & 0.012 &      \\
\noalign{\medskip}
BM Per       & 22.96 &  61 &  8   & 1.43 & 25.79 & 3.086 & 0.322 & 5.944 & 0.152 & 2.240 & 0.096 & 2b   \\
             &       &     &      &      &  0.39 & 0.054 & 0.012 & 0.094 & 0.015 & 0.153 & 0.012 &      \\
\noalign{\medskip}
WZ Car       & 23.02 &  96 &  8   & 0.99 & 24.44 & 3.043 & 0.337 & 5.721 & 0.145 & 2.106 & 0.116 & 2bc  \\
             &       &     &      &      &  0.19 & 0.024 & 0.008 & 0.070 & 0.008 & 0.082 & 0.007 &      \\
\noalign{\medskip}
VZ Pup       & 23.18 &  83 &  7   & 0.65 & 22.71 & 3.015 & 0.289 & 5.719 & 0.107 & 2.018 & 0.066 & 2b   \\
             &       &     &      &      &  0.12 & 0.024 & 0.006 & 0.050 & 0.005 & 0.093 & 0.006 &      \\
\noalign{\medskip}
SW Vel       & 23.43 &  46 & 11   & 0.37 & 25.80 & 3.062 & 0.380 & 5.913 & 0.165 & 2.216 & 0.115 & 2b   \\
             &       &     &      &      &  0.09 & 0.010 & 0.004 & 0.024 & 0.003 & 0.032 & 0.003 &      \\
\noalign{\medskip}
RY Vel       & 28.12 &  61 &  3   & 2.13 & 15.90 & 2.771 & 0.249 & 5.449 & 0.091 &       &       & 2be  \\
             &       &     &      &      &  0.39 & 0.109 & 0.025 & 0.262 & 0.026 &       &       &      \\
\noalign{\medskip}
U Car        & 38.84 & 104 &  7   & 0.59 & 20.88 & 3.056 & 0.387 & 6.134 & 0.216 & 3.011 & 0.133 & 2b   \\
             &       &     &      &      &  0.15 & 0.020 & 0.006 & 0.035 & 0.007 & 0.070 & 0.006 &      \\
\noalign{\smallskip}
\hline

\end{longtable}
\end{center}
\begin{tablenotes}
    \item \textbf{Notes :} Same as Table \ref{Tab:data_funda_quality1}. The last column of the Table indicates the quality classification for each star, with the following flags:
    \item b : unstable fit
    \item c : poor coverage
    \item d : low number of data points
    \item e : standard deviation larger than 1.5$\,$km/s
\end{tablenotes}

\onecolumn
\begin{center}
\begin{longtable}{lSS[table-format=2.0]lcS[table-format=2.2]ccccccl}
\caption{Low-order Fourier parameters of first-overtone Cepheids. \label{Tab:data_fourier_quality1_overtone}} \\
\hline
\hline
\hline
\noalign{\smallskip}
\multicolumn{1}{l}{Star} &
\multicolumn{1}{S}{{Period}} &
\multicolumn{1}{r}{Ndat} &
\multicolumn{1}{l}{$n$} &
\multicolumn{1}{c}{$\sigma_\mathrm{fit}$} &
\multicolumn{1}{c}{$A_1$} &
\multicolumn{1}{c}{$\phi_{21}$} &
\multicolumn{1}{c}{$R_{21}$} &
\multicolumn{1}{c}{$\phi_{31}$} &
\multicolumn{1}{c}{$R_{31}$} &
\multicolumn{1}{c}{$\phi_{41}$} &
\multicolumn{1}{c}{$R_{41}$} &
\multicolumn{1}{l}{flag} \\
\hline
\endfirsthead

\multicolumn{13}{c}%
{{\bfseries \tablename\ \thetable{} -- continued from previous page}} \\
\noalign{\medskip}
\hline
\hline
\noalign{\smallskip}
\multicolumn{1}{l}{Star} &
\multicolumn{1}{S}{{Period}} &
\multicolumn{1}{r}{Ndat} &
\multicolumn{1}{l}{$n$} &
\multicolumn{1}{c}{$\sigma_\mathrm{fit}$} &
\multicolumn{1}{c}{$A_1$} &
\multicolumn{1}{c}{$\phi_{21}$} &
\multicolumn{1}{c}{$R_{21}$} &
\multicolumn{1}{c}{$\phi_{31}$} &
\multicolumn{1}{c}{$R_{31}$} &
\multicolumn{1}{c}{$\phi_{41}$} &
\multicolumn{1}{c}{$R_{41}$} &
\multicolumn{1}{l}{flag} \\
\hline
\endhead

\hline
\noalign{\vskip 0.065truecm}
\multicolumn{2}{r}{{Continued on next page}} \\
\hline
\endfoot

\hline
\hline
\endlastfoot
\noalign{\medskip}
SU Cas       & 1.95 & 298 & 3   & 0.59 &  8.74 & 2.802 & 0.204 & 5.495 & 0.042 &       &       & 1    \\
             &      &     &     &      &  0.05 & 0.030 & 0.006 & 0.136 & 0.006 &       &       &      \\
\noalign{\medskip}
EU Tau       & 2.10 & 122 & 3   & 0.89 &  8.33 & 2.869 & 0.209 & 5.987 & 0.081 &       &       & 1    \\
             &      &     &     &      &  0.11 & 0.075 & 0.014 & 0.178 & 0.013 &       &       &      \\
\noalign{\medskip}
IR Cep       & 2.11 & 181 & 3+t & 0.93 &  9.69 & 2.873 & 0.240 & 5.619 & 0.055 &       &       & 1    \\
             &      &     &     &      &  0.10 & 0.046 & 0.011 & 0.190 & 0.010 &       &       &      \\
\noalign{\medskip}
BP Cir       & 2.40 &  78 & 3   & 0.22 &  7.94 & 2.928 & 0.203 & 5.897 & 0.053 &       &       & 1    \\
             &      &     &     &      &  0.04 & 0.023 & 0.005 & 0.085 & 0.005 &       &       &      \\
\noalign{\medskip}
UY Mon       & 2.40 &  49 & 3   & 0.69 &  9.06 & 2.904 & 0.293 & 5.499 & 0.089 &       &       & 1a   \\
             &      &     &     &      &  0.14 & 0.063 & 0.017 & 0.176 & 0.017 &       &       &      \\
\noalign{\medskip}
LR TrA       & 2.43 &  32 & 3+t & 0.11 &  5.53 & 3.076 & 0.120 & 5.826 & 0.021 &       &       & 1    \\
             &      &     &     &      &  0.03 & 0.049 & 0.005 & 0.263 & 0.006 &       &       &      \\
\noalign{\medskip}
DT Cyg       & 2.50 & 147 & 2+t & 0.54 &  6.72 & 2.936 & 0.161 &       &       &       &       & 1    \\
             &      &     &     &      &  0.06 & 0.061 & 0.009 &       &       &       &       &      \\
\noalign{\medskip}
V351 Cep     & 2.81 &  83 & 3+t & 1.03 &  8.29 & 3.072 & 0.211 & 6.199 & 0.054 &       &       & 1    \\
             &      &     &     &      &  0.18 & 0.107 & 0.021 & 0.429 & 0.020 &       &       &      \\
\noalign{\medskip}
V411 Lac     & 2.91 &  72 & 3+t & 1.14 &  8.38 & 3.386 & 0.254 & 0.658 & 0.058 &       &       & 1    \\
             &      &     &     &      &  0.20 & 0.100 & 0.024 & 0.462 & 0.023 &       &       &      \\
\noalign{\medskip}
AV Cir       & 3.07 & 134 & 3   & 0.26 &  7.16 & 3.005 & 0.199 & 5.694 & 0.045 &       &       & 1    \\
             &      &     &     &      &  0.03 & 0.025 & 0.004 & 0.107 & 0.005 &       &       &      \\
\noalign{\medskip}
EV Sct       & 3.09 & 111 & 3   & 0.58 &  7.74 & 3.303 & 0.200 & 0.454 & 0.022 &       &       & 1    \\
             &      &     &     &      &  0.08 & 0.061 & 0.010 & 0.516 & 0.010 &       &       &      \\
\noalign{\medskip}
VZ CMa       & 3.13 &  56 & 4   & 0.50 &  9.71 & 3.190 & 0.255 & 6.094 & 0.081 & 2.789 & 0.038 & 1    \\
             &      &     &     &      &  0.09 & 0.045 & 0.010 & 0.128 & 0.010 & 0.271 & 0.010 &      \\
\noalign{\medskip}
DX Gem       & 3.14 &  52 & 2+1 & 1.15 &  9.55 & 3.489 & 0.233 &       &       &       &       & 1    \\
             &      &     &     &      &  0.24 & 0.102 & 0.028 &       &       &       &       &      \\
\noalign{\medskip}
SZ Tau       & 3.15 & 146 & 3   & 1.01 &  9.02 & 3.306 & 0.268 & 0.177 & 0.068 &       &       & 1    \\
             &      &     &     &      &  0.12 & 0.051 & 0.014 & 0.201 & 0.013 &       &       &      \\
\noalign{\medskip}
AZ Cen       & 3.21 &  46 & 3   & 0.46 &  7.58 & 3.153 & 0.198 & 5.967 & 0.061 &       &       & 1    \\
             &      &     &     &      &  0.11 & 0.071 & 0.014 & 0.204 & 0.014 &       &       &      \\
\noalign{\medskip}
BY Cas       & 3.22 & 120 & 2+2 & 2.79 &  9.21 & 3.278 & 0.228 &       &       &       &       & 2e   \\
             &      &     &     &      &  0.36 & 0.189 & 0.042 &       &       &       &       &      \\
\noalign{\medskip}
V532 Cyg     & 3.28 & 132 & 4   & 1.19 &  8.68 & 3.306 & 0.226 & 6.221 & 0.083 & 3.129 & 0.021 & 1    \\
             &      &     &     &      &  0.16 & 0.084 & 0.018 & 0.212 & 0.018 & 0.841 & 0.017 &      \\
\noalign{\medskip}
V1334 Cyg    & 3.33 &  70 & 2+3 & 0.08 &  4.09 & 3.233 & 0.099 &       &       &       &       & 1    \\
             &      &     &     &      &  0.01 & 0.041 & 0.003 &       &       &       &       &      \\
\noalign{\medskip}
BG Cru       & 3.34 &  91 & 3   & 0.14 &  4.83 & 3.197 & 0.118 & 5.671 & 0.047 &       &       & 1    \\
             &      &     &     &      &  0.03 & 0.052 & 0.006 & 0.118 & 0.006 &       &       &      \\
\noalign{\medskip}
V950 Sco     & 3.38 &  36 & 3   & 0.33 &  7.93 & 3.286 & 0.217 & 0.061 & 0.055 &       &       & 1    \\
             &      &     &     &      &  0.09 & 0.051 & 0.012 & 0.183 & 0.012 &       &       &      \\
\noalign{\medskip}
FZ Car       & 3.58 &  43 & 3   & 0.76 &  8.07 & 3.508 & 0.221 & 0.177 & 0.039 &       &       & 1    \\
             &      &     &     &      &  0.17 & 0.094 & 0.024 & 0.547 & 0.021 &       &       &      \\
\noalign{\medskip}
QZ Nor       & 3.79 & 125 & 4+t & 0.19 &  7.70 & 3.596 & 0.200 & 0.803 & 0.046 & 4.098 & 0.016 & 1    \\
             &      &     &     &      &  0.02 & 0.017 & 0.003 & 0.070 & 0.003 & 0.218 & 0.003 &      \\
\noalign{\medskip}
$\alpha$ UMi & 3.97 & 167 & 2   & 0.09 &  1.26 & 3.736 & 0.061 &       &       &       &       & 1    \\
             &      &     &     &      &  0.01 & 0.113 & 0.012 &       &       &       &       &      \\
\noalign{\smallskip}
\noalign{\eject}
\noalign{\medskip}
AH Vel       & 4.23 & 101 & 3   & 0.27 &  8.12 & 3.676 & 0.141 & 0.906 & 0.025 &       &       & 1    \\
             &      &     &     &      &  0.05 & 0.070 & 0.006 & 0.257 & 0.006 &       &       &      \\
\noalign{\medskip}
V379 Cas     & 4.31 & 222 & 2   & 1.55 &  8.78 & 3.403 & 0.145 &       &       &       &       & 2e   \\
             &      &     &     &      &  0.15 & 0.118 & 0.018 &       &       &       &       &      \\
\noalign{\medskip}
GI Car       & 4.43 &  50 & 2   & 0.48 &  7.77 & 3.494 & 0.111 &       &       &       &       & 1    \\
             &      &     &     &      &  0.10 & 0.106 & 0.013 &       &       &       &       &      \\
\noalign{\medskip}
FF Aql       & 4.47 &  94 & 4+3 & 0.08 &  7.87 & 3.477 & 0.124 & 0.222 & 0.033 & 3.143 & 0.010 & 1    \\
             &      &     &     &      &  0.01 & 0.012 & 0.002 & 0.047 & 0.002 & 0.158 & 0.002 &      \\
\noalign{\medskip}
V335 Pup     & 4.86 &  84 & 2   & 0.07 &  5.02 & 2.984 & 0.052 &       &       &       &       & 1    \\
             &      &     &     &      &  0.01 & 0.044 & 0.002 &       &       &       &       &      \\
\noalign{\medskip}
X Lac        & 5.44 &  48 & 2   & 0.55 & 10.63 & 2.728 & 0.163 &       &       &       &       & 1    \\
             &      &     &     &      &  0.12 & 0.074 & 0.011 &       &       &       &       &      \\
\noalign{\medskip}
MY Pup       & 5.69 & 143 & 2   & 0.29 &  4.57 & 2.534 & 0.091 &       &       &       &       & 1    \\
             &      &     &     &      &  0.03 & 0.087 & 0.008 &       &       &       &       &      \\
\noalign{\medskip}
GH Car       & 5.73 &  43 & 3   & 0.17 &  8.34 & 2.733 & 0.124 & 5.010 & 0.025 &       &       & 1    \\
             &      &     &     &      &  0.04 & 0.037 & 0.005 & 0.169 & 0.005 &       &       &      \\
\noalign{\medskip}
V495 Cyg     & 6.72 &  83 & 3   & 1.42 & 12.42 & 2.735 & 0.288 & 5.314 & 0.072 &       &       & 1    \\
             &      &     &     &      &  0.25 & 0.075 & 0.018 & 0.250 & 0.019 &       &       &      \\
\noalign{\medskip}
V440 Per     & 7.57 & 158 & 2   & 0.13 &  2.48 & 2.754 & 0.056 &       &       &       &       & 1    \\
             &      &     &     &      &  0.01 & 0.112 & 0.006 &       &       &       &       &      \\
\noalign{\smallskip}
\hline

\end{longtable}
\end{center}
\begin{tablenotes}
    \item \textbf{Notes :} Same as Table \ref{Tab:data_funda_quality1}. The last column of the Table indicates the quality classification for each star,  with the same flags as in previous Tables.
    \item a : slightly unstable fit
    \item e : standard deviation above 1.5km/s
\end{tablenotes}

\subsection{Comments on individual overtone Cepheids}\label{app:overtone}

\textbf{VZ CMa, EV Sct:} the overtone nature of these two Cepheids has been
suspected for a long time \citep{AntonelloPoretti1990}. However, with
periods close to 3.1$\,$day, their light curve Fourier phases $\phi_{21}$ fall
on the intersection of the overtone and the fundamental-mode
progressions. Consequently, their mode of pulsation cannot be
established on the basis of their light curves. On the other hand,
at this period range the pulsation mode can easily be identified
with the radial velocity $\phi_{21}$. In this paper, we show that both VZ~CMa and EV~Sct pulsate in the first-overtone mode. Our results confirm mode identifications of the {\it Gaia} DR3 catalog.
\newline
\newline
\noindent \textbf{V532 Cyg:} another Cepheid ($P=3.3\,$day) for which the pulsation mode
cannot be established with its light curve. The star has been
identified as an overtone pulsator with the radial velocity curve by
\cite{Kienzle1999}, who referred to unpublished results of Krzyt,
Moskalik, Gorynya et al. (1999). We confirm overtone nature of V532~Cyg and provide its RV Fourier parameters for the first time. Our result confirms mode identification of the {\it Gaia} DR3 catalog.
\newline
\newline
\noindent \textbf{V351 Cep:} this is another Cepheid identified as an overtone pulsator
by \cite{Kienzle1999}, who referred to unpublished results of
Krzyt, Moskalik, Gorynya et al. (1999). We confirm overtone nature
of V351~Cep and provide its RV Fourier parameters for the first
time. This result also confirms the mode identification of the {\it Gaia} DR3 catalog.
\newline
\newline
\textbf{V411 Lac:} on the basis of radial velocity Fourier phase $\phi_{21}$, we
classify this Cepheid as an overtone pulsator. This result also confirms the mode identification of the {\it Gaia} DR3 catalog.
\newline
\newline
\noindent \textbf{DT Cyg:} mode identification with the light curve is not possible for this Cepheid,
because its $V$-band light curve is perfectly sinusoidal, with no detectable
harmonic \cite{Poretti1994}. On the basis of radial velocity Fourier phase $\phi_{21}$, we classify DT~Cyg as an overtone pulsator.
\newline
\newline
\noindent \textbf{V335 Pup:} because of varying pulsation amplitude, only data collected after HJD=2456300 are used in the analysis.

\subsection{Comments on individual Cepheids of Table \ref{Tab:unknown}}\label{app:unknown}

\noindent \textbf{CK Cam:} {\it Gaia} DR3 catalog classifies this star as a fundamental-mode Cepheid. This is also supported by the derived value of ABL (see Fig.~\ref{fig:ABL}). The RV Fourier amplitude $A_1$ and amplitude ratios $R_21$ and $R_31$ for this Cepheid are indeed typical for a fundamental-mode pulsator. On the other hand, Fourier phase $\phi_{21}$ fits the overtone progression, although the deviation from the fundamental-mode progression is not large (less than 2$\sigma$). In light of these conflicting evidence we conclude that the pulsation mode of CK Cam cannot be firmly identified with the available RV data.
\newline
\newline

\noindent \textbf{V1726 Cyg:} {\it Gaia} DR3 catalog classifies this star as an overtone Cepheid, which is also supported by the derived value of ABL (see Fig.~\ref{fig:ABL}). The radial velocity curve of V1726~Cyg is perfectly sinusoidal, with no detectable harmonic. This is also the case for the light-curve of this star \cite{Poretti1994}. Thus, pulsation mode of V1726 Cyg cannot be established with the shape of its RV curve or the light curve.
\newline
\newline
\noindent \textbf{V636 Cas:} this low-amplitude variable is usually considered to be a Population I Cepheid. {\it Gaia} DR3 catalog classifies this star as a fundamental-mode pulsator.  The derived value of ABL agrees with this classification (see Fig.~\ref{fig:ABL}). However, our analysis of its radial velocity curve does not support such a classification. The value of RV Fourier phase $\phi_{21}$ is very high, 7.9$\sigma$ above the progression of fundamental-mode Cepheids and 23.8$\sigma$ above the progression of overtone Cepheids. Thus, it does not fit any mode of Cepheid pulsation. The phased RV curve of V636 Cas is distinctly non-sinusoidal, yet it is very symmetric, with descending branch lasting $0.46P$. The curve does not resemble any RV curve of Cepheids of our sample. In our opinion, V636 Cas is not a Pop. I Cepheid. Understanding its true nature requires an in-depth analysis of all available data, which is beyond the scope of this paper.
\newline
\newline
\noindent \textbf{V924 Cyg:} 
This star is classified in {\it Gaia} DR3 catalog as fundamental-mode pulsator, but reclassified by \cite{Ripepi2023DR3} as
overtone pulsator. The derived value of ABL agrees with the latter
   classification (see Fig.~\ref{fig:ABL}). The overtone nature of this Cepheids has been suspected in the past \citep{AntonelloPoretti1990}. V924~Cyg
   is located in the light-curve Fourier plots at the period where
   the first overtone and the fundamental-mode progressions merge.
   Consequently, its pulsation mode cannot be identified with the
   light-curve Fourier phases. Our analysis of the RV curve yields
   Fourier phase $\phi_{21}$, which fits very well to the fundamental-mode progression. Unfortunately, the error is large (0.29 rad) and the
   measured $\phi_{21}$ deviates from the overtone progression by only
   1.7$\sigma$. Therefore, a firm mode identification of V924~Cyg is
   not possible.
\newline
\newline
\noindent \textbf{BD Cas:} {\it Gaia} DR3 catalog classifies this star as an overtone Cepheid, which is also supported by the derived value of
ABL (see Fig.~\ref{fig:ABL}). The star was classified as a first-overtone pulsator by \cite{Poretti1994}. Its mode identification was based on the light-curve Fourier parameters, in particular on $\phi_{21}$.
However, Poretti's classification cannot be regarded as
definitive, because the photometric data was scarce (only 16 points), the light-curve harmonic was essentially insignificant
($A_2=0.016\pm 0.014$) and the resulting error of $\phi_{21}$ was very large (0.72 rad). The light-curve $\phi_{21}$ was consistent within this error with either mode of pulsation. Our analysis of the RV curve yields somewhat better results. The Fourier parameters $A_1$
and $R_{21}$ are both low and consistent with BD~Cas being an overtone pulsator. Nevertheless, we have to keep in mind that low values of $A_1$ and $R_{21}$ do not prove an overtone nature of a Cepheid; fundamental-mode Cepheids can have low amplitudes, too (e.g. V1344~Aql). The value of RV Fourier phase $\phi_{21}$ deviates from the
overtone progression by 2.7$\sigma$ and agrees within 1$\sigma$ with the fundamental-mode progression. 2.7$\sigma$ deviation is somewhat too low for a firm mode identification, but the data indicate that BD Cas might be a fundamental-mode Cepheid, rather than an overtone Cepheid. This conclusion needs to be confirmed with better RV and photometric data. At this point we consider mode identification of BD Cas as uncertain.
\newline
\newline
\noindent  \textbf{CE Cas A:} According to {\it Gaia} DR3 catalogue, this star is as a
   fundamental-mode pulsator. Its ABL value is also consistent with the fundamental-mode (see Fig.~\ref{fig:ABL}) which is also shown from \cite{Reyes2023}. The parallax of this is Cepheid is also robust as it is a member of an open cluster \citep{Reyes2023}. For this Cepheid the RV Fourier amplitude $A_1$ and amplitude
ratio $R_{21}$ are high and rather typical for a fundamental-mode
pulsator. On the other hand, the value of its Fourier phase $\phi_{21}$ is
low. It is 3.5$\sigma$ below the fundamental-mode progression and is
similar to values found for long-period overtone Cepheids. Our mode identification is only tentative, because it is based on radial velocity curve of low quality. Therefore, it needs to be confirmed
with better data. 
\newline
\newline
\noindent  \textbf{GH Lup:} {\it Gaia} DR3 catalog classifies this star as a fundamental-mode Cepheid, which is also supported by the derived value of ABL (see Fig.~\ref{fig:ABL}). The radial velocity curve of this low-amplitude Cepheid is perfectly sinusoidal, with no detectable harmonic. Thus, its pulsation mode cannot be identified with the observed RV curve and remains unknown. However, with the period of 9.3$\,$day, GH~Lup is rather unlikely to pulsate in the first overtone. We recall that the longest pulsation period for securely identified first-overtone Cepheid is 7.57 day (for V440 Per) \citep{Baranowski2009}.

\onecolumn
\begin{center}
\begin{longtable}{lSS[table-format=2.0]lcS[table-format=2.2]ccccll}
\caption{Low-order Fourier parameters of Cepheids with uncertain or unknown pulsation mode.} \label{Tab:unknown} \\
\hline
\hline
\hline
\noalign{\smallskip}
\multicolumn{1}{l}{Star} &
\multicolumn{1}{S}{{Period}} &
\multicolumn{1}{r}{Ndat} &
\multicolumn{1}{l}{$n$} &
\multicolumn{1}{c}{$\sigma_\mathrm{fit}$} &
\multicolumn{1}{c}{$A_1$} &
\multicolumn{1}{c}{$\phi_{21}$} &
\multicolumn{1}{c}{$R_{21}$} &
\multicolumn{1}{c}{$\phi_{31}$} &
\multicolumn{1}{c}{$R_{31}$} &
\multicolumn{1}{c}{flag} &
\multicolumn{1}{c}{Pulsation Mode} \\
\hline
\endfirsthead

\multicolumn{12}{c}%
{{\bfseries \tablename\ \thetable{} -- continued from previous page}} \\
\noalign{\medskip}
\hline
\hline
\noalign{\smallskip}
\multicolumn{1}{l}{Star} &
\multicolumn{1}{S}{{Period}} &
\multicolumn{1}{r}{Ndat} &
\multicolumn{1}{l}{$n$} &
\multicolumn{1}{c}{$\sigma_\mathrm{fit}$} &
\multicolumn{1}{c}{$A_1$} &
\multicolumn{1}{c}{$\phi_{21}$} &
\multicolumn{1}{c}{$R_{21}$} &
\multicolumn{1}{c}{$\phi_{31}$} &
\multicolumn{1}{c}{$R_{21}$} &
\multicolumn{1}{c}{flag} &
\multicolumn{1}{c}{Pulsation Mode} \\
\hline
\endhead

\hline
\noalign{\vskip 0.065truecm}
\multicolumn{2}{r}{{Continued on next page}} \\
\hline
\endfoot

\hline
\hline
\endlastfoot
\noalign{\medskip}
CK Cam    &  3.29 &  46 & 3   & 1.10 & 14.57 & 3.183 & 0.316 & 6.018 & 0.101 & 1  & \,\,\,\,\,\,\,\,\,\, F \\
          &       &     &     &      &  0.25 & 0.063 & 0.018 & 0.167 & 0.017 &    &                            \\
\noalign{\medskip}
BD Cas    &  3.65 &  55 & 2   & 1.46 &  8.82 & 2.812 & 0.142 &       &       & 1  & \,\,\,\,\,\,\,\,\,\, 1O   \\
          &       &     &     &      &  0.28 & 0.263 & 0.030 &       &       &    &                            \\
\noalign{\medskip}
V1726 Cyg &  4.24 &  98 & 1+t & 0.87 &  4.15 &       &       &       &       & 1  & \,\,\,\,\,\,\,\,\,\, 1O     \\
          &       &     &     &      &  0.13 &       &       &       &       &    &                            \\
\noalign{\medskip}
CE Cas A  &  5.14 &  27 & 3   & 1.88 & 13.95 & 2.482 & 0.284 & 5.338 & 0.074 & 2e & \,\,\,\,\,\,\,\,\,\, F  \\
          &       &     &     &      &  0.55 & 0.181 & 0.043 & 0.529 & 0.040 &    &                            \\
\noalign{\medskip}
V924 Cyg  &  5.57 & 196 & 2+t & 1.81 &  7.34 & 3.222 & 0.092 &       &       & 2e & \,\,\,\,\,\,\,\,\,\, 1O  \\
          &       &     &     &      &  0.19 & 0.290 & 0.025 &       &       &    &                            \\
\noalign{\medskip}
V636 Cas  &  8.38 & 128 & 2   & 0.53 &  4.71 & 4.551 & 0.196 &       &       & 1  & \,\,\,\,\,\,\,\,\,\, F     \\
          &       &     &     &      &  0.07 & 0.076 & 0.015 &       &       &    &                            \\
\noalign{\medskip}
GH Lup    &  9.28 &  36 & 1+t & 1.91 &  4.85 &       &       &       &       & 2e & \,\,\,\,\,\,\,\,\,\, F   \\
          &       &     &     &      &  0.49 &       &       &       &       &    &                            \\
\noalign{\smallskip}
\hline
\hline

\end{longtable}
\end{center}
\begin{tablenotes}
\item \textbf{Notes :} In the last column we indicate the most likely mode identification for each star, based when possible, from comparison with ABL-PL relation from \cite{Ripepi2023DR3} (see Fig.~\ref{fig:ABL}).
\end{tablenotes}

\end{appendix}
\end{document}